%% file: TPLP.tex
\documentclass{new_tlp}

\makeatletter
\let\O@argtabularcr\@argtabularcr
\def\O@xtabularcr{\@ifnextchar[\O@argtabularcr{\ifnum 0=`{\fi}\cr}}
\let\O@tabacol\@tabacol
\let\O@tabclassiv\@tabclassiv
\let\O@tabclassz\@tabclassz
\let\O@tabarray\@tabarray
\def\author@tabular{\authorsize\def\@halignto{}\@authortable}
\let\endauthor@tabular=\endtabular
\def\author@tabcrone{{\ifnum0=`}\fi\O@xtabularcr\affilsize\itshape
 \let\\=\author@tabcrtwo\ignorespaces}
\def\author@tabcrtwo{{\ifnum0=`}\fi\O@xtabularcr[-3\p@]\affilsize\itshape
 \let\\=\author@tabcrtwo\ignorespaces}
\def\@authortable{\leavevmode \hbox \bgroup $\let\@acol\O@tabacol
 \let\@classz\O@tabclassz \let\@classiv\O@tabclassiv
 \let\\=\author@tabcrone \ignorespaces \O@tabarray}
\makeatother

\usepackage{array}
\usepackage[table]{xcolor}    
\usepackage{hhline}

\usepackage[export]{adjustbox}

\usepackage{kpfonts}

\usepackage[normalem]{ulem} 

\usepackage[latin1]{inputenc}
\usepackage{mathptmx}
\usepackage{fancyvrb}
\usepackage{amsfonts,amssymb,latexsym}
\usepackage{amsmath} 
\usepackage{graphicx}
\usepackage{algorithm}
\usepackage[noend]{algpseudocode}
\usepackage[all]{xy}
\usepackage{verbatim} 
\usepackage{apalike}
\usepackage{url}
\usepackage[T1]{fontenc}
\usepackage{multirow}
\usepackage[normalem]{ulem}
\usepackage{color}
\newcommand{\ol}[1]{\overline{#1}}

\newtheorem{example}{Example}
\newtheorem{definition}{Definition}

\newtheorem{proposition}{Proposition}[section]

\newtheorem{theorem}{Theorem}[section]

\newcommand\bcmdtab{\noindent\bgroup\tabcolsep=0pt%
  \begin{tabular}{@{}p{10pc}@{}p{20pc}@{}}}
\newcommand\ecmdtab{\end{tabular}\egroup}

\newcommand{\Variables}{\cal X}
\newcommand{\Symbols}{\Sigma}
\newcommand{\narval}{{\sf Narval}}%
\newcommand{\TermsOn}[5]{{{\cal T}^{#4}_{#1}(#2)}_{#3}^{#5}}
\newcommand{\Terms}{\TermsOn{\Symbols}{\Variables}{}{}{}}
\newcommand{\funocc}[1]{\mathit{Pos}_{\Symbols}(#1)}
\newcommand{\subterm}[2]{#1_{|#2}} 
\newcommand{\replace}[3]{#1[#3]_{#2}}
\newcommand{\congr}[1]{=_{\protect #1}}
\newcommand{\idsubst}{\textit{id}}

\newcommand{\cA}{\mathcal{A}}
\newcommand{\cU}{\mathcal{U}}
\newcommand{\cR}{\mathcal{R}}
\newcommand{\cE}{\mathcal{E}}
\newcommand{\cC}{\mathcal{C}}
\newcommand{\cD}{\mathcal{D}}

\newcommand{\cT}{\mathcal{T}}
\newcommand{\cX}{\mathcal{X}}

\newcommand{\cV}{\mathcal{V}}

\newcommand{\var}{Var}

\newcommand{\dom}{Dom}

\newcommand{\pos}{Pos}
\newcommand{\narrow}[1]{\leadsto_{#1}}
\newcommand{\narrowG}[2]{\leadsto_{{#1}}^{#2}}

\newcommand{\trianglelefteqsort}{\breve{\trianglelefteq}}

\newcommand\Tstrut{\rule{0pt}{2.5ex}}       
\newcommand\Bstrut{\rule[-0.9ex]{0pt}{0pt}} 

\def\Presto{\textup{\textsf{Presto}}}
\newcommand{\sort}[1]{\ensuremath{{#1}}}
\newcommand{\TermsS}[1]{\TermsOn{\Symbols,\sort{#1}}{\Variables}{}{}{}}

\newenvironment{ftheorem}[2][]{\vskip\topsep\noindent{\bf
Theorem~#2\ifthenelse{\equal{#1}{}}{}{\ (#1)}.}\em}{\vskip\topsep}
\newenvironment{fproposition}[2][]{\vskip\topsep\noindent{\bf
Proposition #2\ifthenelse{\equal{#1}{}}{}{\ (#1)}.}\em}{\vskip\topsep}

  \title[Partial Evaluation for Rewriting Logic]
                {Symbolic Specialization of Rewriting Logic Theories with \Presto\
        \thanks{
This research was partially supported by TAILOR, a project funded by EU Horizon 2020 research and innovation programme under GA No 952215, 
grant RTI2018-094403-B-C32 funded by MCIN/AEI/10.13039/501100011033 and by ''ERDF A way of making Europe'',
		and
        		 by Generalitat Valenciana under grant PROMETEO/2019/098. 
		 Julia Sapi\~na has been
        		supported by the Generalitat Valenciana APOSTD/2019/127 grant.}}

\author[M. Alpuente, D. Ballis, S. Escobar, J. Sapi\~{n}a]
         {MAR\'{I}A ALPUENTE, SANTIAGO ESCOBAR, JULIA SAPI\~{N}A\\
         VRAIN (Valencian Research Institute for Artificial Intelligence), Universitat Polit\`ecnica de Val\`encia\\
         \email{\{alpuente,sescobar,jsapina\}@upv.es}
         \and
         DEMIS BALLIS\\
         DMIF, University of Udine\\
         \email{demis.ballis@uniud.it}
}
\pagerange{\pageref{firstpage}--\pageref{lastpage}}

\begin{document}

\label{firstpage}

\maketitle
\begin{abstract}
This paper introduces \Presto,\ a symbolic partial evaluator for Maude's rewriting logic theories
 that can  improve  
 system analysis and verification. 
In \Presto,\ the automated optimization of a 
 conditional rewrite theory $\cR$ (whose rules define the concurrent  transitions of a system) 
 is achieved by  partially evaluating, with respect to the rules of $\cR$, an underlying, 
  companion
 equational logic theory $\cE$ that  specifies the 
algebraic structure of the system states of $\cR$.
 This can be particularly useful for specializing     
  an overly general 
  equational theory $\cE$ whose operators may obey 
complex combinations of associativity, commutativity, and/or identity axioms, 
when being plugged  into a 
host rewrite theory $\cR$ as happens, for instance,  in protocol analysis, where sophisticated equational theories for  
cryptography  are used.
   \Presto\   implements 
    different  unfolding  operators  that are   based on \emph{folding variant narrowing} (the 
symbolic  engine of 
Maude's equational theories).
When combined with an   appropriate abstraction algorithm, they allow the specialization to be adapted  
 to the theory termination behavior and
 bring  significant 
improvement 
while ensuring strong correctness and termination of the specialization.
    We demonstrate the effectiveness  of     \Presto\
       in several examples of protocol analysis  where it
        achieves a significant speed-up. 
Actually,  the transformation provided  by  \Presto\ may cut down an infinite folding variant narrowing space 
 to a finite one, 
and moreover, some of the costly  algebraic axioms and rule conditions may be eliminated as well. 
As far as we know, this is the first partial evaluator for Maude that respects the semantics of functional, logic, concurrent, and object-oriented computations.
Under consideration in Theory and Practice of Logic Programming (TPLP).
\end{abstract}


\input{1-intro}
\input{2-prelim}

\input{3-scheme}
\input{4-presto}
\input{5-experiments}
\input{6-conclusion}

\bibliographystyle{apalike}

\newpage 
\appendix
\input{appendix}

\label{lastpage}
\end{document}

%% file: 1-intro.tex
\section{Introduction}\label{sec:intro}
Partial evaluation (PE) is a general and powerful optimization technique that automatically specializes a program to a part of its input that is known statically~\cite{JGS93}. 
PE is motivated by the fact that
dedicated programs tend to operate more efficiently than general-purpose ones.
It has found applications ranging from  compiler optimization to test case generation,
among many others \cite{CS13,CL01}.
 In the context of logic programming (LP), partial evaluation  is often 
 called partial deduction (PD), while the term partial evaluation is often reserved for the specialization of impure logic programs. 
PD  not only allows  input variables to be instantiated with constant values but it also deals with 
 terms that may contain logic variables, thus providing extra  capabilities for program specialization \cite{MG95}.

Rewriting Logic (RWL) 
is a very general \emph{logical} and \emph{semantic framework} 
that is particularly suitable for modeling and analyzing  complex, highly nondeterministic software systems \cite{MM02}.
Rewriting Logic is efficiently implemented in the high-performance\footnote{Maude's rewriting machinery is highly optimized. Recently,  an experimental open platform has been developed  that allows the performance of functional and algebraic programming languages to be compared, including  {\sf CafeOBJ, Clean, Haskell, LNT, LOTOS, Maude, mCRL2, OCaml, Opal, Rascal, Scala, SML, Stratego / XT}, and {\sf Tom} (see references in \cite{GTA18}). In the top 5 of the most efficient tools, Maude ranks second after {\sf Haskell}.} 
language
Maude \cite{Maude07}, which seamlessly integrates functional, logic, concurrent, and object-oriented computations. 
A Maude {\em rewrite  theory}  $\cR=(\Sigma,E\uplus B, R)$ 
   combines  
a  
term rewriting system 
 $R$,  which specifies the concurrent   transitions of a system (for a  signature $\Sigma$ of program operators together with their type definition),
 with an {\em equational theory} $\cE = (\Sigma,E\uplus B)$
that specifies system states as terms of an algebraic datatype. 
The equational theory  $\cE$ contains a set $E$ of equations and a set $B$ of axioms (i.e.,\ distinguished equations that specify algebraic laws such as commutativity, associativity, and unity for some theory operators that are defined on $\Sigma$). The equations of $E$ are implicitly oriented from left to right as rewrite rules and operationally used as simplification rules, while the axioms of  $B$ are mainly used for $B$-matching
so that rewrite steps in $\cR$ are  performed {\em modulo} the equations and axioms of $E\uplus B$. 

For instance, in protocol specification, the equations of $E$ are often used to define the  cryptoarithmetic 
functions   that are used in the protocol  transition rules (e.g.,\ addition, exponentiation, and exclusive-or), while the axioms of $B$ specify properties of the   cryptographic primitives  
(e.g.,\ commutativity of addition and multiplication).
For example, consider the (partial) specification of integer numbers defined by the equations 
$E= \{X+0 = X, ~X+s(Y) = s(X+Y), ~p(s(X)) = X, ~s(p(X)) = X\}$, 
where variables $X$, $Y$ are of {\em sort} \verb!Int! (types are called {\em sorts} in RWL), operators $p$ and $s$ respectively stand for the predecessor and successor functions, and $B$ contains the commutativity axiom $X+Y=Y+X$. 
Also consider  that the program signature $\Sigma$ contains a binary 
state constructor operator $\langle\_,\_\rangle:$ \verb!Int x Int!  $\rightarrow$ \verb!State! for a new sort \verb!State! that models a 
simple network of processes that 
 consume a token from one place (denoted by the first component of the state) and transmit it to another place (the second component),
 keeping  the total amount of tokens invariant. 
The system state $t =\langle s(0),s(0) + p(0)\rangle$ can be rewritten to $\langle 0,s(0)\rangle$ (modulo the equations of $E$ and the commutativity of $+$) using the following rule that specifies the system dynamics:
\begin{equation}
\langle A,B\rangle \Rightarrow \langle p(A),s(B)\rangle, \mbox{ where } A \mbox{ and  } B \mbox{ are variables of  sort  {\tt Int}} \label{eq:rule1}
\end{equation} 

Multiparadigm, functional logic languages  
  combine features from functional programming   (efficient evaluation strategies,   nested expressions, genericity, advanced typing, algebraic data types) 
 and logic programming (logical variables, partial data structures,  nondeterministic   search for solutions, constraint solving  modulo   theories)   \cite{Meseguer92-alp,Hanus97,Escobar14}.  
The operational principle that supports functional  and logic  language integration is  called {\em narrowing} \cite{Fay79,Slagle74}, which is a goal-solving mechanism that subsumes the resolution principle of logic languages and the reduction principle of functional languages. Roughly speaking, narrowing can be seen as a generalization of term rewriting that allows free variables in terms (as in logic programming) and that non-deterministically reduces these  partially instantiated function calls by using unification (instead of  pattern-matching)  at each reduction  step  \cite{Hanus94}.

Besides rewriting with rules modulo equations and axioms, Maude provides full {\em native} support for narrowing computations in rewrite theories since Maude version 3.0 (2020).
This endows Maude with symbolic reasoning capabilities (e.g., symbolic reachability analysis in rewrite theories based on narrowing) and unification modulo user-definable equational theories.
In our example theory, a narrowing reachability goal  from 
$\langle V+V,0+V\rangle$ 
to $\langle p(0),s(0)\rangle$   succeeds (in one  step) with computed substitution  $\{V\mapsto 0\}$\footnote{It is 
calculated by first computing the most general  $\cE$-unifier $\sigma$ of the input term $\langle V+V,0+V\rangle$ and the left-hand side $\langle A,B\rangle$ of rule \eqref{eq:rule1}, $\sigma=\{A\mapsto (V+V), B\mapsto V\}$ 
($\sigma$ is a $\cE$-unifier of $t$ and $s$  if $s\sigma$ is equal to $t\sigma$ modulo the equations and axioms of $\cE$). Second,  an $\cE$-unifier $\sigma'$ is computed between the instantiated right-hand side $\langle p(V+V),s(V)\rangle$ and the
target state  $\langle p(0),s(0)\rangle$,
$\sigma'=\{V\mapsto 0\}$.
Third, the composition $\sigma\sigma'=\{A\mapsto 0+0,B\mapsto 0,V\mapsto 0\}$ is simplified into $\{A\mapsto 0,B\mapsto 0,V\mapsto 0\}$ and finally restricted to the   variable $V$ in the input term,  yielding $\{V\mapsto 0\}$.},
 which   might signal a possible programming error in rule \eqref{eq:rule1} since the number of available tokens in the first component of the state becomes negative, thereby pinpointing a token shortage.
This kind of advanced, LP-like reasoning capability based on narrowing goes beyond standard, rewrite-based equational reasoning that does not support symbolic reachability analysis because   input variables are simply handled as constants.
 Symbolic reasoning methods that are based on narrowing and their   applications  
are generally discussed in 
\cite{Meseguer20-jlamp,Meseguer20-lopstr}. 
For a very brief account of Maude's rewriting and narrowing principles, we refer to \cite{ABES19}.

In this paper, we develop the narrowing-based partial evaluator \Presto\   and show how 
it can have a tremendous impact on the symbolic analysis of concurrent systems that are modeled as Maude rewrite theories.
Traditional partial evaluation techniques typically remove some computation states by performing as much   program computation as possible, hence contracting the search space because some  transitions  are eliminated. However,  in the specialization  of concurrent systems that are specified by Maude rewrite theories, we are only interested in compressing the deterministic, functional computations  (which normalize the system states and are encoded by means of the theory equations) while preserving all concurrent state transitions  (which are defined by means of the rewrite rules). This ensures that all reachable system states  are preserved so that  any reachability property of the original concurrent system can be  correctly analyzed in the specialized one.

Given a rewrite theory  $\cR=(\Sigma,E\uplus B, R)$ our method  proceeds by 
partially evaluating  the
 underlying equational theory  $\cE=(\Sigma,E\uplus B)$ 
  with respect to the function calls in the rules of $\cR$ in such a way that   $\cE$ gets rid of any possible 
  overgenerality. By doing this, only the functional computations in $\cE$ are compressed by partial evaluation, while keeping every  system state in the  search space of the concurrent computations of $\cR$.
 Partial evaluation can  dramatically help in this process by introducing polyvariance (i.e.,\ specializing a given
function into different variants according to distinct invocation
contexts \cite{MG95}). The transformation performed by \Presto\ is non-trivial because, depending on the properties of both theories ($\cE$ and $\cR$), the right unfolding and abstraction operators are needed to efficiently achieve the largest optimization possible while ensuring termination and total correctness of the transformation.

The   partial evaluator \Presto\  provides near-perfect support for the   Maude language; specifically, it deals with all of the features that are currently supported by Maude's narrowing infrastructure. This effort is remarkable for at least two reasons: (i)  Maude has quite sophisticated features (subtype polymorphism,   pattern matching and equational unification  modulo associativity, commutativity and identity axioms, equations,  rules,  
modules, objects, 
{\em etc.}); and (ii) in order  to  efficiently  achieve aggressive specialization that scales to real-world  problems, the key components of the \Presto\ system needed to be thoroughly investigated and highly optimized over the years. This is because equational  problems such as order-sorted equational homeomorphic embedding  and   order-sorted equational least general generalization\footnote{Generalization,  also known as anti-unification, is the dual of unification: a generalization of two terms $t_{1}$ and $t_{2}$  is any term $t$ of which $t_{1}$ and $t_{2}$ are substitution instances \cite{Plotkin70}.}
 are much more costly than their corresponding ``syntactic'' counterparts  and achieving proper formalizations and efficient implementations has required years 
  \cite{AEMO08b,AEMO08a,AEEM14-ic,ABCE+19,ACEM20-jlamp,ACEM20-fi,AEMS21}. 

\subsection{Related work} 
 
Traditional applications of partial evaluation include the optimization of programs that explore a state space by trying to reduce the amount of search. Relevant prior work includes the use of partial deduction to optimize a theorem prover with respect to a given theory \cite{WG94} and to
refine infinite state model checking \cite{LG01}, where both the theorem
prover and the model checker are written as a logic program that searches for a proof, and the experiments use a standard partial evaluator
that is supplemented with abstract interpretation \cite{CC77} so that infinite branches in the search space are removed. Also, the specialization of concurrent processes that are modeled by    (infinite state) Petri nets  or by means of process algebras is performed in \cite{LL00} by a combination of partial deduction 
and abstract interpretation. 

Among the vast literature on 
program specialization,  the  partial evaluation of  functional logic  programs \cite{AFV98,AHV02,HP14} 
is the closest to our work. 
The {\em narrowing-driven Partial Evaluation\/} (NPE) algorithm of \cite{AFV98}, implemented in the {\sf Indy} system \cite{AAFV98}, 
extends to narrowing the classical PD scheme of \cite{MG95} 
 and was proved to be strictly more powerful than the PE of both logic programs and functional programs \cite{AFV98}, 
 with a potential for specialization that is comparable to  conjunctive partial deduction (CPD) and positive supercompilation \cite{SGJL+99}. 
Early instances of this framework implemented partial evaluation algorithms for different narrowing strategies, including lazy narrowing \cite{AFJV97}, innermost narrowing \cite{AFV98}, and needed narrowing \cite{AAHV99,ALHV05}.

NPE was extended  in \cite{ACEM20-jlamp} 
to the specialization of  \emph{order-sorted equational theories} and implemented in the partial evaluator for equational theories {\sf Victoria}.
This was essentially achieved by 
generalizing the key NPE ingredients to work with order-sorted equations and axioms: 
1) a tree {\em unfolding operator} based on folding variant  narrowing  that ensures strong correctness of the   transformation; 2) a novel notion of    order-sorted equational \emph{homeomorphic
embedding}  that  achieves local termination (i.e.,\ finiteness of unfolding); 3)  a suitable notion of  order-sorted equational \emph{closedness} (coveredness of the tree leaves modulo axioms) that ensures  strong completeness;  and 4) an \emph{abstraction
	operator} based on  order-sorted equational  least general generalization 
	that provides global termination of the whole specialization process. 	However, the partial evaluator {\sf Victoria}
	 is only able to specialize  deterministic and terminating equational theories, while the \Presto\ system described in this article  deals with 
	 non-deterministic and non-terminating  rewrite theories that allow concurrent systems to be modeled and model-checked. For a detailed discussion of the  literature related to narrowing-driven partial evaluation, we refer to \cite{ACEM20-jlamp}.

The   partial evaluator \Presto\    implements our most ambitious generic specialization framework for RWL. Currently, there is hardly any system   that can  automatically optimize rich rewrite theories   that include sorts, 
   subsort overloading, rules, equations, and algebraic axioms. Actually, there are very few transformations in the related literature that can improve the analysis of  Maude's rewrite theories and they are only known to the skilled practitioner, and, moreover, they   lack automated tool support. 
   For instance,  the {\em total  evaluation} of \cite{Meseguer20-jlamp}, which  only applies to a restricted class of  equational theories called \emph{constructor finite variant  theories}.
      Moreover,  unlike the optimization that is achieved by \Presto,\  none of the existing transformations respects the {\em narrowing semantics} of the theory  (both values and computed substitutions), 
   just its {\em ground value semantics} or its {\em reduction (normal form) semantics} \cite{ACEI+10,LM16}.

Many kinds of tools are built in Maude that rely on reflection and    theory transformations and preserve   specific properties such as invariants or termination behavior. Full-Maude \cite{maude-manual}, Real-time Maude \cite{OM08}, MTT \cite{DLM08},  and Maude-NPA \cite{EMM09} are prominent examples. Equational abstraction \cite{MPM08,BEM13} reduces an infinite state system to a finite quotient of the original system algebra by introducing some extra equations that preserve certain temporal logic properties.
Explicit coherence (a kind of confluence between rules, equations, and axioms) is necessary for executability purposes and also relies on rewrite theory transformations \cite{Viry02,Meseguer20-jlamp,DMR20}.
Also the semantic $\mathbb{K}$-framework 
 \cite{Rosu17},  the model transformations of \cite{RDRK19}, and the automated program correction technique of \cite{ABS19,ABS20}  are based  on sophisticated   program transformations that  preserve the reduction semantics of the original theory. Nevertheless, none of them aim to achieve program optimization.

 The generic narrowing-driven partial evaluation scheme  was originally formulated in \cite{ABES22-jlamp} for (unconditional) rewrite theories.  An instance of the framework that mimicks the total evaluation of \cite{Meseguer20-jlamp} can be found  in  \cite{ABEMS20-Festschrift}. Experiments with a preliminary version of the {\sf Presto}  system  together with its basic underpinnings can also be found in \cite{ABES22-jlamp}.

\subsection{\bf Contributions}	
The original contributions of this article with respect to the partial evaluation scheme  for unconditional rewrite theories presented in \cite{ABES22-jlamp} are as follows.

\begin{enumerate} 
 \item We extend the generic narrowing-driven partial evaluation scheme of \cite{ABES22-jlamp} to the specialization of   rewrite theories that may contain 
 conditional rules. Given the  rewrite theory $\cR=(\Sigma,E\uplus B, R)$, this is done as follows. First, the equational theory is partially evaluated with respect  to both the function calls in the right-hand sides and the conditions of the rewrite rules of $R$. Then, a post-processing refactoring transformation is  applied to both equations and rules that gets rid of unnecessary symbols and  delivers a more efficient, computationally equivalent rewrite theory where some of   the costly  algebraic axioms  and rule conditions may  simply  get removed.  The new rule condition simplification algorithm is quite inexpensive and does not modify the program semantics.

\item   A novel transformation called \emph{topmost extension} is provided that achieves in one shot the two executability conditions that are required for the completeness of narrowing in Maude rewrite theories: the {\em explicit  coherence\/\footnote{Currently, coherence is implicitly and automatically provided by Maude for rewriting computations. However, for narrowing computations, explicit coherence  must be specifically ensured \cite{maude-manual,Meseguer20-jlamp}.}} of rules with respect to equations and axioms, and the {\em topmost} requirement on rules that forces  all rewrites to happen on the whole state term ---not on its subterms. We   demonstrate that the topmost transformation is correct and preserves 
the solutions 
of the original theory so narrowing-based symbolic reachability  analysis is enabled for all rewrite theories after the  topmost transformation.
 
 \item We  apply  our novel results on the topmost extension to  support narrowing-based 
  specialization for  {\em object-oriented}\/\footnote{Actually, object-oriented modules have been totally redesigned in Maude 3.1 \cite{maude-manual}. They were not handled by the previous version of the  \Presto\/ system in \cite{ACEM20-jlamp}.} specifications. It is worth noting that Maude's  object-oriented   rules are typically non-topmost and they were not covered by \cite{ABES22-jlamp}.

   \item
We  have expanded the  {\sf Presto} system that implements the partial evaluation scheme of  \cite{ABES22-jlamp} both
to cope with conditions in rules and to handle Maude 
    object-oriented   modules that   (in addition to equations, axioms, and rewrite rules)  support classes, objects, messages, multiple class inheritance, and object interaction rules.

 \item We  also endowed {\sf Presto} with an automated checker for:
1) {\em  strong irreducibility}, a  mild condition enforcing that the left-hand sides of the rewrite rules of $R$ cannot be narrowed in $E$ modulo $B$, which is needed for the strong completeness of our specialization method  \cite{ACEM20-jlamp}; and
    2)
 a novel property called {\em U-tolerance}  ensuring that least general
generalization with identity axioms is finitary even if function symbols have  more than one  unit element   \cite{AEMS21}. This scenario can occur when a rewrite theory contains multiple overloaded function symbols associated with distinct  identity elements. 
 Although most commonly occurring theories satisfy these properties, 
 these checks are  an important addition to the system because  checking them manually
 is painful, since they involve term  unification  modulo any combination  of equational axioms, which is difficult to calculate by hand.

\item  We describe the system capabilities and provide an empirical evaluation of   \Presto\ on  a  set of  
protocol analysis problems  where it demonstrates a significant speed-up.

 \end{enumerate}

\subsection{Plan of the paper} 
In Section \ref{sec:prelim}, we provide some preliminary notions on RWL, and we 
 introduce a leading example that will be used throughout the paper to describe the specialization capabilities of \Presto. 
Section \ref{sec:symbolic} recalls the narrowing-based   symbolic principles  of Maude 
and formalizes  an automatic program transformation that generates topmost rewrite theories for which the narrowing machinery for solving symbolic reachability problems can be effectively used.  Specifically it enables symbolic reachability for object-oriented rewrite theories. 
 In Section \ref{sec:nper}, we present our basic specialization scheme for conditional rewrite theories and describe its core functionality. 
 Section \ref{sec:instance} provides two instantiations of the specialization scheme that exploit  two distinct unfolding operators 
  that are suitable for theories that   respectively have  either an infinite behavior or a finite behavior with regard to  narrowing computations.
In Section~\ref{sec:imple}, we present an overview of 
\Presto's main implementation choices and 
some additional tool features that allow the user to inspect the intermediate results of the specialization as well as to visualize all of the narrowing trees that are deployed during the  specialization.  In Section~\ref{sec:exp}, we present experimental evidence
that \Presto\  achieves significant specialization, with some specialized theories running 
 up to two orders of magnitude 
	  faster than the original ones. 
Section \ref{sec:conc} concludes and discusses  future work. Proofs of the main technical results are given in \ref{app:appendix}. 	  

%% file: 2-prelim.tex
\section{Preliminaries}\label{sec:prelim}
Let $\Sigma$ be a \emph{signature} that includes typed operators (also called function symbols) of the form  $f\colon s_1\ldots s_m\to s$, where $s_i$, for  $i=1,\ldots n$, and $s$ are sorts in a poset $(S,<)$ that models subsort relations (e.g.,\ $s < s'$ means that sort $s$ is a subsort of $s'$). 
The connected components of ${(S, <)}$
are the equivalence classes 
corresponding to the least equivalence relation $\equiv_{<}$ containing ${<}$.
As usual in rewriting logic \cite{AEEM14-ic}, we  assume {\em kind-complete}\footnote{Actually, Maude automatically completes any input signature with special kinds  to
ensure uniqueness of top sorts.}
 signatures such that: each connected component in the poset $(S,<)$  has a top sort (also called {\em kind}), and, for each $\sort{s}\in\sort{S}$,
we denote by $[\sort{s}]$ the top
sort in the connected component of \sort{s}
(i.e.,\
if \sort{s} and \sort{s'} are sorts in the same connected component,
then $[\sort{s}]=[\sort{s'}]$); 
and (ii) for each operator declaration 
$f: \sort{s}_1 \times \ldots \times \sort{s}_n \rightarrow \sort{s}$
in $\Symbols$,
there is also a declaration 
$f: [\sort{s}_1] \times \ldots \times [\sort{s}_n] \rightarrow [\sort{s}]$
in $\Symbols$.
Binary operators  in $\Sigma$ may have  an axiom declaration attached  that specifies any combinations of
algebraic laws such as associativity ($A$),  commutativity ($C$),  identity ($U$), left identity ($U_l$),  and right identity ($U_r$). 
We use strings over the alphabet $\{A,C,U,U_l,U_r\}$ to indicate the combinations of axioms satisfied by an operator $f$.  By $ax(f)$, we denote the set of algebraic axioms for the operator $f$. 

We consider an $S$-sorted family $\cX = \{\cX_s\}_{s \in S}$ of disjoint variable sets. $\TermsS{s}$ and $\cT_{{\Sigma},s}$ are the sets of 
terms and ground terms of sort $s$, respectively. By $\cT_{{\Sigma}}(\cX)=\bigcup_{s\in S} \TermsS{s}$, we denote the usual non-ground term algebra built over $\Sigma$ and the variables in $\cX$. By $\cT_{\Sigma}=\bigcup_{s\in S} \cT_{{\Sigma},s}$, we denote the ground term algebra over $\Sigma$. 
By notation $x:s$, we denote a variable $x$ with sort $s$.
The set of variables that appear in a term $t$ is denoted by $\var(t)$.

We assume 
\emph{pre-regularity} of the signature $\Sigma$:
for each operator declaration
$f\colon s_1\ldots s_m\to s$,
and for 
the set $S_f$ containing all sorts $s'$  that appear in 
operator declarations of the form 
: $f\colon s'_1\ldots s'_m\to s'$ in $\Sigma$ 
such that $s_i < s'_i$ for $1 \leq i \leq m$,
then the set $S_f$
has a least sort.
Given a term $t\in\cT_{{\Sigma}}(\cX)$, $ls(t)$ denotes the {\em least sort} of $t$ in the poset $(S,<)$.
An expression $\ol{t_n}$ denotes a finite sequence of terms $t_1\ldots t_n$, $n\geq 0$.
A \emph{position} $w$ in a term $t$ is represented by a sequence of
natural numbers that addresses a subterm of $t$ ($\Lambda$ denotes the empty sequence, i.e., the
root position). Given a term $t$, we let $\pos(t)$ denote the set of
positions of $t$. We denote  the usual prefix preorder over positions by $\leq$.
By $t_{|w}$, we denote the \emph{subterm} of $t$ at position $w$.  By $root(t)$, we denote the operator of $t$ at position $\Lambda$.

A \textit{substitution} $\sigma$ is a sorted mapping from a finite
subset of $\Variables$ to $\Terms$.
Substitutions are written as 
$\sigma=\{X_1 \mapsto t_1,\ldots,X_n \mapsto t_n\}$.
The identity
substitution is denoted by $\idsubst$.  
Substitutions are homomorphically extended
to $\Terms$. 
The application of a substitution $\sigma$ to a term $t$ is
denoted by $t\sigma$. 
The restriction of $\sigma$ to a set of variables
$V \subset \cX$ is denoted $\subterm{\sigma}{V}$.
Composition of two substitutions is denoted by $\sigma\sigma'$ so that $t(\sigma\sigma')=(t\sigma)\sigma'$.

 A  \textit{$\Symbols$-equation} (or simply equation, where $\Sigma$ is clear from the context) is an unoriented pair  $\lambda = \rho$, where 
 $\lambda,\rho \in \TermsS{s}$ for some sort $\sort{s}\in\sort{S}$, where  $\TermsS{s}$ is the set of terms of sort $s$ built over $\Sigma$ and $\cX$.  An equational theory $\cE$ is a pair  $(\Sigma,E\uplus B)$ that consists of a signature $\Sigma$, a set $E$ of $\Sigma$-equations,  and a set $B$ of 
 algebraic axioms 
 (e.g.,\ associativity, commutativity, and/or identity)  that are 
expressed by means of equations for some binary operators in $\Sigma$.  
The equational theory $\cE$ induces a congruence relation $\congr{\cE}$ on  $\Terms$.

A term $t$ is more general than (or at least as general as) $t'$ modulo $\cE$, 
denoted by $t \leq_{\cE} t'$, if 
there is a substitution $\gamma$ such that 
$t' =_\cE t\gamma$.
We also define
 $t \simeq_{\cE} t'$ iff $t \leq_{\cE} t'$ and $t' \leq_{\cE} t$.
By abuse of notation,  we write $\leq_B$ and $\simeq_B$ when  $B$ is an axiom set.

A substitution $\theta$ is more  general than (or at least as general as) $\sigma$ modulo $\cE$, 
denoted by $\theta \leq_{\cE} \sigma$, if there is a substitution $\gamma$ such that $\sigma =_\cE \theta\gamma$, i.e.,\  
for all $x \in \cX, x\sigma =_{\cE}  x\theta\gamma$.
Also, $\theta \leq_{\cE} \sigma \ [V]$ iff 
there is a substitution $\gamma$ such that, 
for all $x \in V, ~x\sigma =_{\cE}  x\theta\gamma$.

An \textit{$\cE$-unifier} for a $\Symbols$-equation  
$t = t'$ is a
substitution $\sigma$ such that $t\sigma =_\cE t'\sigma$.

We consider three different kinds of expressions that may appear in a  conditional rewrite theory:
1) an \emph{equational condition}, which is
 is any (ordinary\footnote{A boolean equational condition $b=true$, with $b\in \cT_{{\Sigma}}(\cX)$ of sort Bool, can be abbreviated 
as  $b$, although it is internally represented as the equation $b=true$.})
 equation  $t=t'$, with  
$t, t' \in \TermsS{s}$ for some sort $\sort{s}\in\sort{S}$;
2) a \emph{matching condition}, which is a pair $t:=t'$, with 
$t, t' \in \TermsS{s}$ for some sort $\sort{s}\in\sort{S}$; and 
3) a \emph{rewrite expression}, which is a pair $t \Rightarrow t'$, with 
$t, t' \in \TermsS{s}$ for some sort $\sort{s}\in\sort{S}$.
A \emph{conditional} rule is an expression of the form  $\lambda \Rightarrow \rho\ \mathit{if}\ C$, where 
$\lambda,\sigma\in\TermsS{s}$ for some sort $\sort{s}\in\sort{S}$, 
and $C$ 
is a (possibly empty,  with identity symbol  $\mathit{nil}$) sequence $c_1 \wedge\ldots \wedge c_n$, where each $c_i$ is an equational condition, a matching condition, or a rewrite expression. 
Conditions are evaluated\footnote{Given a parameter-passing substitution $\sigma$, the interpretation of equational conditions $t'=t$ consists in the joinability of the normal forms $t\sigma$ and $t'\sigma$  (in $E$ modulo $B$); pattern matching equations $t:=t'$ are evaluated by computing the equational pattern matchers, within the instantiated pattern $t\sigma$, of the normal form of $t'\sigma$  (in $E$ modulo $B$); and   rewrite expressions $t  \Rightarrow t'$ are interpreted as rewriting-based reachability goals in $R$ (modulo $E\uplus B$).} 
 from left to right, and therefore the order in which they appear, although mathematically inessential, is operationally important.
When the condition $C$ is empty, we simply write $\lambda \Rightarrow  \rho$. 
A {\em rewrite theory} is a triple $\cR = (\Sigma,E\uplus B,R)$, where $(\Sigma,E\uplus B)$ is an  equational theory and $R$ is a set of  rewrite rules.
A rewrite theory $(\Sigma,E\uplus B,R)$ is called \emph{topmost} if there is a sort \textit{State}  such that: (i) for each rewrite  rule $\lambda \Rightarrow \rho\ \mathit{if}\ C$, 
$\lambda$ and $\rho$ are of sort \textit{State};
and
(ii)  for each  $f\colon [s_1]\ldots [s_n]\to s \in \Sigma$ and $i\in\{1,\ldots,n\}$,  $[s_i] \neq \mathit{State}$. 
\footnote{Note that \textit{State} names an arbitrary sort for which  conditions (i) and (ii) are satisfied.}

In a rewrite theory $\cR=(\Sigma,E\uplus B, R)$,  computations evolve by rewriting   states 
using  the {\it equational rewriting} relation  $\rightarrow_{R,\cE}$, which applies the rewrite rules in $R$ to states {\it modulo the equational theory} $\cE=(\Sigma, E\uplus B)$ 
\cite{Meseguer92-tcs}. The Maude interpreter implements 
equational rewriting  $\rightarrow_{R,\cE}$ by means of two 
simple relations, namely $\rightarrow_{\vec{E},B}$ and 
$\rightarrow_{R,B}$. 
These allow rules and (oriented) equations 
to be intermixed in the rewriting process by simply using both an algorithm of matching modulo $B$.
The relation $\rightarrow_{\vec{E},B}$  uses $\vec{E}$ (the explicitly oriented version of the equations in $E$) for term simplification. Thus, for any term $t$, 
by repeatedly applying the equations as simplification rules, we eventually reach a term $t\!\downarrow_{\vec{E},B}$ to which no further equations can be applied.  The term $t\!\downarrow_{\vec{E},B}$ is  called  \emph{canonical form} (also called irreducible or normal form) of $t$ w.r.t.\ $\vec{E}$ modulo $B$.
On the other hand, the relation $\rightarrow_{R,B}$  implements rewriting with the rules of $R$, which  might be nonterminating and nonconfluent, whereas   
$\cE$ is required to be convergent (i,.e., $\rightarrow_{\vec{E},B}$ is terminating and confluent modulo $B$) and 
$B$-coherent
(a kind of non-interference between $E$ and $B$) in  order to guarantee the existence and unicity (modulo $B$) of a canonical form w.r.t.\  $\vec{E}$  for any term. 
Also, the set $R$ of rules must be coherent w.r.t.\ $\cE$, ensuring that any rewrite step with $\rightarrow_{R,B}$ can always be postponed in favor of  deterministically rewriting with  $\rightarrow_{\vec{E},B}$.

Formally, for a set of rules $P$ 
(with $P$ being either $R$ or $\vec{E}$), the rewriting relation $\rightarrow_{P,B}$ in $P$ modulo $B$  is defined as follows. Given a rewrite rule $r=(\lambda\Rightarrow\rho \mathit{\:if\:} C)\in P$, a substitution $\sigma$, a term $t$,  and a position $w$ of $t$, 
  $t \stackrel{r,\sigma,w}{\rightarrow}_{\!\!P,B} t'$ 
iff  $\lambda\sigma =_B t_{|w}$, $t'=t[\rho\sigma]_w$, and $C$ holds (the specific evaluation strategy for rule conditions of Maude can be found in  \cite{Maude07}).
When no confusion arises, we simply write $t \rightarrow_{P,B} t'$  instead of $t \stackrel{r,\sigma,w}{\rightarrow}_{\!\!P,B} t'$.

Under these conditions,  a $(R,E\uplus B)$-rewrite step $\rightarrow_{R,E\uplus B}$ 
on a term $t$  in the rewrite theory $\cR=(\Sigma,E\uplus B,R)$ can be implemented  by applying the following rewrite strategy:
(i) reduce $t$ w.r.t.\ $\rightarrow_{\vec{E},B}$ to the canonical form $t\!\downarrow_{\vec{E},B}$; and
(ii) rewrite $t\!\downarrow_{\vec{E},B}$ w.r.t.\  $\rightarrow_{R,B}$.
This strategy is still complete in the sense that, rewriting of congruence classes induced by $=_\cE$, with $\cE=(\Sigma,E\uplus B)$, can be mimicked by  $(R,E\uplus B)$-rewrite steps.

A rewrite sequence $t \rightarrow_{R,\cE}^{*} t'$
  in the rewrite theory $\cR=(\Sigma,E\uplus B,R)$  is then deployed as the (possibly infinite) rewrite sequence (with  $t_{0}=t$ and $t_n\!\!\downarrow_{\vec{E},B}= t'$)
$$t_0\rightarrow_{\vec{E},B}^*t_0\!\!\downarrow_{\vec{E},B}\: \rightarrow_{R,B} t_1\rightarrow_{\vec{E},B}^*t_1\!\!\downarrow_{\vec{E},B}\rightarrow_{R,B}\ldots \rightarrow_{R,B} t_n\!\!\downarrow_{\vec{E},B} $$ 
that interleaves  $\rightarrow_{\vec{E},B}$ rewrite steps and 
$\rightarrow_{R,B}$ rewrite steps following the strategy mentioned above. 
Note that,  after each rewrite step    using $\rightarrow_{R,B}$, generally the resulting term $t_{i}$, $i=1,\ldots, n$, is not in canonical form  
and is thus normalized before the subsequent rewrite step 
using $\rightarrow_{R,B}$ is performed. 
Also, in the precise strategy adopted by Maude, the last term of a finite computation 
is  finally normalized  before the result is delivered. 

\subsection{Concurrent Object-Oriented Specifications: a Communication Protocol Model with {\em Caesar} ciphering}\label{sec:oo-spec}
Rewrite theories provide a natural 
 computation model for concurrent  object-oriented  systems \cite{Meseguer92-tcs}. Indeed, Maude fully supports the object-oriented  programming paradigm by means of pre-defined
 data structures that identify the main building blocks of this paradigm such as classes, objects, and messages \cite{maude-manual}. The essential facts about concurrent object-oriented configurations are all formalized in the {\tt CONFIGURATION} module, which provides special notation and avoids boilerplate {\tt class} declarations.
  More specifically,  
the  sorts  \texttt{Oid}  and \texttt{Cid} respectively define object  and class identifiers, while the sort \texttt{Msg} represents messages (which are typically used to model object communication). It is worth noting that class names are just sorts. Therefore, class inheritance is directly supported by rewriting logic order-sorted type structure. A subclass declaration is thus an expression of the form {\tt subsort C < C'}
where \texttt{C} and \texttt{C'} are  two sorts representing object classes. Multiple inheritance is also supported, allowing a class {\tt C} to be defined as a subsort of many several sorts (each of which represent a different 
class).

The partial evaluation framework that we define in this work is able to deal with objects that are represented as terms of the following form:
$${< O\ :\ C\ |\ a_{1}\ :\ v_{1},~...~,~a_{n}\ :\ v_{n} >}$$
\noindent where $O$ is a term of sort $Oid$, $C$ is a term of sort $Cid$, and $a_{1}\ :\ v_{1},\ldots, a_{n}\ :\ v_{n}$ is a list of attributes of sort {\it Attribute}, each consisting of an identifier 
$a_{i}$ followed by its respective value  
$v_{i}$. 

The concurrent state of an object-oriented system is  a multiset (of sort {\it Configuration}) of objects and messages that
are built  using
the empty syntax (juxtaposition) ACU operator  {\it \_\;\_ : Configuration Configuration }$\to$ {\it Configuration} whose identity is {\it none}.
Thus, an object-oriented configuration will be either {\it none} (empty configuration) or  $Ob_1\ \dots\ Ob_k\ Mes_1\ \ldots\ Mes_n$, where $Ob_1$, \ldots,$Ob_k$ are objects, and 
$Mes_1$,\ldots,  $Mes_n$ are messages.

Transitions between object-oriented configurations are specified by means of rewrite rules of the form 
$$Ob_1\ \dots\ Ob_k\ Mes_1\ \ldots\ Mes_n
 \Rightarrow  Ob'_1\ \dots\ Ob'_j\  Ob_{k+1}\ \dots\  Ob_{m}\ Mes'_1\ \ldots\ Mes'_p\ if\ Cond .
$$ 
where $Ob'_1\ \dots\ Ob'_j$ are updated versions of $Ob_1\ \dots\ Ob_j$ for  $j\leq k$,  $Ob_{k+1}\ \dots\  Ob_{m}$ 
are newly created objects,  $Mes'_1\ \ldots\ Mes'_p$ are new messages, and
 {\it Cond} is a rule condition.

An important special case are rules with a single object and at most one message on the lefthand side. These rules directly model asynchronous distributed interactions. Rules involving multiple objects are called synchronous and are used to model higher-level communication abstractions.

\begin{example}\label{ex:handshakeNOFVP}
Let us consider a rewrite theory $\cR=(\Sigma,E\uplus B,R)$ 
that encodes an object-oriented\footnote{
A non-object-oriented, topmost version of this protocol can be found in \cite{ABES22-jlamp}.} specification for a  
client-server communication  protocol in an asynchronous medium where messages can arrive out-of-order. 
The theory signature $\Sigma$ 
includes several operators and sorts that model the protocol entities. The constant operators {\tt Cli} and {\tt Serv} (of respective sorts {\tt Client} and {\tt Server})  are used to specify two classes that respectively identify clients and servers. 
Both sorts inherit the class behavior of the  built-in abstract class {\tt Cid} via the subsort relations {\tt Client < Cid} and {\tt Server < Cid}.

\begin{figure}[h!]
{\footnotesize
\begin{Verbatim}[frame=single]  
crl [req] : < C : Cli | server : S , 
                       data : M , 
                       key : s(K) , 
                       status : mt >
           =>  
				
           < C : Cli | server : S , 
                       data : M , 
                       key : s(K) , 
                       status : mt > 
           ( S <- { C , enc(M,s(K)) } ) if s(K) < len /\ 
                                           enc(M,s(K)) = enc(enc(M,K),s(0)) .
	
rl [reply] :  < S : Serv | key : K > 
              (S <- {C,M}) 
				  
              =>
				  
               < S : Serv | key : K > 
               (C <- {S, dec(M,K)}) [narrowing] .
				  	
rl [rec] : < C : Cli | server : S , 
                       data : M , 
                       key : K , 
                       status : mt > 
           (C <- {S,M})
				
           =>
				
           < C : Cli | server : S , 
                       data : M , 
                       key : K , 
                       status: success > [narrowing] .
\end{Verbatim}
}
\caption{Rewrite rules for the client-server communication protocol.}\label{fig:rules}
\end{figure}

Data exchanged between clients and servers is encoded as non-empty, associative sequences ${\tt s_1\ldots s_n}$, where, for the sake of simplicity, each $\tt s_i$ is a term of sort {\tt Symbol} in the alphabet {\tt \{a,b,c\}}. We assume that {\tt Symbol} is a subsort of {\tt Data}; hence, any symbol is also a (one-symbol) data 
sequence.

Clients are represented as objects of the form 
$$\tt 
<\ C\ :\ Cli\ |\ server\ :\ S\ ,\ data\ :\ M\ ,\ key\ :\ K\ ,\ status\ :\ V >
$$
where {\tt C} is an object identifier for a client of the class {\tt Cli}, {\tt S} is the   server object  that {\tt C} wants to communicate with, {\tt M} is a piece of data representing a client request, {\tt K} is a natural number (specified in Peano's notation) that determines an encryption/decryption key for messages, and {\tt V} is a constant value that models the client status. Initially, the status is set to the empty value {\tt mt}, and it changes to {\tt success} whenever a server acknowledges message reception.

Servers are simple objects of the form 
$$\tt
 <\ S\ :\ Serv\ |\ key\ :\ K\ > 
$$ 
where {\tt S} is an object identifier for a server, and {\tt K} is an encryption/decryption key. 

Network packets are naturally modeled by object messages that can be exchanged between servers and clients. 
More specifically a network packet is a pair of the form {\tt Host <- CNT} of sort {\tt Msg}, where {\tt Host} is a client or server recipient, and {\tt CNT} specifies
the packet content. Specifically, {\tt CNT} is a term {\tt \{H,M\}}, with {\tt H} being the sender's name
 and {\tt M} being a data item that represents either a client request or a server response, expressed as a list of symbols in the alphabet {\tt\{a,b,c\}}.

System states are represented by object-oriented configurations which may include  clients, servers, and network packets.  

The protocol dynamics is specified by the term rewriting system $R$ in $\cR$ of Figure \ref{fig:rules}. It consists of three rules, where clients and servers agree on a shared key $\tt K$.
\begin{figure}[h!]
{\footnotesize
\begin{Verbatim}[frame=single]  
eq len = s(s(s(0))) [variant] .  --- alphabet cardinality is hardcoded 
                                 --- via the constant len = 3

--- Function toNat(s) takes an alphabet symbol s as input and returns the
--- corresponding position in the alphabet.
eq toNat(a) = 0 [variant] .
eq toNat(b) = toNat(a) + s(0) [variant] .
eq toNat(c) = toNat(b) + s(0) [variant] .
    
--- Function toSym(n) takes a natural number n as input and returns the
--- corresponding alphabet symbol.
eq toSym(0)= a [variant] .
eq toSym(s(0)) = b [variant] .
eq toSym(s(s(0))) = c [variant] .
    
var M : Data .
vars K X Y  : Nat .
  
--- Function shift(k) increments (modulo the alphabet cardinality) 
--- the natural number k.
eq shift(X) = [ s(X) < len,s(X), 0 ] [variant] .
    
--- Function unshift(k) decrements (modulo the alphabet cardinality) 
--- the natural number k.
eq unshift(0) = s(s(0)) [variant] .
eq unshift(s(X)) = X [variant] .
    
--- Function e(n,k) increments the natural number n by k units 
--- (modulo the alphabet cardinality).
eq e(X,0) = X [variant] .
eq e(X,s(Y)) = e(shift(X),Y) [variant] .
	
--- Function d(n,k) decrements the natural number n by k units 
--- (modulo the alphabet cardinality).
eq d(X,0) = X [variant] .
eq d(X,s(Y)) = d(unshift(X),Y) [variant] .

--- Function enc(m,k)  (resp. dec(m,k)) takes a data item m and a 
--- natural number k as input and returns the corresponding encrypted (resp. decrypted) 
--- data item using the Caesar cipher with key K
eq enc(S:Symbol,K) = toSym(e(toNat(S:Symbol),K)) [variant] .
eq enc(S:Symbol M,K) = toSym(e(toNat(S:Symbol),K)) enc(M,K) [variant] .
eq dec(S:Symbol,K) = toSym(d(toNat(S:Symbol),K))[variant] .
eq dec(S:Symbol M,K) = toSym(d(toNat(S:Symbol),K)) dec(M,K) [variant] .
\end{Verbatim}    
}
\caption{Equations of the equational theory encoding the {\em Caesar} cipher.}\label{fig:caesar}
\end{figure}

\end{example}

\noindent More specifically, the rule $\tt req$ allows a client {\tt C}  to initiate a transmission request with a server {\tt S} by sending  
 the data item {\tt M} that is encrypted by   function {\tt enc(M,s(K))} using a positive client's key {\tt s(K)}.  Note that this rule is conditional and is enabled if and only if
 the key  {\tt s(K)} is less than the constant {\tt len}, which represents the cardinality of the chosen alphabet of symbols.
Furthermore, to initiate a client request, we require the  satisfaction of the property {\tt enc(M,s(K)) = enc(enc(M,K),s(0))}, which enforces
 an additive property on the {\tt enc} function.
 
 The rule {\tt reply} lets the server $\tt S$  consume a client request packet {\tt S <- \{C,M\}} by first decrypting the incoming data item
{\tt M}  with the  server key  and then sending a response packet back to {\tt C} that includes the decrypted request data.
The rule {\tt rec} successfully completes  the data transmission  between {\tt C} and {\tt S} whenever the server response packet  
{\tt C <- \{S,M\}} includes a data item {\tt M} that is equal  to the initial client request message. In this case, the  status of the client
is changed from {\tt mt}  to  {\tt success}. \ 
Note that the  transmission  succeeds when the client and server use the same key {\tt K}, as this guarantees that the client plain message {\tt M} is equal to  {\tt dec(enc(M,K),K)} as required by the {\tt rec} rule.

Encryption and decryption functionality is implemented by two functions (namely, {\tt enc(M,K)} and {\tt dec(M,K)}) that are  specified by the equational theory $\cE$ in $\cR$. The equational theory $\cE$ implements a {\em Caesar} cipher with key {\tt K}, which is a simple substitution ciphering where each symbol in the plaintext data item  {\tt M} is replaced by the symbol that  appears {\tt K} positions later in the alphabet (handled as the list {\tt a,b,c}). The cipher is circular, i.e., it works modulo the cardinality of the alphabet. For instance, {\tt enc({\tt a b},s(0))}  delivers {\tt (b c)}, and {\tt dec(a b,s(0))} yields {\tt (c a)}.
The equational theory $\cE$ includes the equations\footnote{
For the sake of simplicity, we omitted the definition of the operators {\tt [\_,\_,\_]}, {\tt \_<\_}, and {\tt \_+\_}  that respectively implement the usual {\em if-then-else} construct, the {\em less-than} relation, and the associative and commutative addition over natural numbers.} 
in Figure \ref{fig:caesar}.

\section{A Glimpse into Narrowing-based Symbolic Computation in Maude}\label{sec:symbolic}

 Similarly to {\it rewriting modulo an equational theory $\cE$}, where syntactic pattern-matching is replaced with matching modulo $\cE$ (or $\cE$-matching),  in 
{\it narrowing modulo an equational theory}, 
syntactic unification is replaced by {\em equational} unification (or $\cE$-unification). 

Given a rewrite theory $\cR=(\Sigma,E\uplus B, R)$,  with $\cE=(\Sigma,E\uplus B)$,
 the 
 \emph{conditional narrowing}
 \footnote{Conditional narrowing is not currently supported by Maude but can be easily mimicked  through an unraveling transformation that removes the rule conditions  \cite{Meseguer20-jlamp}.} 
 relation $\narrow{R,\cE}$ on $\Terms$ is defined by
	$t \narrow{\sigma,p,R,\cE} t'$  (or simply $t \narrow{R,\cE} t'$) 
	if and only if there is a non-variable position $p \in \funocc{t}$, 
	a (renamed apart) rule $\lambda \Rightarrow \rho\ \mathit{if}\ C$ in $R$,
	and 
		 an $\cE$-unifier of $\subterm{t}{p}$ and $\lambda$ such that
	 $t' = \replace{t}{p}{\rho}\sigma$ and condition $C\sigma$ holds \cite{AMPP14}.
	Furthermore, we write $t \narrow{\sigma,p,r,\cE} t'$, when we want to highlight the rewrite rule $r$ that has been used in the narrowing step.
A term $t$ is called $(R,\cE)$-\emph{strongly irreducible} (also called a rigid normal form \cite{AEI09})  iff there is no term $u$ such that $t \narrow{\sigma,p,R,\cE} u$ for any position $p$, which amounts to say that    no subterm of $t$ unifies modulo $B$ with the left-hand side of any equation of $E$.

   Narrowing derivations correspond to  sequences  $t_0~  \narrowG{\sigma_{0},p_0,r_{0},B}
    ~t_1 \allowbreak~ \narrowG{\sigma_{1},p_1,r_{1},B}{}
  \ldots 
  \narrowG{\sigma_{n},p_{n-1},r_{n-1},B}~ t_n$.
 The composition $\sigma_0 \sigma_1\ldots \sigma_{n-1}$ of all the unifiers along a narrowing sequence leading to $t_n$ (restricted to the variables of $t_0$) is called {\em  computed  substitution}.
 
 The search space of all narrowing derivations for a term $t$ in $\cR=(\Sigma,E\uplus B, R)$  is represented as a tree-like structure called {\em $(R,E\uplus B)$-narrowing tree} (or simply narrowing tree).

 \subsection{Two-level narrowing in Maude}
A rewrite theory $\cR=(\Sigma,E\uplus B, R)$   
can be symbolically executed in Maude  by using narrowing at \emph{two levels}:
(i)~narrowing with 
  oriented equations $\vec{E}$ (i.e., rewrite rules obtained by explicitly orienting the equations in $E$) 
  modulo the axioms $B$; and 
(ii) narrowing with the (typically non-confluent and non-terminating) rules of $R$ modulo $\cE=(\Sigma,E\uplus B)$.

\paragraph{Narrowing with $\vec{E}$ modulo $B$.} This form of narrowing 
is typically used for 
 $\cE$-unification and it is implemented in Maude via  the {\em folding variant narrowing} (or simply  fV-narrowing) strategy  of \cite{ESM12}.
fV-narrowing is centered around the notion of equational variant of a term.
Given ${\cal E} = (\Sigma, E \uplus B)$, the set of all pairs  $(t',\sigma)$, where $t'$  is 
  $(t\sigma)\!\!\downarrow_{\vec{E},B}$, i.e.,\ the canonical form of $t\sigma$  
for a substitution $\sigma$, is called the set of \emph{equational variants} (or simply variants) of a term $t$ \cite{ESM12}. 
Intuitively, the variants of $t$ are the ``irreducible patterns'' to which $t$ can be symbolically evaluated by applying  $\vec{E}$ modulo $B$.
A variant $(t,\sigma)$ is {\em more general}  than a variant $(t',\sigma')$ w.r.t. an equational theory $\cE$ (in symbols, $(t,\sigma)\leq_\cE (t',\sigma')$) iff there exists a substitution $\gamma$ such that $t\gamma =_\cE t'$ and $\sigma\gamma =_\cE \sigma'$.

An equational theory $\cE$ has the {\em finite variant property} (FVP)  (or $\cE$ is called a {\em finite variant theory}) iff there is a finite, complete, and minimal set of most general equational variants for each term. 
 It is generally undecidable whether an equational theory has the FVP ~\cite{BGLN13}; a semi-decision procedure is given in \cite{Meseguer15} that works well in practice.
The FVP can be semi-decided  by checking whether there is a finite number of 
 most general variants for all \emph{flat} terms $f(X_1,\ldots,X_n)$ for any $n$-ary operator $f$ in the theory and pairwise-distinct variables $X_1,\ldots,X_n$ (of the corresponding sort).
 For instance,  the specification of the 
 exclusive-or
  operator on natural numbers given by  $E= \{N\oplus N = 0, N\oplus 0 = N \}$,
  with $\oplus$ being commutative, is a finite variant theory, since the (flat) term $N\oplus M$
 has four most general variants: 
$(N\oplus M, \{\})$, 
$(M, \{N \mapsto 0\})$, 
$(N, \{M \mapsto 0\})$,
and  $(0, \{N \mapsto M\})$. 
 Instead, the specification of the auxiliary function \verb!d! in Figure \ref{fig:caesar} does not satisfy the  FVP, since 
the term \verb!d(X,Y)! has an infinite number of most general variants (\verb!X!, $\{$\verb!Y! $\mapsto$\verb!0!$\}$),  (\verb!unshift(X)!, $\{$\verb!Y! $\mapsto$\verb!s(0)!$\}$), $\ldots$, 
(\verb!unshift!$^k$\verb!(X)!, $\{$\verb!Y! $\mapsto$ \verb!s!$^k$\verb!(0)!$\}$).  Therefore, the equational theory of Example \ref{ex:handshakeNOFVP} does not have the FVP either.

The main idea of folding variant narrowing is to ``fold''\footnote{This notion is quite different from the classical folding operation of Burstall and Darlington's fold/unfold transformation scheme \cite{BD77}. The idea of {\em folding} in fV-narrowing   is essentially a \emph{subsumption} check that is applied to the tree   leaves so that less general leaves are subsumed (folded into)  most general ones.} 
the
$(\vec{E},B)$-narrowing tree by subsumption modulo $B$. 
 That is, 
folding variant narrowing  shrinks the search space and  avoids computing any variant that is a substitution instance modulo $B$ of a more general variant.

 A fV-narrowing derivation is a sequence  
$t_0~  \narrowG{\sigma_{0},p_0,\vec{e}_{0},B}
    ~t_1 \allowbreak~ \narrowG{\sigma_{1},p_1,\vec{e}_{1},B}{}
  \ldots 
  \narrowG{\sigma_{n},p_{n-1},\vec{e}_{n-1},B}~ t_n,$ where 
 $ t   \narrowG{\sigma,p,\vec{e},B}~ t'$
(or simply $ t   \narrowG{\sigma}~ t'$
when no confusion can arise)
 denotes a transition (modulo the axioms $B$) from term $t$ to $t'$ via the {\em variant equation} $e$ (i.e., an  oriented equation $\vec{e}$ that is enabled to be used for fV-narrowing thanks to the attribute \texttt{variant}) using the  $B$-unifier $\sigma$. Assuming that the initial term $t$ is normalized, each  step 
 $ t   \narrowG{\sigma,p,\vec{e},B}~ t'$
 (or folding variant narrowing step) is followed by the simplification of the term into its normal form by using all equations in the theory, which may include not only the variant equations in the theory but also (non-variant) equations (e.g., built-in equations in Maude). 
In order for finite variant narrowing to be effectively applicable the equational theory  $\cE=(\Sigma,E\uplus B)$ must be convergent, $B$-coherent, finite variant, and $B$-unification must be decidable. 

\paragraph{Narrowing with $R$ modulo $\cE$. } It is mainly used for solving \emph{reachability goals}  \cite{MT07} and 
\emph{logical model checking} \cite{EM07}.

Given a non-ground term $t$, $t$ represents an abstract characterization of the (possibly) infinite set of all of the concurrent states $[\![t]\!]$ (i.e.,\, all the ground substitution instances of $t$, or, more precisely, the $\cE$-equivalence classes associated to such ground instances) within $\cR$.
In this scenario, each narrowing derivation $\cD$  subsumes all of the rewrite computations that are ``instances'' of $\cD$ modulo $\cE$ \cite{maude-manual}. 
Therefore, an exhaustive exploration of the $(R,E\uplus B)$-narrowing tree of $t$ allows one to prove existential reachability goals of the form
\begin{equation}\label{eq:reach}
(\exists X)\; t \longrightarrow^{*} t' 
\end{equation}
with $t$ and $t'$ being two (possibly) non-ground terms ---called the {\em input} term and {\em target} term, respectively--- and $X$ being the set of variables appearing in both $t$ and $t'$. 

A {\em solution}  for the reachability goal (\ref{eq:reach}) in a rewrite theory $\cR=(\Sigma,E\uplus B,R)$, with $\cE=(\Sigma, E\uplus B)$, is a substitution $\sigma$ such that 
$t\sigma\rightarrow_{R,\cE}^* t'\sigma$; thus, any computed substitution $\theta$ of a narrowing derivation $t\leadsto_{R,\cE}^*t'$ is a {\em solution} for the reachability goal (\ref{eq:reach}) since $t\theta\rightarrow_{R,\cE}^* t'\theta$.

In practice, solving a reachability goal $(\exists X)\; t \longrightarrow^{*} t'$ 
means searching for a symbolic solution 
within the $(R,E\uplus B)$-narrowing tree that originates from $t$ in a hopefully \emph{complete} way (so that, for any existing solution, a more general answer modulo $\cE$ will be found).
Completeness of narrowing with $R$ modulo $\cE$ holds for topmost rewrite theories, provided that a finitary and complete
$\cE$-unification algorithm exists\footnote{$\cE$-unification can be effectively and completely computed by folding variant narrowing under the  executability   conditions presented in this section.}.
 
 We note that only rewrite rules with the {\tt  narrowing} attribute are used by $\narrow{R,\cE}$, while any rule without this attribute is only considered for rewriting modulo $\cE$. By $\mathit{att(r)}$ we denote the attributes\footnote{Other available statement attributes in Maude include {\tt label}, {\tt metadata},  {\tt nonexec} and {\tt print} \cite{maude-manual}.} associated with the rewrite rule $r$. 
For instance, in Example \ref{ex:handshakeNOFVP},
{\tt rec} and {\tt reply} rules can be used by $\narrow{R,\cE}$, while {\tt req} only supports rewriting modulo $\cE$, since ${\tt narrowing}\not\in \mathit{att({\tt req})}$.

 Unfortunately, not all the rewrite theories are topmost or admit finitary and complete $\cE$-unification. For instance, object-oriented specifications of Section
 \ref{sec:oo-spec} are typically non-topmost, since their rewrite rules might apply to inner state positions, thereby implementing local state changes. 
 This fact may compromise  the effectiveness of narrowing-based symbolic reachability, because existing solutions of a given reachability goal may be missed due to incompleteness of $\leadsto_{R,\cE}$. Let us see an example.

\begin{example}\label{ex:missed}
Consider  a 
simpler version of the client-server protocol given in 
 Example \ref{ex:handshakeNOFVP},
 where the shared key 
 {\tt K} is set to a fixed value (for simplicity, {\tt s(0)}) and messages consist of just one symbol. In this setting, we can greatly simplify the equational theory of Figure \ref{fig:caesar} 
into an equational theory $\cE$ that only contains the encryption and decryption equations:
\begin{verbatim}
eq enc(a,s(0)) = b [variant] .
eq enc(b,s(0)) = c [variant] .
eq enc(c,s(0)) = a [variant] .
eq dec(a,s(0)) = c [variant] .
eq dec(b,s(0)) = a [variant] .
eq dec(c,s(0)) = b [variant] .
\end{verbatim}
 Note that $\cE$-unification can be performed by fV-narrowing, since $\cE$ has the FVP (in contrast to the equational theory of Example \ref{ex:handshakeNOFVP}, which did not have the FVP) and fV-narrowing  executability  conditions are  met 
 (in particular, $\cE$ is convergent
  and $B$-unification is  trivially decidable since $B=\emptyset$).

The protocol dynamics is implemented by the  rewrite rules {\tt reply},  {\tt rec}, and 
 {\tt req} that are specified in Example \ref{ex:handshakeNOFVP}.
 Such rules are clearly not topmost since they may be applied to (strict) fragments of an object-oriented configuration.

Now, consider the reachability goal $G$

{\footnotesize
\begin{align*}
\exists {\tt S:Symbol}&\\
&\mbox{\tt < Cli-A : Cli\ |\ server : Srv-A , data : S , key : s(0) , status : mt >}\\
&\mbox{\tt < Srv-A : Serv\ |\ key : s(0) >  (Srv-A  <-  \{ Cli-A , c \} )}\\
&								\longrightarrow^*\\									  
&\mbox{\tt < Cli-A : Cli | server : Srv-A , data : S , key : s(0) , status : success > }\\
&\mbox{\tt < Srv-A : Serv | key : s(0) > }
\end{align*}									  
}
\noindent
The goal admits the  solution ${\tt \{S\mapsto b\}}$ and  proves  that
the handshake between client {\tt Cli-A} and server {\tt Srv-A}  succeeds when {\tt Cli-A} sends the encrypted message {\tt c} to {\tt Srv-A} if the original
unencrpyted data {\tt Q} is equal to {\tt b}.

Nonetheless, since completeness of the narrowing relation $\leadsto_{R,\cE}$  is not ensured for rewrite theories that are not topmost,  the solution ${\tt \{S\mapsto b\}}$ for $G$ cannot be found.  Indeed, executing $G$ by means of the {\tt vu-narrow} built-in Maude command, which implements narrowing-based reachability search in Maude,  produces the following output:  

{\footnotesize
\begin{verbatim}
vu-narrow [1, 100] in CLI-SERV-PROTOCOL-OBJECT-ORIENTED-ONESYMBOL : 
    < Cli-A : Cli | server : Srv-A , data : S:Symbol , key : s(0) , status : mt > 
    < Srv-A : Serv | key : s(0) > (Srv-A <- {Cli-A,c})
    =>* 
    < Cli-A : Cli | server : Srv-A , data : S:Symbol , key : s(0) , status : success > 
    < Srv-A : Serv | key : s(0) > .

No solution.
rewrites: 3361 in 409ms cpu (413ms real) (8212 rewrites/second)
\end{verbatim}
}
\end{example}

\subsection{Relaxing the topmost requirement for object-oriented specifications}\label{sec:relax}

Object-oriented specifications, albeit not topmost, fall into the more general category of topmost {\em modulo $Ax$} rewrite theories, for which the topmost property can be easily recovered by means of an automatic program transformation. 
Topmost {\em modulo $Ax$} rewrite theories are rewrite theories where $Ax$ consists of any of the combinations of axioms  $ACU$, $ACU_l$, $ACU_r$, $AC$, $AU$, $AU_l$, $AU_r$, or  $A$
  for a given binary symbol $\otimes : \mathit{Config} \: \mathit{Config} \rightarrow \mathit{Config}$   of the signature. The operator $\otimes$ is used to build {\em system configurations} that obey the {\em structural axioms} of $\otimes$ given by $Ax$.
\begin{definition}[topmost {\em modulo} $Ax$ rewrite theory]\label{def:topmostAx}
Let $\cR=(\Sigma,E\uplus B,R)$ be a rewrite theory with a finite poset of sorts ${(S, <)}$. Let $\mathit{Config}$ be a top sort in $S$, and let $\otimes : \mathit{Config}\:\mathit{Config}\to\mathit{Config} \in \Sigma$ be a binary operator that obeys a combination of associativity, commutativity, and/or identity axioms 
$Ax\in\{ACU, ACU_l, ACU_r, AC, AU, AU_l, AU_r, A\}$.

The theory $\cR$  is said to be topmost {\em  modulo} $Ax$  if 
\begin{enumerate}
\item for each rule $(\lambda\Rightarrow\rho\mbox{ if } C)$ in $R$,  $\lambda$ and $\rho$ are terms of sort {\it Config}  and $C$ does not contain terms of sort 
 $\sort{s}$
such that its top sort 
$[\sort{s}] =\mathit{Config}$;
\item $\otimes$ is the only operator in $\Sigma$ whose 
arguments include a sort 
$\sort{s}$
such that its top sort  $[\sort{s}] =\mathit{Config}$.
\end{enumerate}
The rewrite rules in $R$ are also said to be topmost modulo $Ax$.
\end{definition}
\begin{example}
The rewrite theory in Example \ref{ex:handshakeNOFVP} is topmost modulo $ACU$.
To show this,  it suffices to interpret the juxtaposition ACU operator  {\it \_\; \_ : Configuration Configuration }$\to$ {\it Configuration}, which is used to build object-oriented configurations,
as the operator $\otimes : \mathit{Config}\:\mathit{Config}\to\mathit{Config}$ of Definition \ref{def:topmostAx}. 
\end{example}

We are able to convert topmost modulo $Ax$ rewrite theories   into topmost theories automatically  by means of the following solution-preserving theory
transformation.

\begin{definition}[topmost extension of $\cR$]\label{def:transfTopmost} 
Let  $\cR=(\Sigma,E\uplus B,R)$ be a topmost modulo $Ax$ rewrite theory and $Ax\in B$.
 Let $X$, $X_1$, and $X_2$ be variables of sort {\it Config} not occurring in either $R$ or $E$. We define the topmost rewrite theory  $\hat{\cR}=(\hat{\Sigma},E\uplus B,\hat{R})$ 
 where
 $\hat{\Sigma}$ extends $\Sigma$  by adding a new top sort {\it State}, and a  new operator $\{\_ \} \colon \mathit{Config}\to\mathit{State}$; and  $\hat{R}$ is obtained by transforming $R$ according to $Ax$ as follows. 
For each $(\lambda \Rightarrow \rho\  \mathit{if}\: C )\in R$\\

\fbox{\begin{tabular}{ll}
 {\bf Case} $Ax=ACU$. & $(\{X\otimes \lambda\} \Rightarrow \{X\otimes \rho\} \mathit{if}\: C) \in \hat{R}$;\\
 {\bf Case} $Ax=ACU_l$. & $(\{X\otimes \lambda\} \Rightarrow \{X\otimes \rho\} \mathit{if}\: C) \in \hat{R}$;\\
{\bf Case} $Ax=ACU_r$. & $(\{X\otimes \lambda\} \Rightarrow \{X\otimes \rho\}  \mathit{if}\: C) \in \hat{R}$;\\
{\bf Case} $Ax=AC$.& $(\{X\otimes \lambda\} \Rightarrow \{X\otimes \rho\}  \mathit{if}\: C) \in \hat{R}$, \\
& $(\{\lambda\} \Rightarrow \{\rho\}  \mathit{if}\: C ) \in \hat{R}$;\\
{\bf Case} $Ax=AU$. &$(\{X_1\otimes \lambda\otimes X_2\} \Rightarrow \{X_1\otimes \rho\otimes X_2\}  \mathit{if}\: C) \in \hat{R}$;\\
{\bf Case} $Ax=AU_l$. &$(\{X_1\otimes \lambda\otimes X_2\} \Rightarrow \{X_1\otimes \rho\otimes X_2\}  \mathit{if}\: C) \in \hat{R}$;\\
&$(\{X_1\otimes \lambda \} \Rightarrow \{X_1\otimes \rho \}  \mathit{if}\: C) \in \hat{R}$;\\
{\bf Case} $Ax=AU_r$. &$(\{X_1\otimes \lambda\otimes X_2\} \Rightarrow \{X_1\otimes \rho\otimes X_2\}  \mathit{if}\: C) \in \hat{R}$;\\
&$(\{\lambda\otimes X_2\} \Rightarrow \{\rho\otimes X_2\}  \mathit{if}\: C) \in \hat{R}$;\\
{\bf Case} $Ax=A$.& $(\{X_1\otimes \lambda\otimes X_2\} \Rightarrow \{X_1\otimes \rho\otimes X_2\}  \mathit{if}\: C) \in \hat{R}$, \\
&$(\{X_1\otimes \lambda\} \Rightarrow \{X_1\otimes \rho\}  \mathit{if}\: C) \in \hat{R}$,\\
&$(\{\lambda\otimes X_1\} \Rightarrow \{\rho\otimes X_1\}  \mathit{if}\: C) \in \hat{R}$,\\
&$(\{\lambda\} \Rightarrow \{\rho\}  \mathit{if}\: C) \in \hat{R}$.
\end{tabular}
}
\end{definition}
We call $\hat{\cR}$ the {\em topmost extension} of $\cR$.

The topmost extension of a rewrite theory $\cR$ preserves the rewriting semantics of $\cR$, as shown in the following proposition.

\begin{proposition}[correctness and completeness of the topmost extension]\label{pr:TopmoduloAXTransf}
Let $\cR=(\Sigma,E\uplus B,R)$, with $\cE=(\Sigma, E\uplus B)$, be a topmost modulo $Ax$ theory and  let $\hat{\cR}$ be
the topmost extension of $\cR$.
For any terms $t_i$ and $t_f$ of sort $\mathit{Config}$, $t_i\rightarrow_{R,\cE}^{*}t_f$ iff  
$\{t_i\}\rightarrow_{\hat{R},\cE}^{*}\{t_f\}$.
\end{proposition}

The above proposition implies that the set of solutions of a reachability goal $G=(\exists X)\; t\longrightarrow^* t'$ in $\cR$ is the same as 
the set of solutions of $\hat{G}=(\exists X)\; \{t\}\longrightarrow^* \{t'\}$ in the topmost extension $\hat{\cR}$ of $\cR$. Thus, to find a complete set of solutions in 
$\cR$  for $G$, we can just find a complete set of solutions for $\hat{G}$ in $\hat{\cR}$. 
Furthermore, as $\hat{\cR}$ is topmost, narrowing with $\hat{R}$ modulo $\cE$ is complete and can thus be effectively used to search for all solutions of $G$ even if they could not be computed in Maude by narrowing in $\cR$.

\begin{theorem}[strong completeness of the topmost extension]\label{theo:sc}
Let  $\cR=(\Sigma,E\uplus B,R)$ be a topmost modulo $Ax$ rewrite theory, with $Ax\in B$, and let $\cE=(\Sigma,E\uplus B)$
be  a convergent, finite variant equational theory where $B$-unification is decidable.  Let $\hat{\cR}=(\hat{\Sigma},E\uplus B,\hat{R})$
be the topmost extension of $\cR$ and let $G=(\exists X)\; t \longrightarrow^{*} t'$ be a reachability goal in $\cR$. 
If $\sigma$ is a solution for $G$ in $\cR$, then there exists a term $t''$ such that $\{t\}\leadsto_{\hat{R},\cE}^* \{t''\}$ with computed substitution $\theta$ and
$\sigma=_\cE\theta\eta$, where $\eta$ is an $\cE$-unifier of $t''$ and $t'$.
\end{theorem}

\begin{example}\label{ex:extension}
Consider the  rewrite theory $\cR$ for the simplified client-server protocol and the reachability goal $G$ of  Example \ref{ex:missed}. 
 Let  {\tt EXT} be a variable that we use as a placeholder for arbitrary object-oriented configurations to generate an ACU topmost extension $\hat{\cR}$ of ${\cR}$.
The topmost extension $\hat{\cR}$  includes  the three rewrite rules of Figure \ref{fig:missed-CS}.
\begin{figure}[h!]
{\footnotesize
\begin{Verbatim}[frame=single]  
crl [req'] : { < C : Cli | server : S , 
                         data : M , 
                         key : s(K) , 
                         status : mt > EXT }
            =>  
				
            { < C : Cli | server : S , 
                         data : M , 
                         key : s(K) , 
                         status : mt > ( S <- { C , enc(M,K) } )
                         EXT } if s(K) < len /\ enc(M,s(K)) = enc(enc(M,K),s(0)) .
	
rl [reply'] : { < S : Serv | key : K > 
                (S <- {C,M}) EXT }
				 
               =>
				  
               { < S : Serv | key : K > 
                 (C <- {S, dec(M,K)}) EXT } [ narrowing ] .
				  	
rl [rec'] : { < C : Cli | server : S , 
                          data : M , 
                          key : K , 
                          status : mt > 
             (C <- {S,M}) EXT }
				
            =>
				
            { < C : Cli | server : S , 
                          data : M , 
                          key : K, 
                          status : success > EXT }[ narrowing ] .   
\end{Verbatim}
} 
\caption{Topmost extension of the simplified client-server protocol of Example \ref{ex:missed}}\label{fig:missed-CS}
\end{figure}

In contrast to Example \ref{ex:missed}, here the solution ${\tt \{S\mapsto b\}}$ for $G$   can be automatically obtained in Maude by searching the solution 
for $\hat{G}$ in $\hat{\cR}$ via narrowing. This is because $\hat{\cR}$ is topmost and hence $\cE$-unification is decidable. In fact, executing $\hat{G}$ in $\hat{\cR}$ via the {\tt vu-narrow} command 
{\footnotesize
\begin{verbatim}
vu-narrow [1, 100] in CLI-SERV-PROTOCOL-OBJECT-ORIENTED-ONESYMBOL-EXTENDED : 
    { < Cli-A : Cli | server : Srv-A , data : S:Symbol , key : s(0) , status : mt >  
      < Srv-A : Serv | key : s(0) > (Srv-A <- {Cli-A,c}) }
    =>*
    { < Cli-A : Cli | server : Srv-A , data : S:Symbol , key : s(0) , status : success > 
      < Srv-A : Serv | key : s(0) > } .
 \end{verbatim}   
}
\noindent  produces
{\footnotesize
\begin{verbatim}
Solution 1
rewrites: 55 in 10ms cpu (11ms real) (5257 rewrites/second)
state: { < Srv-A : Serv | key : s(0) > < Cli-A : Cli | server : Srv-A,data : b,
    key : s(0),status : success > }
accumulated substitution:
S:Symbol --> b
\end{verbatim}
}
\end{example}

It is worth noting that the program transformation in Definition \ref{def:transfTopmost} is not only able to transform a topmost modulo $Ax$ rewrite theory $\cR$ into a semantically equivalent topmost rewrite theory, but it also 
ensures coherence of the rewrite rules of $\hat{\cR}$ w.r.t. $Ax$ for free. 
Indeed, the transformation algorithm  of Definition \ref{def:transfTopmost} 
extends to  object-oriented specifications 
the Maude (explicit) coherence completion algorithm that  
  automatically produces coherent rewrite theories w.r.t. any given combination of ACU axioms  \cite{maude-manual}.

 In the following section, we show that  our partial evaluation algorithm
 can deal with finite variant  theories as well as theories that do not satisfy the FVP, and it is powerful enough to transform them  into specialized theories that, in many cases, do satisfy the FVP, thus enabling the use of both levels of narrowing on the theory. In Section \ref{sec:instance}, we show how  the communication protocol of Example $\ref{ex:handshakeNOFVP}$ can be automatically transformed into a specialized theory that meets the FVP and in which reachability goals of the form 
$(\exists X) ~t \longrightarrow^* t'$   can be symbolically solved by means of narrowing after having applied the topmost extension described above.

%% file: 3-scheme.tex
\section{Specializing Rewrite Theories with \Presto\/}\label{sec:nper}
	In this section, we present the 
	 specialization procedure 
	 {\sc NPER}$^{\cU}$, which allows a rewrite theory $\cR=(\Sigma,E\uplus B, R)$   to be optimized 
	by  specializing the underlying equational theory $\cE=(\Sigma,E\uplus B)$ with respect to the 
 calls   in the rewrite rules of $R$. 
 The    {\sc NPER}$^{\cU}$ procedure extends the  equational, narrowing-driven partial evaluation 
 algorithm  {\sc EqNPE}$^{\cU}$   of \cite{ACEM20-jlamp}, 
  which applies to equational theories 
  and is parametric  w.r.t.\ 
  an unfolding operator $\cU$ that is used to construct finite narrowing trees for a given expression.

\subsection{Partial Evaluation of Equational Theories}\label{sec:EqNPE}

Given   $\cE=(\Sigma, E \uplus B)$, we consider  a natural partition of the signature as $\Sigma = \cD \uplus \cC$, where $\cC$ are the {\em constructor} symbols, which are used to define the (irreducible) values of the theory, and  $\cD$ are the {\em defined} symbols, which are evaluated away by 
equational rewriting.
Given a set $Q$
 of   calls (henceforth called \emph{specialized calls}), the main goal of  {\sc EqNPE}$^{\cU}$   is to derive a new  equational theory $\cE'$ that computes the same substitutions
(and values) as $\cE$ for any input term $t$ that is a recursive instance
(modulo $B$) of the terms in $Q$ (i.e.,\ $t$ is an instance of some term in $Q$ and, recursively, every non-variable subterm of $t$ that is rooted by a defined symbol is an instance of some term in $Q$). 
The algorithm follows the classic control strategy of logic specializers \cite{MG95}, with two separate components: 1) local control (managed by the unfolding operator), which avoids infinite evaluations and is responsible for the construction of the residual rules for each specialized call; 2) global control or control of polyvariance (managed by an abstraction operator), which avoids infinite iterations of the partial evaluation algorithm and decides which specialized functions appear in the transformed theory.  Abstraction  guarantees that only finitely many expressions are evaluated, thus ensuring global termination. 
 
The transformation of \cite{ACEM20-jlamp} is done by iterating two steps: 

\begin{enumerate}
\item Symbolic execution ({\em Unfolding}).
A finite, possibly partial
folding variant narrowing	tree   for each input  term $t$  of $Q$ is generated. This is done by using the unfolding operator $\cU(Q,\vec{\cE)}$ that determines when and how to stop the  derivations in the narrowing tree.\\ 
		
\item Search for regularities ({\em Abstraction}).
In order to ensure that  
all calls that may occur at runtime are  \emph{covered}
	by the specialization, 
	it must be guaranteed that every (sub-)term in 
	any leaf of the tree is {\em equationally closed} w.r.t.\ $Q$. Equational closedness
	\footnote{A term $u$ is closed modulo $B$  w.r.t.\ $Q$
	iff either: 
	(i) it does not contain defined function symbols of $\cD$, or (ii) there exists a 
	substitution $\theta$ and a (possibly renamed) $q \in Q$ such that $u=_B q\theta$, and 
	the terms in $\theta$ are recursively $Q$-closed. 
	For instance, given a defined  binary  symbol $\bullet$ that does not obey any structural axioms, 
	the term $t = a \bullet (Z\bullet  a)$  is 
	closed w.r.t. $Q=\{a \bullet  X, Y \bullet  a \}$ or
	$\{X \bullet  Y\}$, 
	but it is not closed
		with $Q$ being $\{a \bullet  X\}$; however, it would be closed if $\bullet$ were commutative.}
				  extends the  classical PD closedness \cite{LS91}  by: 1) considering  $B$-equivalence of terms; and 2) recursing over the term structure (in order to handle
nested function calls).  
To  properly add the non-closed (sub-)terms  to the set of  already partially evaluated calls, an abstraction operator ${\cA}$ is applied that yields a new set of terms which may need further evaluation.

\end{enumerate}

Steps  (1) and (2) are iterated 
as long as new terms are generated until a fixpoint is reached, and the augmented, final set $Q'$  of 
specialized calls is yielded.
Note that the algorithm does not explicitly compute a partially evaluated equational theory.
It does so implicitly;
the desired partially evaluated equations $E$ are generated
as the set of {\em resultants}  
$t\sigma = t'$   	associated with the  derivations in the narrowing tree  from the root  $t\in Q'$ to the leaf $t'$  
 with computed substitution $\sigma$, such that the closedness condition modulo $B$ w.r.t. $Q'$ is satisfied for all function calls that appear in the right-hand sides of the 
 equations in $E'$.
 We assume the existence of  a function  {\sc GenTheory}$(Q',(\Sigma, E \uplus B))$
that delivers the partially evaluated equational theory
$\cE'=(\Sigma,E'\uplus B)$
that is uniquely determined by $Q'$ and  the original equational theory $\cE=(\Sigma, E \uplus B)$. 
Formally,
\begin{align*}
\mbox{{\sc GenTheory}}(Q',\cE) = (\Sigma,\{t\sigma = t'\,\mid\, & t\in Q', t'\in\cU(Q',\vec{\cE}),\: t \mbox{ fV-narrows to } t' \\
&\mbox{with computed substitution } \sigma\} \uplus B).
\end{align*}

\subsection{The {\sc NPER}$^{\cU}$ 
Scheme for the Specialization of Rewrite Theories }\label{sec:NPER}

The specialization of the rewrite theory $\cR=(\Symbols,E\uplus B,R)$ is achieved by partially evaluating the  hosted equational theory $\cE=(\Sigma,E\uplus B)$ w.r.t.\ the rules  of $R$, which is done by using 
the 
partial evaluation procedure  {\sc EqNPE}$^{\cU}$   of Section \ref{sec:EqNPE}. 
By providing suitable unfolding (and abstraction) operators, different instances of the specialization scheme can be defined.

Before going into the details, 
let us summarize the advantages of the final, specialized rewrite theory  $\cR_{\sf sp}=(\Symbols_{\sf sp},E_{\sf sp}\uplus B_{\sf sp},R_{\sf sp})$ over the original 
$\cR=(\Symbols,E\uplus B,R)$:
\begin{enumerate}
\item The transformed rewrite theory $\cR_{\sf sp}$ generally performs more efficiently than $\cR$ for two reasons. On the one hand,  the partially evaluated  equational theory $\cE_{\sf sp}=(\Symbols_{\sf sp},E_{\sf sp}\uplus B_{\sf sp})$ has been optimized to fulfill the needs of the host rewrite rules of $R$ (that have been correspondingly transformed into $R_{\sf sp}$). On the other hand, the specialization process may remove unnecessary equational axioms and rule conditions, thereby significantly speeding up  
 the rewriting and narrowing computations  in the specialized theory at the level of both equations and rules.
\item The specialization may cut down an infinite folding variant narrowing space to a finite one.
Actually, the transformed equational theory $\cE_{\sf sp}$  may have the FVP whilst the original one may not, so that symbolic analysis based on narrowing with $R_{\sf sp}$ modulo $\cE_{\sf sp}$  is enabled, whereas narrowing with $R$ modulo $\cE$ is not. 
\end{enumerate} 

The   {\sc NPER}$^{\cU}$   procedure is outlined in Algorithm \ref{alg:NPER-Algorithm} and  
it consists of  two phases.\\ 

\begin{algorithm}[!h]\caption{Symbolic Specialization of Rewrite Theories {\sc NPER}$^{\cU}\!\!(\cR)$ }\label{alg:NPER-Algorithm}
{ 			\begin{algorithmic}[1]
				\Require
				\Statex 
				A rewrite theory $\cR = (\Sigma,E\uplus B,R)$, 
				 an unfolding 
operator ${\cU}$ 

				\Function{{\sc NPER}$^{\cU}$ }{$\cR$} 
				\Statex {\it Phase 1. Partial Evaluation}
				\State $R' \gets \{(\lambda\downarrow_{\vec{E},B}) \Rightarrow (\rho\!\downarrow_{\vec{E},B}) ~\mathit{if}\ C\!\downarrow_{\vec{E},B}~|~ (\lambda \Rightarrow \rho  ~\mathit{if}\ C) \in R 
								\}$
				\State $Q \gets mcalls(R')$
				\State $Q' \gets \Call{EqNPE$^{\cU}$}{(\Sigma,E\uplus B), Q}$ 
				\State $(\Sigma',E'\uplus B') 
				 \gets \Call{GenTheory}{Q',(\Sigma, E \uplus B)}$
								\Statex {\it Phase 2. Compression}
				\State $\cR_{ren} \gets \Call{Rename}{(\Sigma,E\uplus B,R'),(\Sigma',E'\uplus B'),Q'}$
               \State $(\Sigma_{\sf sp}, E_{\sf sp}\uplus B_{\sf sp},R_{\sf sp}) \gets \Call{Simplify}{\cR_{ren}}$
				\EndFunction
				\State \Return $(\Sigma_{\sf sp}, E_{\sf sp}\uplus B_{\sf sp},R_{\sf sp})$
\end{algorithmic}
}
		\end{algorithm}

\noindent{\bf Phase 1 (Partial Evaluation).}
It applies the {\sc EqNPE}$^{\cU}$  algorithm to specialize 
the equational theory $\cE=(\Sigma, E \uplus B)$
w.r.t.\ a set $Q$ of specialized calls   that consists of 
 all of the {\em maximal function calls}
that appear in  the $(\vec{E},B)$-normalized version $R'$ of the  conditional rewrite rules of $R$, where  $C\!\downarrow_{\vec{E},B}$ denotes $c_1\!\downarrow_{\vec{E},B}\wedge\ldots\wedge c_n\!\downarrow_{\vec{E},B}$ for any rule condition $C=c_1\wedge\ldots\wedge c_n$. 
Given   $\Symbols = (\cD \uplus \cC)$, a maximal function call in a term $t$ is any outermost subterm of $t$ that is rooted by a defined function symbol of $E$.
 By $mcalls(R)$, we denote the set of all maximal calls in both the right-hand sides and  the conditions of the 
 rules of $R$.
This phase produces the new set of specialized calls $Q'$ from which the partial evaluation  $\cE'=(\Sigma', E' \uplus B')$ of $\cE$ w.r.t.\ $Q$ is   unambigously  derived by 
 {\sc GenTheory}($Q',(\Sigma,E\uplus B)$).\\

\noindent{\bf  Phase 2 (Compression).}  
It consists of a 
refactoring transformation
 that computes a new, simplified rewrite theory 
   $\cR_{\sf sp}=(\Sigma_{\sf sp},E_{\sf sp}\uplus B_{\sf sp},R_{\sf sp})$
 by taking as input the $({\vec{E},B})$-normalized rewrite theory $\cR'=(\Sigma,E\uplus B,R')$, the computed partially evaluated theory $\cE'=(\Sigma', E' \uplus B')$, and the final set of specialized calls $Q'$ from which $\cE'$ derives.
 
\begin{algorithm}[!h]\caption{Renaming algorithm}\label{alg:compress}
			\begin{algorithmic}[1]
				\Require
				\Statex 
				A rewrite theory $\cR' = (\Sigma,E\uplus B,R')$, a partial evaluation $\cE' =(\Sigma',E'\uplus B')$
				of $(\Sigma, E\uplus B)$ w.r.t. a set of specialized calls $Q$
				
				\Function{{\sc Rename}}{$\cR',\cE',Q$}
				\State Let $\delta$ be an independent renaming for $Q$ in
                \State \hspace{1cm}$E_{ren} \gets {\bigcup_{t \in Q}} \{ \delta(t)\theta = {\sc RN}_{\delta}(t')\mid  t\theta = t' \in E'\}$ 
                \State \hspace{1cm}$R_{ren}\gets \{{\sc RN}_{\delta}(\lambda)\Rightarrow {\sc RN}_{\delta}(\rho) ~\mathit{if}\ ~{\sc RN}_{\delta}(C)  \mid \lambda \Rightarrow \rho ~\mathit{if}\ C\in R' \}$
	       \State \hspace{1cm}$\Sigma_{ren} \gets (\Sigma' \setminus\{f\mid f \mbox{ occurs in } ((E\uplus B)\setminus(E'\uplus B')) \})
\cup\{root(\delta(t))\mid t\in Q\}$
	        	\State \hspace{1cm}$B_{ren} \gets\{ax(f)\in B' \mid f\in\Sigma'\cap\Sigma_{ren}\}$			
	   
	   	\State  \hspace{1cm}$\cR_{ren} \gets (\Sigma_{ren}, E_{ren}\uplus B_{ren},R_{ren})$
				
				\EndFunction
				\State \Return $\cR_{ren}$
				\Statex where
\vspace{-4ex}				
$$	\begin{array}{l}
		{\sc RN}_{\delta}(t) = 
		\left\{ \begin{array}{ll@{}}
		c(\ol{{\sc RN}_{\delta}(t_{n})}) & \mbox{if }  t=c(\ol{t_n}) \mbox{ with } c : \ol{s_{n}} \rightarrow s \in \Sigma ~s.t.\ 
		c \in \cC,   ~ ls(t)=s, ~ n\geq 0
		\\[1ex]
		\delta(u)\theta'  & \mbox{if } \exists \theta, \exists u \in Q \mbox{ s.t. }
		t =_B u\theta \mbox{ and } 
				\theta' = \{ x \mapsto {\sc RN}_{\delta}(x\theta)
		\mid x \in \dom(\theta) \} \\[0.25ex]
		t & \mbox{otherwise}
		\end{array} \right.
		\end{array}$$
			\end{algorithmic}
\vspace{-2ex}				
		\end{algorithm} 

The compression phase consists of two subphases: 
{\it renaming} and {\it rule condition simplification}. 
The renaming (sub-)phase 
 aims at yielding a much more compact equational theory 
$\cE_{ren}=(\Sigma_{ren}, E_{ren} \uplus B_{ren})$ where unused symbols and unnecessary repetitions of variables are removed and  equations of $E_{ren}$ are simplified by recursively renaming all expressions that are $Q'$-closed modulo $B$
by using an independent renaming function that is derived from the set of specialized calls $Q'$. 
Formally,  an {\em independent renaming} $\delta$  for   a set of terms $T$ and a signature $\Sigma$ is a mapping  from terms to terms, which is defined as
follows.
For each	$t$ of sort $s$ in $T$ with $root(t)=f$, 
				$\delta(t)={f_{t}(\ol{x_{n}:s_{n}})}$, 
				where $\ol{x_{n}}$ are the distinct variables in $t$ in the
				order of their first occurrence and $f_{t}\colon \overline{s_n}\to s$ is a new function symbol
				that does not occur in $\Sigma$ or $T$ and
				is different from 
				the root symbol of any other $\delta(t')$,
				with $t'\in T$ and $~t'\neq t$. Abusing notation, we let $\delta(T)$ denote
				the set 
				$\{\delta(t)\mid t\in T\}$ for a given set of terms $T$.
\begin{example}
Consider the rewrite theory in Example \ref{ex:handshakeNOFVP} together with the set of specialized calls  $$Q= \{\tt dec(enc(M:Data,s(s(0)))), enc(enc(M,K1:Nat),K2:Nat)\}$$
An 
independent renaming $\delta$ for $Q$ is given by
\begin{align*}
\delta =\{&{\tt dec(enc(M:Data,s(s(0)))) \mapsto f0(M:Data),} \\
        &{\tt  enc(enc(M:Data,K1:Nat),K2:Nat) \mapsto f1(M:Data,K1:Nat,K2:Nat)}\}
\end{align*}      
where {\tt f0: Data -> Data} and {\tt f1: Data Nat Nat -> Data} are new function symbols. 
\end{example} 
 Renaming is performed  by the 
{\sc Rename} function given in Algorithm \ref{alg:compress}.			

		  Essentially, the  {\sc Rename} function recursively  computes (by means of the function ${\sc RN}_{\delta}$) a new equation set $E_{ren}$ by replacing each call in $E'$ 
		by a call  to the corresponding renamed function according to $\delta$.
		Furthermore, a new rewrite rule set  $R_{ren}$ is also produced by consistently applying ${\sc RN}_{\delta}$ to the conditional rewrite rules of $R'$. Formally, each conditional rewrite rule $\lambda \Rightarrow \rho ~\mathit{if}~ C$ in $R'$ is  transformed into the rewrite rule 
				${\sc RN}_{\delta}(\lambda)\Rightarrow {\sc RN}_{\delta}(\rho) ~\mathit{if}~ {\sc RN}_{\delta}(C)$, in which every maximal function call $t$ in the 
		rule is recursively renamed according to the independent renaming $\delta$.
  The renaming algorithm also computes the specialized signature $\Sigma_{ren}$ and  restricts the set $B'$ to those axioms obeyed by the function symbols in  $\Sigma'\cap\Sigma_{ren}$, delivering  the rewrite theory 
$(\Sigma_{ren}, E_{ren}\uplus B_{ren},R_{ren})$.

Note that, while the independent renaming suffices to rename the left-hand sides of the equations in $E'$
(since they are mere instances
of the specialized calls), the right-hand sides are renamed by means of the auxiliary function ${\sc RN}_{\delta}$, which recursively replaces each call in the given expression by a call to the corresponding renamed function (according to $\delta$).
\begin{algorithm}\caption{Simplification algorithm}\label{alg:simplify}
			\begin{algorithmic}[1]
				\Require
				\Statex 
				A rewrite theory $\cR = (\Sigma,E\uplus B,R)$, with  equational theory  $\cE=(\Sigma,E\uplus B)$
				\Function{{\sc Simplify}}{$\cR$}
								     \State $R'\gets \mathit{SimpConds}(\cR)$
				                     \State $E' \gets E \setminus \{ l=r \mid root(l)  \mbox{ does not occur  in either }   R' \mbox{ or } (E \setminus \{ l=r \}) \}$ 
                 \State$\Sigma' \gets \{f \in \Sigma\mid f \mbox{ occurs in } R' \mbox{ or } E' \}$
	        	\State $B' =\{ax(f)\in B \mid f\in\Sigma'\}$			
				\EndFunction				
				\State \Return $(\Sigma',E'\uplus B',R')$ 	
			\end{algorithmic}		
		\end{algorithm}

The simplification subphase is undertaken by the {\sc Simplify} procedure of  Algorithm 
 \ref{alg:simplify} that  further refines the rewrite theory $\cR_{ren}$. Indeed, the partial evaluation process may produce specialized rewrite rules that include 
conditions that can be safely removed  without changing the original program semantics. This transformation is implemented by function  $\mathit{SimpConds}(\cR)$ and may have a huge impact on program efficiency  since  Maude executes 
unconditional rules much faster than conditional rules because the evaluation of rule conditions is notoriously speculative in Maude. This is because, besides the fact that many equational matchers for the rule conditions may exist, solving rewriting conditions requires non-deterministic search 
\cite{SR06,maude-manual}.  The rule simplification function $\mathit{SimpConds}(\cR)$ is left generic in  Algorithm 
 \ref{alg:simplify} and  can be specialized for different kinds of  rewrite theories.

After simplifying rule conditions, there might exist spurious equations $l=r$ in $E_{ren}$ such that $root(l)$ does not occur  in either  the 
simplified rewrite rules of $R'$ or  in $(E_{ren}\setminus \{l=r\})$. Hence, 
{\sc Simplify}  removes such equations, and  the theory signature is   cleaned up by deleting any operators (and their corresponding axioms) that do not occur in the transformed equations and rules.

		A concrete implementation  $\mathit{SimpConds}^*(\cR)$ of the generic rule simplication function \linebreak
	$\mathit{SimpConds}(\cR)$ of Algorithm \ref{def:simplify2}
		 that works for the theories that are considered in this article  is given in the following definition:	 

		 \begin{definition}[Rule Condition Simplification]\label{def:simplify2}
		Let  $\cR = (\Sigma,E\uplus B,R)$ be a rewrite theory with an equational theory $\cE=(\Sigma,E\uplus B)$. Then,  $\mathit{SimpConds}^*(\cR)$  is given by the set		 		
        \begin{align*}
		\bigcup_{\lambda \Rightarrow \rho \;\mathit{if} \;C  \;\in  \;R} \{\lambda \Rightarrow \rho \;\mathit{if}\ 
	 \mathit{sCond}_{\cE}(\lambda,\rho,C,\mathit{nil}) \mid &~  \mbox{no condition  in } C \mbox{ is } ~\mathit{false}  
\}
				      		\end{align*}	
where  the definition of the auxiliary function $\mathit{sCond}_{\cE}(\lambda,\rho,C,\sharp C)$ is given in Figure \ref{fig:scond}. 
\end{definition}

 		       \begin{figure}[h!]
        {\small
		$$\mathit{sCond}_{\cE}(\lambda,\rho,C,\sharp C)=\begin{array}{l} 
				\left\{ \begin{array}{ll@{}}
		 \sharp C & \mbox{if (1) } C=\mathit{nil}\\
		\mathit{sCond}_{\cE}(\lambda,\rho,C'',\sharp C) &  \mbox{if (2) } C= (C' \wedge C''), \mbox{ with }  C'  \mbox{ being a maximal,}\\
		& ~~~~~~~~~\mbox{non-empty subsequence of } $C$ \mbox{ s.t.\ either}\\
		& \mbox{~~~~(a) $C'$ \mbox{consists of expressions of the form} }\\[-0.10ex]
                  &~~~~~~~~~t=t' \mbox{ (or } t:=t' \mbox{ or } t  \Rightarrow t') \mbox{ s.t.\ } t=_B t' ; \mbox{ or}  \\[-0.10ex]
		  &\mbox{~~~~(b) } C' \mbox{ consists of } n \mbox{ equations } t_{i} = t'_{i},  \\[-0.10ex]
		   &~~~~~~~~~~\mbox{with } t_{i} \neq_B t'_{i} \mbox{ for } i=1,\ldots, n, \\[-0.10ex]
                     &~~~~~~~~~~\cE \mbox{ has the FVP,}\\[-0.10ex]
                     &~~~~~~~~~~{\footnotesize \var(C')   \cap  \var(\lambda \Rightarrow \rho \;\mathit{if}\    \sharp C \wedge C'')=\emptyset},\\[-0.10ex]
		   &~~~~~~~~~~\mbox{and exists } \theta\  \mbox{s.t.\ }~t_i\theta =_\cE t'_i\theta \mbox{ for } i=1,\ldots,n  \\  [-0.10ex]
		\mathit{sCond}_{\cE}(\lambda,\rho,C',\sharp C) &  \mbox{if (3) } C= (c \wedge C'), \mbox{ with } c \mbox{ being either }
		\\
		  		   &  \mbox{~~~~(a) }t_1=t_2 (\mbox{ resp.\ } t_1 := t_2 \mbox{ or } t_2 \Rightarrow t_1)\mbox{ s.t.\ }\\[-0.10ex]
				    &~~~~~~~~~ t_1 ~B\mbox{-unifies with } t_2   \mbox{ (resp.\ } t_2 \mbox{ is a $B$-instance of}~ t_1), \\[-0.10ex]				   		    &~~~~~~~~~\mbox{and }   {\footnotesize(\var(c)  \cap   \var(\lambda \Rightarrow \rho \;\mathit{if}\    \sharp C \wedge C')=\footnotesize \emptyset}; \mbox{ or}\\
		& \mbox{~~~~(b) an equation } t _{1}= t_{2}, \mbox{with } t_{1} \neq t_{2},\\[-0.10ex]
                   &~~~~~~~~~~\cE \mbox{ has the CFVP}, 		CV_\cE(t_1) = CV_\cE(t_2),  \mbox{and} \\[-0.1ex]
		  &~~~~~~~~~~ 
{\tt narrowing}\not\in \mathit{att}(\lambda \Rightarrow \rho \;\mathit{if}\ C) \\  
		\mathit{sCond}_{\cE}(\lambda,\rho,C',\sharp C\wedge c) & \mbox{if (4) } C= (c \wedge C'), \mbox{~and Cases 2) and 3) do not apply}
		\end{array} \right.
		\end{array}$$
		}
			\caption{Definition of the $\mathit{sCond}$ function.}\label{fig:scond}
		\end{figure}

\noindent
Roughly speaking, for any conditional rule   $\lambda \Rightarrow \rho ~\mathit{if}~ C$ in $R$  such that   none of the conditions in $C$ is false\footnote{This is because a rule condition that is generally satisfiable (e.g.,\ $f(Y)=g(Y)$) can specialize into a contradiction w.r.t.\ the   equational theory that is considered   (e.g.,\ $\cE=\{f(X)=a, ~g(X)=b$)\}.},
the auxiliary function $\mathit{sCond}_{\cE}$ 
gets rid of 

\begin{itemize}
\item[-] (Case 2) those maximal subsequences\footnote{We consider ordered subsequences of not necessarily consecutive elements of $C$.}  $C'$ of $C$ such that 
\begin{itemize}
\item[(2.a)] all of its  elements are   tautological  expressions of the form $t=t$, $t':=t'$, or $t''\Rightarrow t''$ (so that $t$ and $t'$ unify with the empty substitution by using fV-narrowing). Note that this  optimization cannot be done by standard Maude  Boolean simplification, which does not apply to  non-ground, rule condition expressions; or 
\item[(2.b)] $C'$ consists of $n$ equations of the form   $t_{i} = t'_{i}$,  with $t_{i} \neq t'_{i}$, such that  there exists an $\cE$-unifier $\theta$ for all of the equations in the sequence. To safely perform this  optimization two extra conditions must hold: (i) $\cE$ must have the FVP 
to make $\cE$-unification decidable; (ii) the variables occurring in $C'$  must not occur in the rest of the rule to be simplified. The constraint on the variables in $C'$ is required  to preserve the narrowing semantics since we must ensure that 
removing the equationally satisfiable sequence $C'$ does not affect the overall computation.
\end{itemize}

\item[-](Case 3) 
\begin{itemize}
\item[(3.a)] those equations (resp.\ matching equations $t_1 := t_2$ and rewrite expressions $t_2 \Rightarrow t_1$) such that  $t_1$ and $t_2$ are $B$-unifiable (resp.\ $t_2$ is an instance of $t_1$ modulo $B$), and the variables in $t_1$ and $t_2$  do not occur in  the rest of the rule.
\item[(3.b)] those equations $t _{1}= t_{2}$,  with $t_{1} \neq t_{2}$, such that the set of (most general) {\em constructor variants} (i.e.,\ equational variants that only include constructor symbols and variables  \cite{Meseguer20-jlamp}) of $t_{1}$ and $t_{2}$ are the same (in symbols, $CV_{\cE}(t_1) = CV_{\cE}(t_2)$), which implies that $t _{1}$ and $t_{2}$ compute the very same set of constructor  values and answers. However, since variables in $t _{1}= t_{2}$ are not required to be apart from the variables in the rest of the rule, this simplification only applies when  the rule to be simplified is not enabled for narrowing but only for rewriting.
Otherwise, removing the condition $t _{1}= t_{2}$ might not preserve the overall computed answers. Furthermore,  to ensure that constructor variant sets are computable, two extra conditions on $\cE$ are required 
\cite{Meseguer20-jlamp}: (i) $\cE$ must have  the FVP, and (ii) $\cE$ must be {\em sufficiently complete modulo axioms} (i.e.,\  
it specifies total functions).
 We say that  equational theories satisfying (i) and (ii) have the {\em constructor finite variant property}\footnote{The (unconditional)  {\sc NPER}$^{\cU}$ scheme of  \cite{ABES22-jlamp} is instantiated  in \cite{ABEMS20-Festschrift} for theories that satisfy the CFVP.} (CFVP).
\end{itemize} 
 \end{itemize}  
Note that Cases 1 and 4 of the $\mathit{sCond}_{\cE}$ function are simply used to incrementally process the rule condition to be simplified, yet they do not remove any expression.

The renaming subphase of the compression transformation can be seen as a qualified  extension to the Rewriting Logic setting of the classical renaming\footnote{Renaming can
also be presented as a new definition in Burstall-Darlington style
unfold-fold-new definition transformation frameworks \cite{BD77}. The removal of unnecessary structure by renaming is discussed in \cite{AFV98b,GB90,Gallagher93}.}
 that is applied in the partial evaluation of logic programs and
implemented in well-known partial deduction tools such as Logen and ECCE \cite{LEVC+06},
which greatly lessens the program structure. 
The   rule condition simplification transformation is novel and  further contributes to lighten the specialized rewrite theory. Although more aggressive rule simplification is possible, our transformation is relatively straightforward and inexpensive. Actually, neither it propagates   answers from the removed   expressions to the rest of the rule, nor does it require replicating rules due to several equational unifiers.

Let us illustrate the rule condition simplification given by  $\mathit{SimpConds}^*(\cR)$ 
by means of two examples.

\begin{example}\label{ex:cond-simplification1}
Let $\cE=(\Sigma,E\uplus B)$ be the equational theory of Figure \ref{fig:caesar} modeling the {\em Caesar} cypher, and let $\cR$ be a rewrite theory $(\Sigma,E\uplus B,\{r\})$,
where $r$ is a conditional rewrite rule whose condition $C$ is  ${\tt enc(a,s(K:Nat)) = enc(b,K:Nat)}$. 
 By applying the partial evaluation algorithm {\sc NPER}$^{\cU}$ to $\cR$, first the rule condition  $C$ 
is normalized in $\vec{E}$ (modulo $B$) into 
\[ {\tt toSym(e(s(0),K))  = toSym(e(s(0),K)). }\]
After partially evaluating the equational theory, the rule condition   is transformed  
into a trivial equation {\tt f(K:Nat) = f(K:Nat)},
with independent renaming  
$${\tt toSym(e(s(0), K:Nat)) \mapsto f(K:Nat)}$$
for the call  {\tt toSym(e(s(0), K:Nat))}.
Then, the trivial equation {\tt f(K:Nat) = f(K:Nat)} is removed  (via Case 2.a of the {\it sCond} function), thereby  delivering an unconditional version of $r$ as final outcome.

\end{example}

\begin{example}\label{ex:cond-simplification2}
Let $\cE=(\Sigma,E\uplus B)$ be the equational theory of Figure \ref{fig:caesar} modeling the {\em Caesar} cypher, and let $\cR$ be a rewrite theory $(\Sigma,E\uplus B,\{r\})$,
where $r$ is a conditional rewrite rule whose condition $C$ is  ${\tt enc(S:Symbol,s(0)) = enc(enc(S:Symbol,0),s(0))}$. 
We note that this condition does not generally hold for every symbol {\tt S} in an arbitrary encryption function,  but it certainly does when 
considering the implementation of the {\tt enc}  function defined by the equational theory of Example \ref{ex:handshakeNOFVP}.
 We also assume that the variable {\tt S} only appears in the condition $C$ of the rule $r$. 
 By specializing in \Presto\
 the two (already normalized) maximal function calls in rule $r$, {\tt enc(S:Symbol,s(0))} and {\tt enc(enc(S:Symbol,0),s(0))}, we get an equational theory $\cE_{ren}$ containing the following equations 
\begin{verbatim}
  eq f0(a) = b [ variant ] .   eq f1(a) = b [ variant ] .
  eq f0(b) = c [ variant ] .   eq f1(b) = c [ variant ] .
  eq f0(c) = a [ variant ] .   eq f1(c) = a [ variant ] .
\end{verbatim}
where {\tt f0} and {\tt f1} are two new operators that are introduced to rename   the initial calls \linebreak
{\tt enc(S:Symbol,s(0))} and {\tt enc(enc(S:Symbol,0),s(0))}, and actually denote  the very same encryption function.

Furthermore, the condition {\tt C} is renamed as 
\[{\tt f1(S:Symbol) = f0(S:Symbol).}\] 
\noindent Since $\cE_{ren}$ has the finite variant property, equational unification of  $\tt f1(S:Symbol)$ and $\tt f0(S:Symbol)$ is decidable and  delivers
three unifiers, namely,  $\tt \{S \mapsto a\}$, $\tt \{S \mapsto b\}$ and $\tt \{S \mapsto c\}$. 
Hence, the condition {\tt f1(S:Symbol) = f0(S:Symbol)} can be safely removed from the specialized version of $r$  (via Case 2.b of the $\mathit{sCond}_{\cE}$ function), since variable $\tt S$ does not appear elsewhere in the rule.
\end{example}
 An example of condition simplification that exploits Case 3.b of the $\mathit{sCond}_{\cE}$ function is shown in Example \ref{ex:NoFVP2FVP}.

Given the   rewrite theory $\cR=(\Symbols,E\uplus B,R)$  and its specialization ${\cR_{\sf sp}}=${\sc NPER}$^{\cU}\!\!(\cR)$,  all of the executability conditions that are satisfied by $\cR$  are also satisfied by  ${\cR_{\sf sp}}=(\Symbols_{\sf sp},E_{\sf sp}\uplus B_{\sf sp},R_{\sf sp})$,   
 including the fact that 
$R_{\sf sp}$ is 
 coherent w.r.t.\ $E_{\sf sp}$ modulo $B_{\sf sp}$. This is proved in Theorem 1 of \cite{ABES22-jlamp} for the case of unconditional rewrite theories such that the left-hand sides of the rules in $R$ are $(\vec{E},B)$-\emph{strongly irreducible}, and straightforwardly extends to conditional rewrite theories  since 1)  the conditional version of {\sc NPER}$^{\cU}$ only augments the initial set $Q$ of specialized calls with  the maximal calls in the rule  conditions, 
while the core equational partial evaluation algorithm is kept untouched; and 2)   only  
 conditions that do not modify any semantic property are removed by the rule condition simplification that is performed  by the compression phase. 

Similarly, the strong semantic correspondence between  both narrowing and rewriting computations in $\cR$ and  $\cR_{\sf sp}$ was proved  in Theorem 2 of \cite{ABES22-jlamp} for   unconditional rewrite theories and immediately extends to the conditional 
case for the same reasons and under the above-mentioned strong irreducibility requirement.

\section{Instantiating the Specialization Scheme for Rewrite Theories}\label{sec:instance}
In the following, we outline two instances
 of the {\sc NPER}$^\cU$ scheme that are obtained by choosing  two distinct implementations of the unfolding operator $\cU$.
 More precisely, we present the unfolding operators  
 $\cU_{\mathit{FVP}}$  and $\cU_{\mathit{\ol{FVP}}}$ that allow $\cR$ to be specialized  when $\cE$ is (respectively, is not) a finite variant theory. Both operators have been implemented in \Presto.  

With regard to global control, 
 \Presto\ implements  the same abstraction operator for both scheme instantiations. This abstraction operator is described in \cite{ACEM20-fi} and
 relies on the  equational least general generalization algorithm of
\cite{AEMS21} that improves our original calculus in \cite{AEEM14-ic} by providing a finitary, minimal and complete set of order-sorted generalizers  modulo any combinations of associativity and/or commutativity and/or identity axioms for any unification problem. This  property holds even when the theory contains function symbols with an arbitrary number $n$ of distinct identity elements, $e_{1}, \ldots, e_{n}$, provided  
 their  respective \emph{least sorts} 
 are incomparable at the kind level. In symbols, 
 $[ls(e_{i})] \neq [ls(e_{j})]$ 
  for any $i\neq j$, with $i,j \in 1,\ldots, n$.
 Theories that satisfy this property are called $U$-$\mathit{tolerant}$ theories.
Note that the traditional one-unit condition  that was implicitly assumed in \cite{AEEM14-ic} implies $U$-tolerance. 

\begin{example} \label{LIST+MSET}
Consider the following Maude functional module that defines lists and multisets of
natural numbers:

{\small
\begin{verbatim}
fmod LIST+MSET-NO-U-tolerant is
 sorts Nat List MSet Top .
 subsorts Nat < List < Top .
 subsorts Nat < MSet < Top .
 op nil : -> List [ctor] . 
 op _;_ : List List -> List [ctor assoc id: nil] .
 op null : -> MSet [ctor] . 
 op _,_ : MSet MSet -> MSet [ctor assoc comm id: null] .
 op 0 : -> Nat [ctor] .
 op s : Nat -> Nat [ctor] .
endfm
\end{verbatim}
}
\noindent where the \texttt{ctor} declarations specify that
all of these operators are \emph{data constructors}. 
Note that, since \verb+Nat < List+ and \verb+Nat < MSet+,
 $0$ is both a list of length one and a singleton multiset,
but the list  \verb+0 ; s(0) ; s(s(0))+
and the multiset \verb+0 , s(0) , s(s(0))+
have incomparable least sorts \texttt{List}
and \texttt{MSet}.  
Moreover these are also the least sorts of \texttt{nil} and \texttt{null}, respectively.  
 Therefore, this theory is not $U$-\emph{tolerant}, since $[ls({\tt nil})] = [ls({\tt null})] = {\tt Top}$. 

Automated ways to achieve 
$U$-tolerance 
in a \emph{semantics-preserving manner} are discussed in \cite{AEMS21}.
\end{example}

\paragraph{\bf Case 1: $\cE$ is not a finite variant theory.}
When $\cE$ is not a finite variant theory, the fV-narrowing strategy may lead to the creation of an infinite fV-narrowing tree for some specialized calls in $Q$. 
In this case,  an \emph{equational order-sorted extension}  $\trianglelefteqsort_B$  \cite{ACEM20-jlamp} of the classical homeomorphic embedding relation\footnote{A notion of homeomorphic embedding for a typed language was presented in \cite{AGGP09}.}  $\trianglelefteq$  
is used to detect the risk of non-termination. 
Roughly speaking, a homeomorphic embedding relation is a  structural
preorder under which a term $t$ is greater than or equal to another
term $t'$  (i.e., $t$ embeds $t'$), written as 
$t \trianglerighteq t'$ , if $t'$ can be obtained from $t$ by deleting some
parts, e.g., $s(s(X+Y)\ast(s(X)+Y))$ embeds $s(Y\ast(X+Y))$).
Embedding gets much less intuitive when dealing with axioms. For example, let us write natural numbers in classical decimal notation and
    the addition operator $\verb!+!$ for natural numbers in prefix notation. Due to associativity and commutativity of symbol   
	$\verb!+!$, there can be many   equivalent terms such as $+(+(1,2),+(3,0))$ and $+(+(1,+(3,2)), 0)$, 
	all of  which are internally represented in Maude
	by  the flattened term $+(0,1,2,3)$. 
Actually, 	flattened terms like $+(1,0,2,3)$ can be further simplified into a single
		{\em canonical representative} $+(0,1,2,3)$, hence   $+(1,2) \trianglelefteqsort_B +(0,1,2,3)$ and also $+(2,1) \trianglelefteqsort_B +(0,1,2,3)$.
	In contrast, if $+$ was associative (and not commutative), then $+(2,1) \not\!\!\!\trianglelefteqsort_B +(1,0, 3, 2)$. A more detailed description  that   takes into account the extra complexity given by sorts and subsorts can be found in \cite{ACEM20-fi}.

When iteratively computing a sequence $t_1, t_2, \ldots,t_n$,  finiteness of the sequence can be guaranteed by using the well-quasi order relation $\trianglelefteqsort_B$  \cite{ACEM20-fi} as a whistle \cite{Leuschel98-sas}: whenever a new expression $t_{n+1}$ is to be added to the sequence, we first check whether $t_{n+1}$ embeds any
of the expressions already in the sequence. If that is the case, we say that  
the homeomorphic embedding test $\trianglelefteqsort_B$ whistles, i.e., it has detected (potential) non-termination and the computation has to be stopped. Otherwise, $t_{n+1}$ can be safely added to the sequence and the
computation can proceed.

The operator $\cU_{\mathit{\ol{FVP}}}$ implements such an embedding check to guarantee the termination of the unfolding phase. 
 Specifically, the generation of a fV-narrowing tree fragment is stopped when each fV-narrowing derivation (i.e.,\ each branch of the fV-narrowing tree) is stopped because 
  a term is reached that is either unnarrowable or it  embeds (w.r.t. $\trianglelefteqsort_B$) a previously narrowed 
term in  the same derivation.

\begin{example}\label{ex:handshakeNOFVP-specialized}
Consider  the  specific instance of the 
rewrite theory of Example \ref{ex:handshakeNOFVP}
where  servers and clients use a pre-shared fixed key  {\tt K=s(0)} but messages can have any length, in contrast to Example \ref{ex:missed}. 

  Let $\cR=(\Sigma,E\uplus B,R)$ be such a rewrite theory, where $\cE=(\Sigma,E\uplus B)$ is the equational theory of $\cR$.
 In $\cE$, the fV-narrowing  trees associated with encryption and decryption functionality 
 may be infinite since $\cE$ does not have the FVP, as shown in Section \ref{sec:symbolic}. For instance, 
 terms of the form  ${\tt (t_1\ \ldots\ t_n\ enc(M',s(0)))}$ derive
from ${\tt enc(M,k1)}$ by fV-narrowing, where $\tt enc(M',s(0))$
can be further narrowed to unravel an unlimited sequence of identical terms modulo renaming. 
Nonetheless, equational homeomorphic embedding  detects this non-terminating behavior since  $\tt enc(M',s(0))$ embeds  $\tt enc(M,s(0))$. 

By using the unfolding operator $\cU_{\ol{FVP}}$,  the first phase of 
 the {\sc NPER}$^{\cU_{\ol{FVP}}}(\cR)$ algorithm  first normalizes the rewrite rules in $R$. The normalization process produces a new  set of rules $R'$ where the {\tt reply} and {\tt rec} rules are left unchanged,
while the {\tt req} rule is simplified into

{\footnotesize
\begin{verbatim}
crl [req] : < C : Cli | server : S , 
                        data : M , 
                        key : s(0) , 
                        status : mt >
           =>  
				
           < C : Cli | server : S , 
                       data : M , 
                       key : s(0) , 
                       status : mt > 
           ( S <- { C , enc(M,s(0)) } ) if true = true /\ enc(M,s(0)) = enc(enc(M,0),s(0)).
\end{verbatim}
}
\noindent because of the normalization of the equational condition  {\tt (s(0) < len) = true} and the instantiation of the condition
 {\tt  enc(M,s(0)) = enc(enc(M,0),s(0))} to the specific case where {\tt K}  is equal to {\tt s(0)}.

Subsequently,  the specialization algorithm computes
the initial set 
$${\tt Q=\{enc(M,s(0)),dec(M,s(0)),enc(enc(M,0),s(0))\}}$$
of the (normalized) maximal function calls
in $R'$, where {\tt M} is a variable of sort {\tt Data},
and the equational theory $\cE$ is partially evaluated by {\sc EqNPE}$^{\cU_{\ol{FVP}}}$ w.r.t. {\tt Q}. During the partial evaluation process, $\cU_{\ol{FVP}}$ only unravels finite fragments of the fV-narrowing trees that are rooted by the specialized calls, and the final set $Q'$ of specialized calls is

\begin{align*}
\{&{\tt dec(X:Data, s(0))},  {\tt enc(X:Data, s(0))}, {\tt enc(enc(X:Data, 0), s(0))},\\
  &{\tt toSym([toNat(Y:Symbol) < s(s(0)),s(toNat(Y:Symbol)),0])},{\tt toSym(toNat(Y:Symbol))},\\
&{\tt toSym(unshift(toNat(Y:Symbol)))}\}.
\end{align*}

\begin{figure}[h!]
{\footnotesize
\begin{Verbatim}[frame=single]  
eq dec(a, s(0)) = c [ variant ] .
eq dec(b, s(0)) = a [ variant ] .
eq dec(c, s(0)) = b [ variant ] .
eq dec(S:Symbol M:Data, s(0)) = toSym(unshift(toNat(S:Symbol))) 
                                dec(M:Data, s(0)) [ variant ] .
eq enc(a, s(0)) = b [ variant ] .
eq enc(b, s(0)) = c [ variant ] .
eq enc(c, s(0)) = a [ variant ] .
eq enc(S:Symbol M:Data, s(0)) = toSym([toNat(S:Symbol) < s(s(0)),s(toNat(S:Symbol)),0]) 
                                enc(M:Data, s(0)) [ variant ] .
eq enc(enc(a, 0), s(0)) = b [ variant ] .
eq enc(enc(b, 0), s(0)) = c [ variant ] .
eq enc(enc(c, 0), s(0)) = a [ variant ] .
eq enc(enc(S:Symbol M:Data, 0), s(0)) = toSym([toNat(toSym(toNat(S:Symbol))) < s(s(0)),
                                                    s(toNat(toSym(toNat(S:Symbol)))),0]) 
                                        enc(enc(M:Data, 0), s(0)) [ variant ] .
eq toSym([toNat(a) < s(s(0)),s(toNat(a)),0]) = b [ variant ] .
eq toSym([toNat(b) < s(s(0)),s(toNat(b)),0]) = c [ variant ] .
eq toSym([toNat(c) < s(s(0)),s(toNat(c)),0]) = a [ variant ] .
eq toSym(toNat(a)) = a [ variant ] .
eq toSym(toNat(b)) = b [ variant ] .
eq toSym(toNat(c)) = c [ variant ] .
eq toSym(unshift(toNat(a))) = c [ variant ] .
eq toSym(unshift(toNat(b))) = a [ variant ] .
eq toSym(unshift(toNat(c))) = b [ variant ] .
\end{Verbatim}
}
\caption{{\sc NPER}$^{\cU_{\ol{FVP}}}$ Phase 1: Partial evaluation of $\cE$ w.r.t. {\tt Q} output by \Presto\/}\label{fig:phase1} 
\end{figure}

The resulting partial evaluation $\cE'$ of $\cE$, which is generated from $Q'$, is given in Figure \ref{fig:phase1}.
The second phase (compression) renames  
$\cE'$ into the equational theory $\cE_{ren}$ of Figure \ref{fig:phase2}
by computing the following renaming for the  specialized calls:

$$ 
\begin{array}{l}
{\tt dec(X:Data, s(0))} \mapsto {\tt f0(X:Data)} \\
{\tt enc(X:Data, s(0))} \mapsto {\tt f1(X:Data)}\\
{\tt enc(enc(X:Data, 0), s(0))} \mapsto {\tt f2(X:Data)}\\
{\tt toSym([toNat(Y:Symbol) < s(s(0)),s(toNat(Y:Symbol)),0])}\mapsto {\tt f3(Y:Symbol)}\\
{\tt toSym(toNat(Y:Symbol))} \mapsto {\tt f4(Y:Symbol)}\\
{\tt toSym(unshift(toNat(Y:Symbol)))} \mapsto {\tt f5(Y:Symbol)}
\end{array}
$$

\noindent
which eliminates complex nested calls and redundant arguments in $\cE_{ren}$  computations.

It is worth noting that the resulting specialization $\cE_{ren}$ provides a highly optimized version of $\cE$ for  the   arbitrarily fixed key {\tt s(0)}, 
where  both functional and structural compression are achieved. Specifically, data structures  in $\cE$  for natural numbers and their associated operations  for message
encryption and decryption 
are totally removed from $\cE_{ren}$. 
Also, code reuse is automatically achieved in $\cE_{ren}$. Indeed,  the obtained specialization reuses 
the  renamed specialized  call   {\tt f3(X:Data)} in the specialization of both {\tt enc(X:Data,s(0))} (given by {\tt eq f1(S:Symbol M:Data)} {\tt = f3(S:Symbol) f1(M:Data) [variant]}) 
and  {\tt enc(enc(X:Data,0),s(0))} \linebreak (given by  
{\tt eq  f2(S:Symbol M:Data) = 
f3(f4(S:Symbol)) f2(M:Data) [variant]}). 

Note that the {\tt \_+\_} operator, together with its associative and commutative axioms, disappears from $\cE_{ren}$, thereby avoiding expensive matching operations  modulo axioms. 
This transformation power cannot be achieved by existing functional, logic,  or  functional-logic partial evaluators for a language that implements native matching modulo axioms.
\ Decryption  (resp., encryption) in $\cE_{ren}$ is now the direct mapping {\tt f0} (resp., {\tt f1}) that associates messages to their corresponding  decrypted (resp. encrypted) counterparts, avoiding a huge amount of computation in the profuse domain of natural numbers.
\begin{figure}
{\footnotesize
\begin{Verbatim}[frame=single]
eq f0(a) = c [variant] .    eq f0(b) = a [variant] .    eq f0(c) = b [variant] .
eq f2(a) = b [variant] .    eq f2(b) = c [variant] .    eq f2(c) = a [variant] .
eq f1(a) = b [variant] .    eq f1(b) = c [variant] .    eq f1(c) = a [variant] .
eq f3(a) = b [variant] .    eq f3(b) = c [variant] .    eq f3(c) = a [variant] .
eq f4(a) = a [variant] .    eq f4(b) = b [variant] .    eq f4(c) = c [variant] .
eq f5(a) = c [variant] .    eq f5(b) = a [variant] .    eq f5(c) = b [variant] .

eq f0(S:Symbol M:Data) = f5(S:Symbol) f0(M:Data) [variant] .
eq f1(S:Symbol M:Data) = f3(S:Symbol) f1(M:Data) [variant] .
eq f2(S:Symbol M:Data) = f3(f4(S:Symbol)) f2(M:Data) [variant] .
\end{Verbatim}
}
\caption{{\sc NPER}$^{\cU_{\ol{FVP}}}$ Phase 2: Renamed theory $\cE_{ren}$ output by \Presto\/}\label{fig:phase2} 
\end{figure}
The computed renaming is also applied to  $\cR$ by respectively replacing, into the rewrite rules of $\cR$, the maximal function calls 
{\tt dec(M,s(s(0))}, {\tt enc(M,s(s(0))}, and {\tt enc(enc(M,0),s(0))} with {\tt f0(M)}, {\tt f1(M)}, and {\tt f2(M)}, respectively.
This allows the renamed rewrite rules to be able to access the new specialized encryption and decryption functionality provided by $\cE_{ren}$.
Finally, condition simplification is run, 
thereby 
removing the trivial condition {\tt true = true}  from the renamed version of the {\tt req} rule and  delivering  the set of specialized rules
of Figure \ref{fig:finalout} as outcome. Note that condition simplification is not able to remove condition {\tt f1(M) = f2(M)} from the specialized rewrite
rule {\tt req-sp} since the partially evaluated equational theory $\cE_{ren}$ does not have the FVP 
  and thus  the {\sc Simplify} Algorithm can only remove trivial rule conditions.
\begin{figure}
{\footnotesize
\begin{Verbatim}[frame=single]  
crl [req-sp] : < C : Cli | server : S , 
                       data : M , 
                       key : s(0) , 
                       status : mt >
             
               =>  		
             
               < C : Cli | server : S , 
                           data : M , 
                           key : s(0) , 
                           status : mt > 
               ( S <- { C , f1(M) } ) if  f1(M) = f2(M) .
	
rl [reply-sp] : < S : Serv | key : s(0) > 
                (S <- {C,M}) 
				  
                =>
				  
                < S : Serv | key : s(0) > 
                (C <- {S, f0(M)}) .
				  	
rl [rec-sp] : < C : Cli | server : S , 
                          data : M , 
                          key : s(0) , 
                          status : mt > 
              (C <- {S,M})
				
              =>
				
              < C : Cli | server : S , 
                          data : M , 
                          key : s(0) , 
                          status: success >
\end{Verbatim}
}

\caption{{\sc NPER}$^{\cU_{\ol{FVP}}}$ Phase 2: Specialized rewrite rules output by \Presto\/}\label{fig:finalout} 
\end{figure}
\end{example}

Example \ref{ex:handshakeNOFVP-specialized} shows that a great degree of simplification  can be achieved by the specialization technique of \Presto\ for theories that do not have the FVP. Furthermore,   in many cases,
the algorithm {\sc NPER}$^{\cU_{\ol{FVP}}}$   is also able to transform an equational theory that does not meet the FVP into a specialized one that does. This typically happens when the function calls to be specialized 
 can  only be unfolded a finite number of times.  Let us see an example.

\begin{example}\label{ex:NoFVP2FVP}
Consider a slight variant of the 
protocol theory of Example \ref{ex:handshakeNOFVP-specialized} in which
 messages  consist of one single symbol instead of  arbitrarily long sequences of symbols.  This variant
 can be obtained by simply modifying the sort of the messages from {\tt Data} to {\tt Symbol}
 in the protocol rewrite rules.
 In this scenario, the set of (normalized) maximal function  calls becomes
 $$\tt Q=\{enc(M,s(0))\downarrow_{\mathit{\vec{E},B}}, dec(M,s(0))\downarrow_{\mathit{\vec{E},B}}, enc(enc(M,0),s(0))\downarrow_{\mathit{\vec{E},B}}\}$$
\noindent where {\tt M} is a variable of sort {\tt Symbol}.
 The rewrite theory 
can be  automatically specialized by \Presto\ for this use case  by using  {\sc NPER}$^{\cU_{\ol{FVP}}}$.
The intermediate, partially evaluated equational theory $\cE_{ren}$ is shown in Figure \ref{fig:noFVP2FVP}. 

The obtained specialization gets rid of the associative data structure that is needed to build messages of arbitrary size since only one-symbol messages are allowed in the specialized program. 
Also, note that $\cE_{ren}$ clearly  meets the FVP because it specifies three non-recursive functions
(namely, {\tt f0}, {\tt f1}, {\tt f2}),
 which all work over the finite domain {\tt \{a,b,c\}}. 
 Additionally, $\cE_{ren}$ is sufficiently complete (indeed, {\tt f1}, {\tt f2}, and {\tt f3} are total functions that respectively implement the behaviour of the calls {\tt dec(M,s(0))}, {\tt enc(M,s(0))}  and {\tt enc(enc(M,0),s(0))} for each $\tt M\in\{a,b,c\}$).
Therefore $\cE_{\sf ren}$ also has  the CFVP, which means that the {\sc Simplify} algorithm can fully apply its  simplification strategy  
to the (renamed) specialized conditional rule for protocol request

\begin{verbatim} 
crl [req-ren] : < C : Cli | server : S , 
                       data : M , 
                       key : s(0) , 
                       status : mt >            
               =>  		
               < C : Cli | server : S , 
                           data : M , 
                           key : s(0) , 
                           status : mt > 
               ( S <- { C , f1(M) } ) if  true = true /\ f1(M) = f2(M) .
\end{verbatim}

\noindent This rule is  completely deconditionalized  by removing  both the trivial condition {\tt true = true} (as in Example \ref{ex:handshakeNOFVP-specialized})  and the condition  {\tt f1(M) = f2(M)}. The latter condition is removed by applying Case 3.b of the $\mathit{sCond}_{\cE}$ function  since the {\tt req-ren} rule has not the {\tt narrowing} label  and  {\tt f1(M)} and {\tt f2(M)} have the same (most general) constructor variants in $\cE_{ren}$:
$$CV_{\cE_{ren}}({\tt f1(M)}) = {\tt \{(\{M\mapsto a\},b),(\{M\mapsto b\},c),(\{M\mapsto c\},a)\}} = CV_{\cE_{ren}}({\tt f2(M)}).$$   
\begin{figure}
{\footnotesize
\begin{Verbatim}[frame=single]
eq f0(a) = c [variant] .     eq f0(b) = a [variant] .     eq f0(c) = b [variant] .  
eq f1(a) = b [variant] .     eq f1(b) = c [variant] .     eq f1(c) = a [variant] .   
eq f2(a) = b [variant] .     eq f2(b) = c [variant] .     eq f2(c) = a [variant] .
\end{Verbatim}
}
\caption{Equations of the specialized equational theory for one-symbol messages and key {\tt s(0)}.}\label{fig:noFVP2FVP}
\end{figure}
Finally, note that the satisfaction of the FVP allows narrowing-based reachability problems in the specialized rewrite theory $\cR_{\sf sp}$ to be effectively solved in the topmost extension of $\cR_{\sf sp}$, which can be automatically computed by the program transformation of Section \ref{sec:relax}.  Therefore, reachability goals such as the one of 
Example \ref{ex:extension} can be solved in the topmost extension of the specialized rewrite theory obtained in Example \ref{ex:NoFVP2FVP}.
 \end{example}

\paragraph{\bf Case 2: $\cE$ is a finite variant theory.}
When $\cE$ is a finite variant theory, fV-narrowing trees are always finite objects that can be effectively constructed in finite time.  
Therefore, it is possible to construct the complete fV-narrowing tree for any possible 
specialized call.
The unfolding operator $\cU_{\mathit{FVP}}$ 
develops such a complete fV-narrowing tree for a given input term in $\cE$.
 		
The advantage of using $\cU_{\mathit{FVP}}$ instead of $\cU_{\mathit{\ol{FVP}}}$ is twofold.
First, $\cU_{\mathit{FVP}}$ disregards any embedding check, which can be extremely time-consuming when $\cE$ includes  
several operators that obey complex modular combinations of 
algebraic axioms\footnote{Actually, given an AC operator $\circ$, if we want to check whether a term   $t=t_1\circ t_2\ldots\circ t_i$ is embedded into another term with a similar form, all possible permutations of the elements of both terms must be tried.}. 
Second, $\cU_{\mathit{FVP}}$ exhaustively explores the whole fV-narrowing tree of a term, while $\cU_{\mathit{\ol{FVP}}}$ does not. This leads to a lower degree of specialization  when $\cU_{\mathit{\ol{FVP}}}$ is applied  to a finite variant theory, as shown in the following (pathological) example.
\begin{example}\label{ex:mkEven}
Consider the equational theory that is encoded by the following Maude functional module:
{\small
\begin{verbatim}
fmod PATHOLOGICAL is sort Nat .
  ops 0 1 : -> Nat [ctor] .  op mkEven : Nat Nat -> Nat .
  op_+_ : Nat Nat -> Nat [ctor assoc comm id: 0] .
  vars X Y : Nat .
  eq mkEven(X + X + 1,1 + Y) = mkEven(X + X + 1 + 1,Y) [variant] .
  eq mkEven(X + X,0) = X + X [variant] .
endfm
\end{verbatim}
}
The equational theory specifies the encoding for natural numbers in Presburger's style\footnote{Using this encoding,  a natural number can be the constant  {\tt 0} or a sequence of the form {\tt 1 + 1 ... + 1}.} and uses this encoding to define the function {\tt mkEven(X,Y)} that makes {\tt X} even (if {\tt X} is odd) by ``moving'' one unit from {\tt Y} to {\tt X}. Otherwise, if {\tt X} is even, {\tt X} is left unchanged.
The partial evaluation of the given theory  w.r.t. the call {\tt mkEven(X,Y)} yields different outcomes that depend on the chosen unfolding operator.
On the one hand, by using  $\cU_{\mathit{\ol{FVP}}}$, the output is the very same input theory, thus no ``real'' specialization is achieved. On the other hand, the unfolding operator $\cU_{\mathit{FVP}}$ produces the specialized definition of {\tt mkEven}  given by
{\small
\begin{verbatim}    
eq mkEven(X + X,0) = X + X [variant] .
eq mkEven(1 + X + X, 1) = 1 + 1 + X + X [variant] .
\end{verbatim}
}
\noindent where the second equation is  generated 
by fully exploring the fV-narrowing derivation
{\small $${\tt mkEven(Z,W)}\leadsto_{\tt\{Z\mapsto X + X + 1,W\mapsto 1 + Y\}} {\tt mkEven(1\ +\ 1\ +\ X\ +\ X, Y)} \leadsto_{\tt \{Y\mapsto 0\}} {\tt  1\ +\ 1\ +\ X\ +\ X}$$}
where $t \narrowG{\sigma}{} t$ denotes a fV-narrowing step from $t$ to $t'$ with substitution $\sigma$. In contrast, $\cU_{\mathit{\ol{FVP}}}$ stops the 
sequence at {\tt\small mkEven(1 + 1 + X + X, Y)} since {\tt\small  mkEven(1 + 1 + X + X, Y)}  embeds 
{\tt\small mkEven(Z,W)} and hence yields a specialized equation that is equal (modulo associativity and commutativity of {\tt +}) to the original equation  
{\tt\small  mkEven(X + X + 1,1 + Y) = mkEven(X + X + 1 + 1,Y)}. 
\end{example}

%% file: 4-presto.tex
\section{The \Presto\ System}\label{sec:imple}
\Presto\ is a program specializer for Maude programs whose core engine consists of approximately  1800  lines of Maude. The system is available,  together with a quick start guide and a number of examples 
 at {\tt http://safe-tools.dsic.upv.es/presto}. The \Presto\/ distribution package also contains   \Presto\  source code  for local installation,
which is updated to the latest Maude release.

 Similarly to {\sf Victoria}, the earlier   partial evaluation tool for equational theories  \cite{ACEM20-jlamp}, \Presto\ follows the on-line approach to partial evaluation that makes 
	control decisions about specialization on the fly. This  offers better opportunities for powerful 
	automated strategies than off-line partial evaluation \cite{JGS93}, where decisions are made before 
	specialization by using 
	abstract data descriptions  that are represented as program annotations. 
	Compared to   {\sf Victoria}, the whole \Presto\ implementation is totally new code that is  cleaner and much simpler, 
 including the equational least general generalization component of the system that  is 
based on the improved least general generalization algorithm of \cite{AEMS21}.
 Also, \Presto\ can deal with Maude system modules (i.e.,\ rewrite theories) and multi-module Maude specifications that cannot be managed by {\sf Victoria}, which only handles  equational theories  encoded by single Maude functional modules. 

  \Presto\ can be accessed via a user-friendly web interface
that has been developed by using CSS, HTML5, and JavaScript technologies  in addition to a command-line interface. This enables users to
specialize Maude rewrite theories using a web browser, without the need
for a local installation.   The Maude core component is connected to the client web user interface through 10 RESTful Web Services that are implemented by using  the JAX-RS API for developing Java RESTful Web Services (around 500
 lines of Java source code).   Since \Presto\ is a fully automatic online specializer, its usage is simple and accessible  for inexperienced users. 
With regard to the security of the system, in order to prevent the execution of unsafe code submitted by users, every call to a Maude function is first checked against a blocklist of unsafe functions so that the process is aborted if there is a call that   belongs to the blocklist and a warning message is triggered. 
We consider unsafe functions to be all of those Maude system-level commands that may remotely control the system-level activities in the server hosting Presto. Also, spawned  processes  are automatically killed when a time limit is exceeded. 

 \Presto\ comes with a set of preloaded Maude specifications that includes all of the examples shown in this paper. Preloaded examples can be selected from a pull-down menu, or new specifications can be written from scratch using a dedicated text area. Another pull-down menu allows the user to select the unfolding operator (either ${\cU_{\mathit{\ol{FVP}}}}$ or ${\cU_{\mathit{FVP}}}$)   
to automatically optimize the input rewrite theory by specializing  its underlying guest equational theory and hardcoding the obtained specialized functions into the rewrite rules.
 \Presto\ also  supports the partial evaluation of equational theories w.r.t. a given set of user-defined calls by using either ${\cU_{\mathit{\ol{FVP}}}}$ or ${\cU_{\mathit{FVP}}}$. 

\begin{figure}[h!]
\centering
\includegraphics[width=\textwidth]{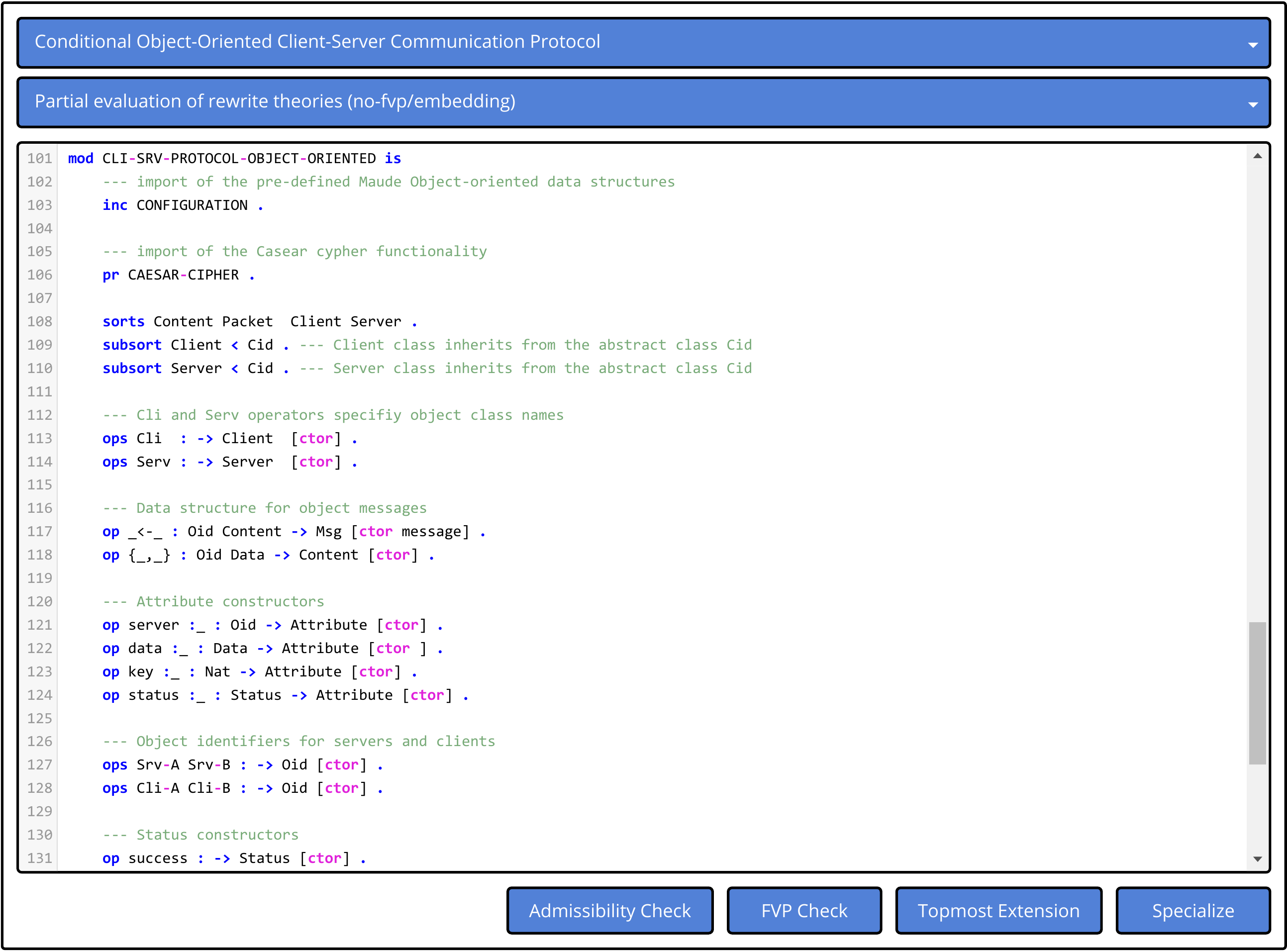}
\caption{Loading specifications and selecting the unfolding strategy in \Presto.}\label{fig:load}
\end{figure}

Figure \ref{fig:load} shows a screenshot of the \Presto\ interface  where the object-oriented client-server communication protocol of Example \ref{ex:handshakeNOFVP-specialized} has been selected  among the preloaded examples and the  ${\cU_{\mathit{\ol{FVP}}}}$ operator has been chosen  for the specialization of the rewrite theory under consideration (the {\sf Partial evaluation of rewrite theories (no-FVP/embedding)} option in the pull-down menu).

\noindent Finally,  \Presto\ includes the following useful additional features.\\
 
\begin{figure}[h!]
\centering
\includegraphics[width=\textwidth]{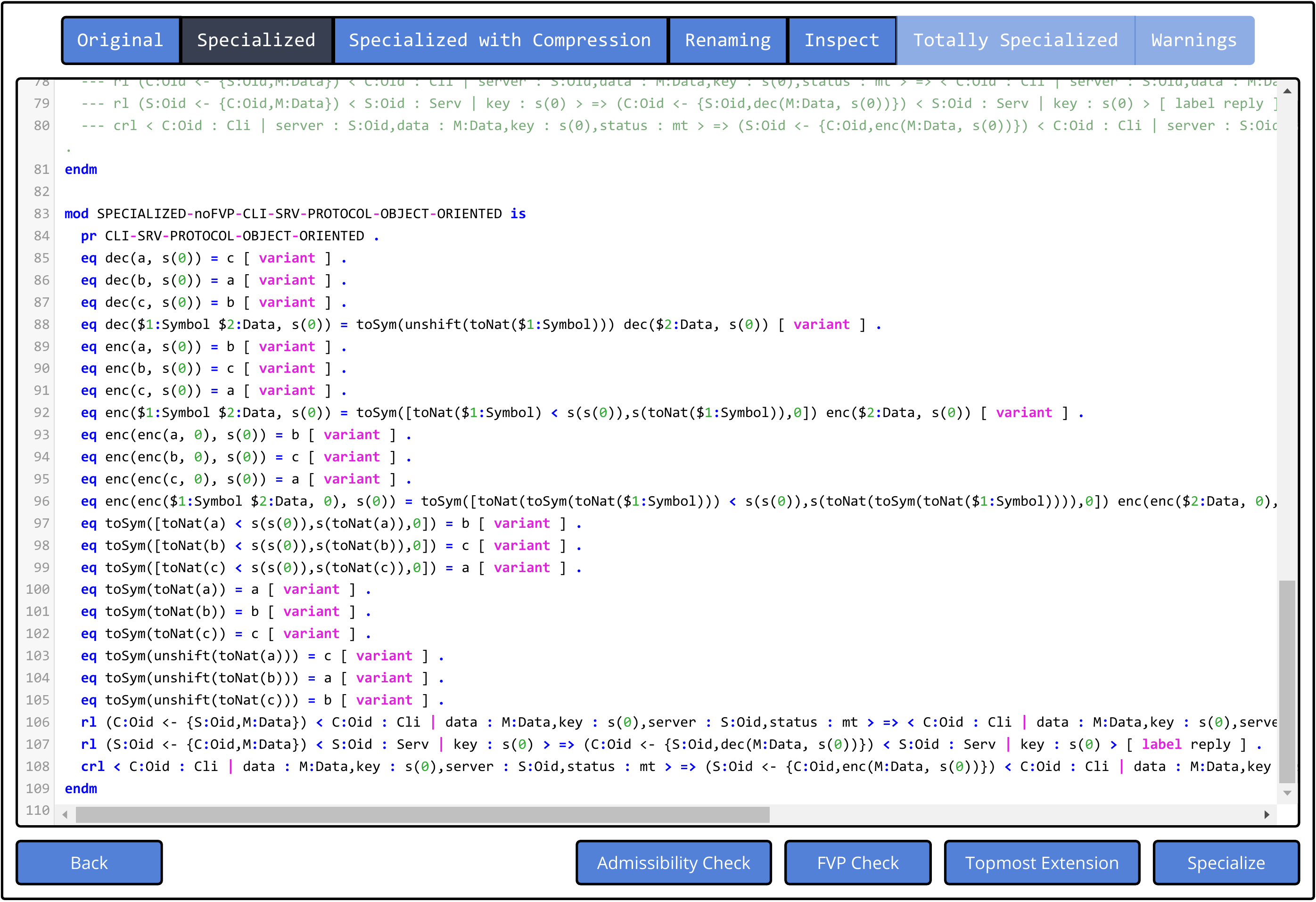}
\caption{Specialized client-server communication protocol before the compression phase.}\label{fig:spec}
\end{figure}

\begin{figure}[h!]
\centering
\includegraphics[width=\textwidth]{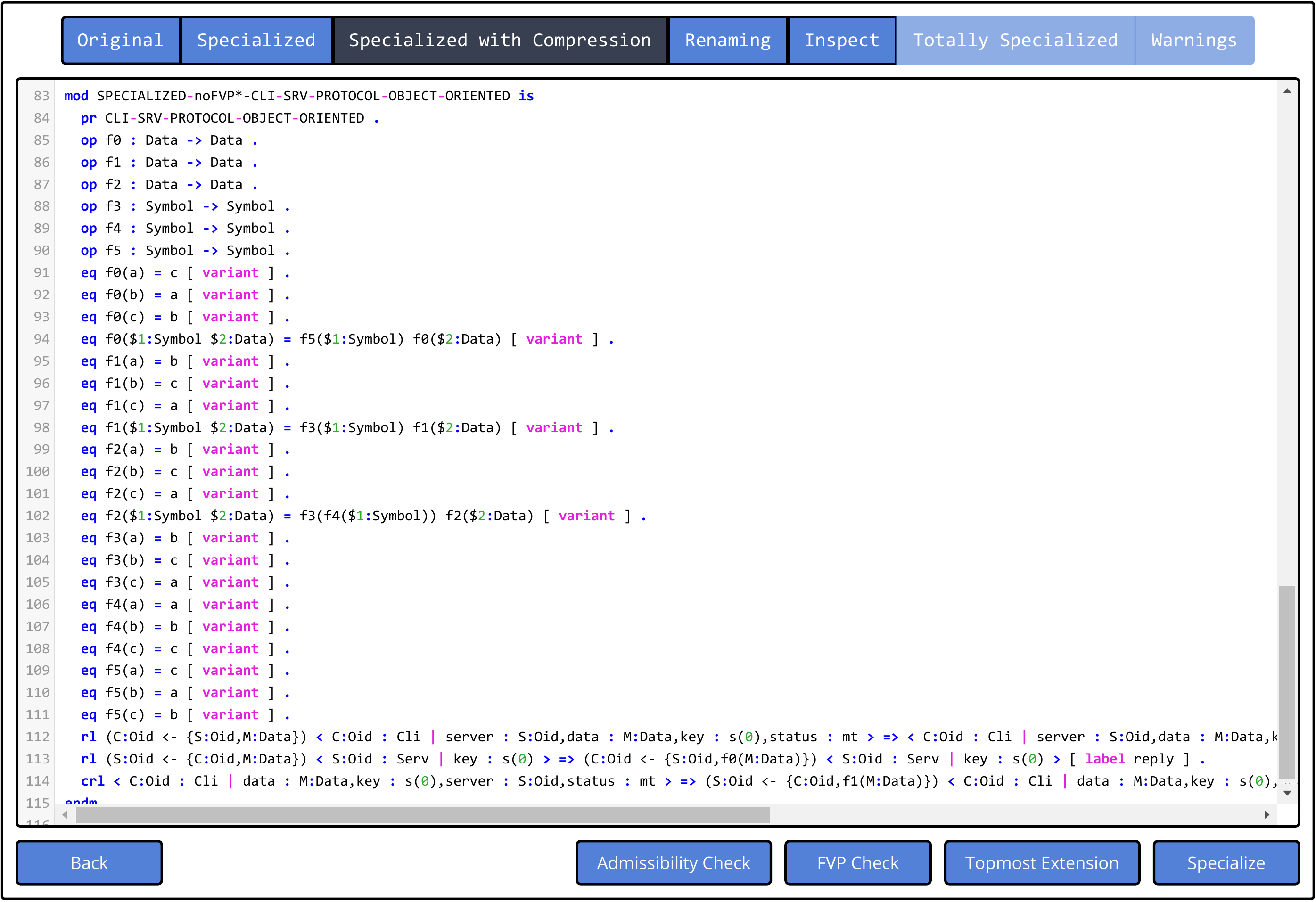}
\caption{Specialized client-server communication protocol after the compression phase.}\label{fig:compressed}
\end{figure}

\noindent {\bf Specialization views}. Two possible views of the resulting specialization are delivered, with or without the compression phase enabled. The user can switch back and forth between the two representations by respectively selecting the {\sf Specialized} and {\sf Specialized with Compression} tabs.   
Figures \ref{fig:spec} and \ref{fig:compressed} respectively show the specialization of the client-server communication protocol before and after the compression phase. 
Furthermore, by accessing the {\sf  Renaming} tab, it is possible to explicitly inspect the renaming computed by the compression phase and see how 
 nested calls  are dramatically compacted by using the corresponding fresh function symbols.   
\\

\begin{figure}[t]
\centering
\includegraphics[width=\textwidth]{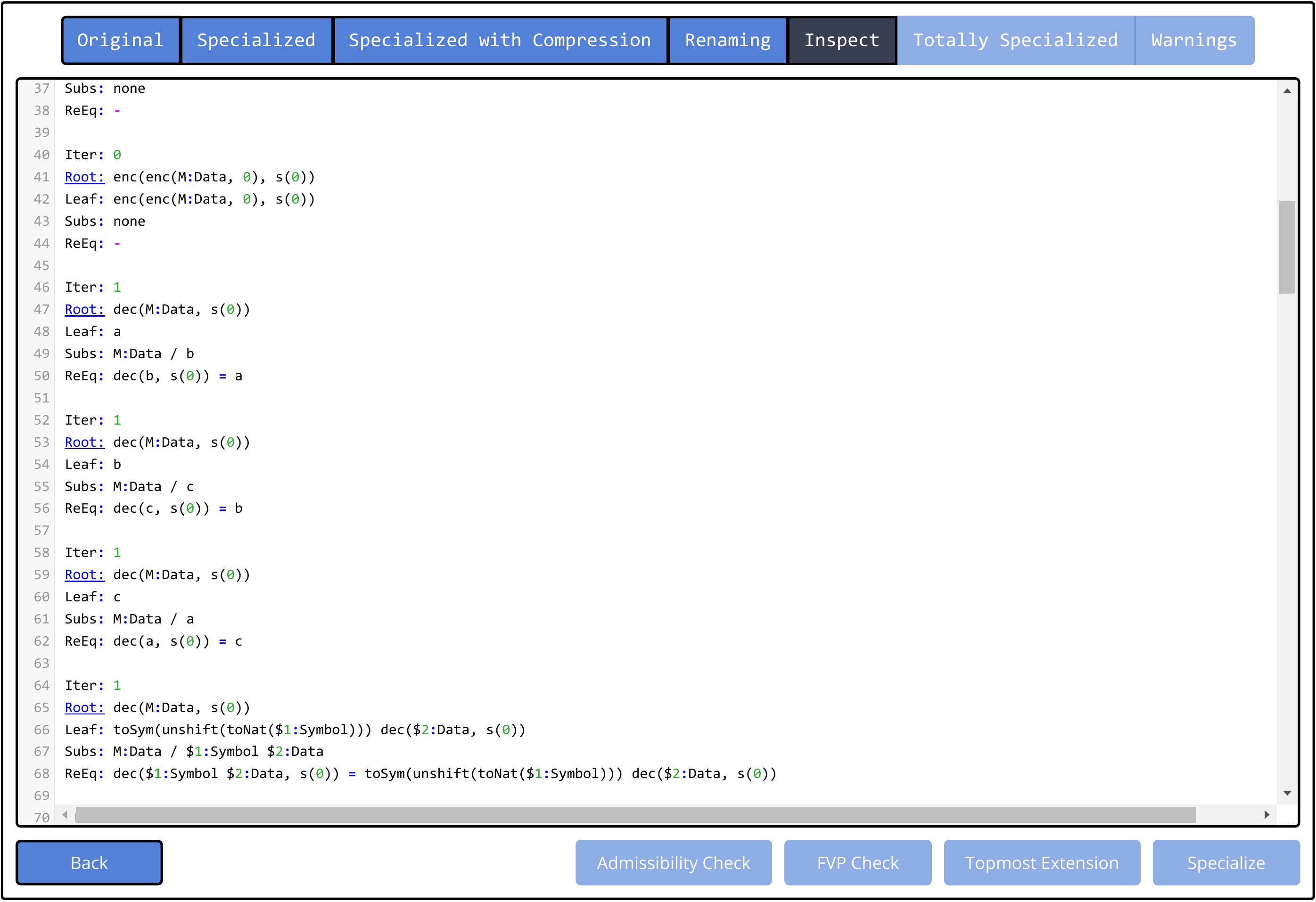}
\caption{Inspecting the specialization process.}\label{fig:inspect}
\end{figure}

\begin{figure}[t]
\centering
\includegraphics[width=\textwidth]{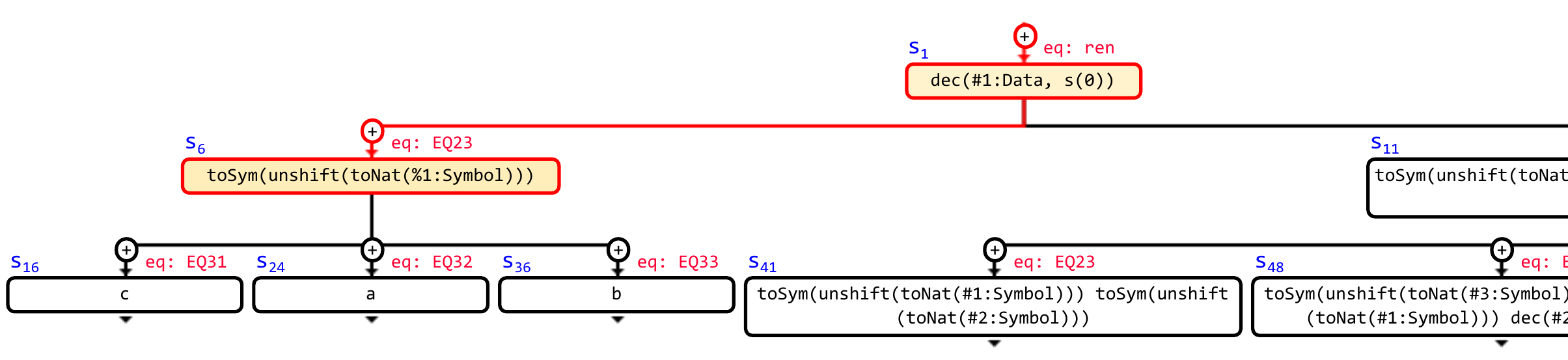}
\caption{Fragment of the fV-narrowing tree for {\tt dec(M:Data,s(0))}.}\label{fig:tree}
\end{figure}

\noindent {\bf Inspection mode}. By activating the {\sf inspect} tab, \Presto\ shows detailed information of the specialization process that can be particularly useful 
for a user who wants to experiment with different specializations for analysis and verification purposes. Specifically, 
for each iteration $i$ of the {\sf EqNPE}$^{\cU}$ algorithm (invoked in Line 4 of Algorithm \ref{alg:NPER-Algorithm}),
 \Presto\ shows all of the leaves (term variants) in the deployed fV-narrowing trees 
   for  the calls that were specialized at the iteration $i$,
together with their corresponding  computed substitution (i.e.,\ the  substitution component of a term variant).
Also, thanks to the  interconnection with the narrowing stepper {\sc Narval} \cite{ABES19}, each  fV-narrowing tree computed at iteration $i$ can be graphically visualized and interactively navigated. This is automatically done by simply selecting the desired tree root term  to be 
expanded 
from   the displayed list of   calls that were specialized at any iteration (including the last one). This can be done by simply clicking the hyperlink  associated with the desired call. By doing this, the user can  perform a fine-grained and stepwise inspection of all narrowing steps,  computation of equational unifiers, and interactive exploration of different representations of Maude's narrowing and rewriting search spaces (tree-based or graph-based, source-level or meta-level).
For instance, Figure \ref{fig:inspect} shows that, at iteration 1,  a fV-narrowing tree was expanded whose root term is the specialized call {\tt dec(M:Data,s(0))} (Line 47). The root can be narrowed to the leaf {\tt a} (Line 48) with the computed 
substitution  {\tt \{M $\mapsto$ b\}} (Line 49), which allows the  resultant equation {\tt dec(b,s(0)) = a} to be generated  (Line 50). 
Furthermore, Figure  \ref{fig:tree} shows (a fragment of) the fV-narrowing tree for the specialized call {\tt dec(M:Data,s(0))} of Line 47 that has been generated by the \narval\ system\footnote{Variables in \narval\ are 
standardized apart by using fresh variable names  of the form {\tt \#n}. In our example, {\tt \#1} is a renaming for variable {\tt M}.}.\\


\noindent{\bf Topmost extension transformation}. The topmost extension transformation of Section \ref{sec:relax} has been fully implemented 
in  \Presto.\ Since the topmost property is not required to perform the partial evaluation, it  can be applied either  {\em ex-ante} to the original input program   (and it is preserved by the specialization) or {\em ex-post} to the specialized program. The transformation enables narrowing-based search reachability for all topmost modulo $Ax$ programs, and particularly for object-oriented specifications. 
Reachability goals can be solved in \Presto\ through the connection to the {\sc Narval} system \cite{ABES19}, which provides a graphical environment
for symbolic analysis in Maude. This feature is illustrated in the system quick start guide.
 \\

\begin{figure}[t]
\centering
\includegraphics[width=\textwidth]{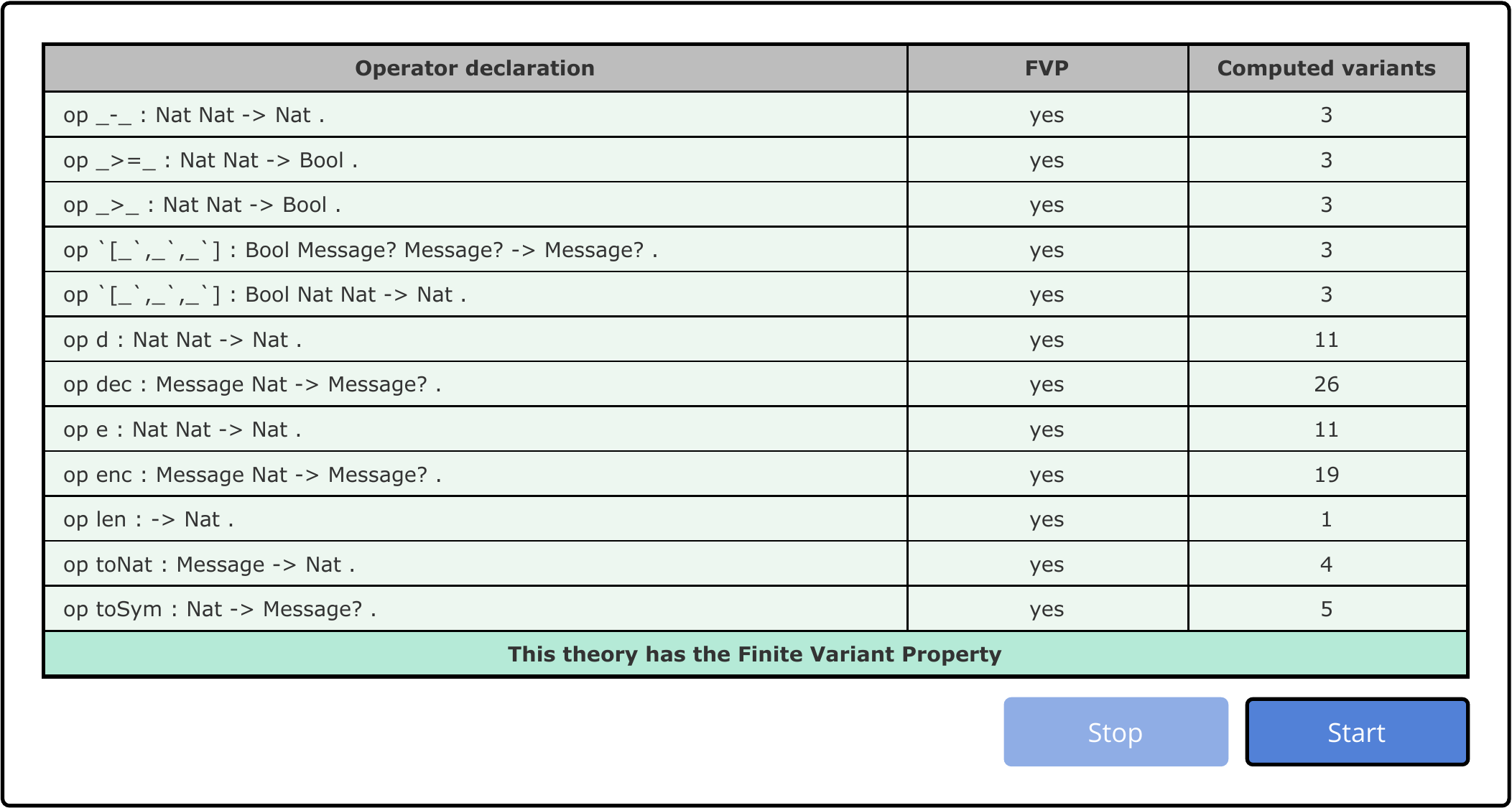}
\caption{ FVP checking of the Client-Server Communication Protocol with FVP.}\label{fig:FVPcheck}
\end{figure}

\noindent{\bf Finite Variant Property  and Strong irreducibility checkers}. \Presto\ includes a FVP checker for equational theories that is based on the checking procedure described in \cite{Meseguer15}. All of the theories that satisfy the FVP can be specialized by using the unfolding operator $\cU_{\mathit{FVP}}$ instead of the typically costlier operator $\cU_{\ol{\mathit{FVP}}}$.
As discussed in Section \ref{sec:symbolic}, the checker implements a semi-decidable procedure that terminates when the FVP holds for the equational theory under examination.
If the FVP does not hold, the user has to manually stop the checking process or wait for the time-out; otherwise, the process would produce an infinite number of most general variants for some terms.
For instance, consider the  {\it Client-Server Communication Protocol with FVP} that is available in the \Presto\ preloaded examples. This Maude specification is a slight mutation of the client-server communication protocol  of Example \ref{ex:handshakeNOFVP-specialized} that models the {\it Caesar} cypher by using a finite variant equational theory. 
By clicking the {\sf Check FVP} button, the checking process starts and, in this case, the check succeeds since the flat terms associated with the program operators have  only  a finite number of most general variants, as illustrated in Figure \ref{fig:FVPcheck}. Hence, the specification could be safely specialized by using the $\cU_{\mathit{FVP}}$ unfolding operator without the risk of jeopardizing termination.

\begin{figure}[t]
\centering
\includegraphics[width=\textwidth]{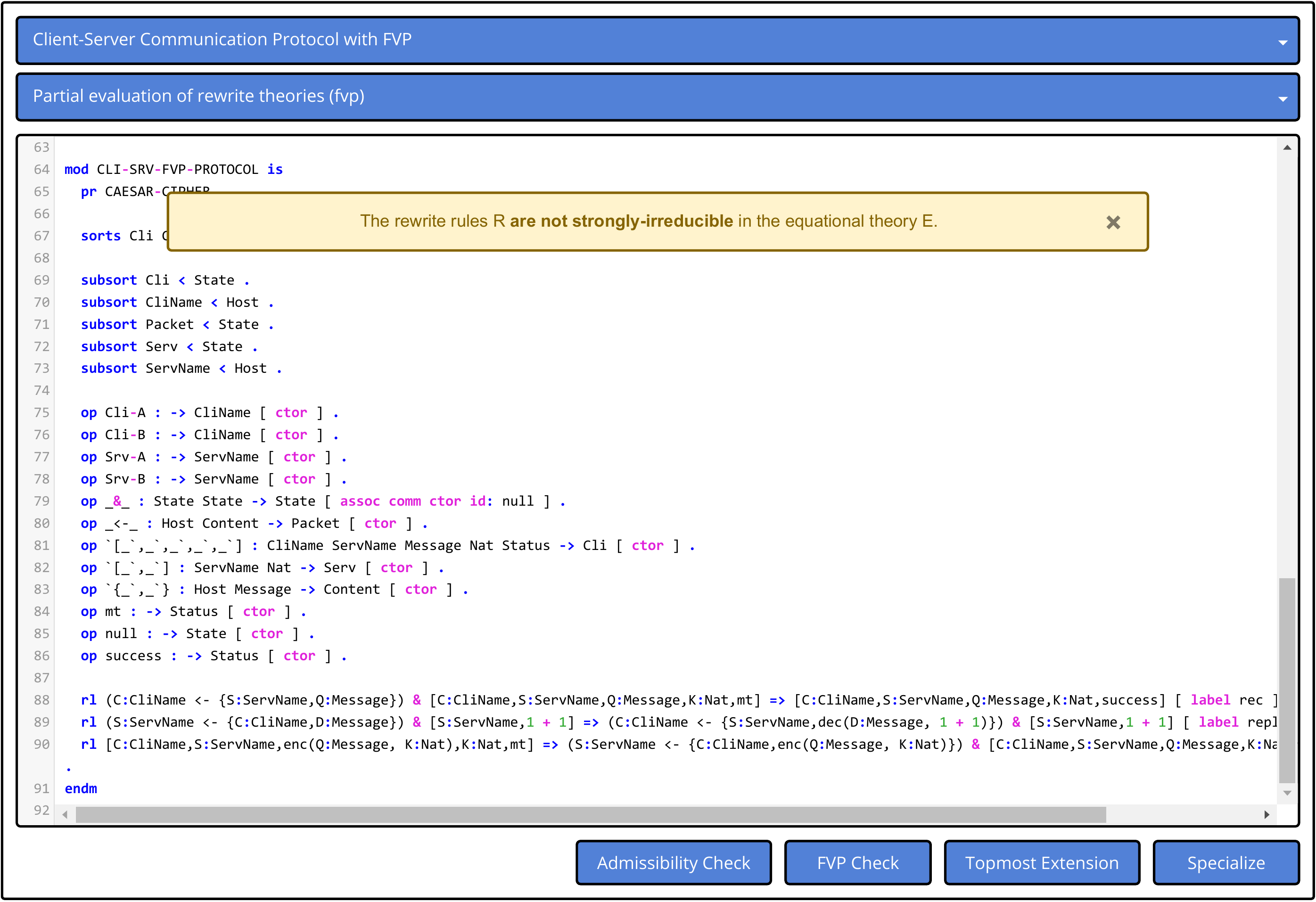}
\caption{ Strong irreducibility failure for a variant of the Client-Server Communication Protocol with FVP.}\label{fig:SIcheck}
\end{figure}

Moreover, \Presto\  implements an efficient checker for the $(\vec{E},B)$-strong irreducibility of the left-hand side of the rewrite rules of the input program. 
As discussed at the end of Section \ref{sec:NPER}, strong irreducibility is an essential requirement for the completeness of the specialization algorithm.  Figure \ref{fig:SIcheck} illustrates the outcome of the  strong irreducibility check for a  variant of the Client-Server Communication Protocol with FVP 
that includes the rewrite rule

{\footnotesize
\begin{verbatim}
rl [C:CliName,S:ServName,enc(Q:Message,K:Nat),K:Nat,mt] 
                      => (S:ServName <- {C:CliName,enc(Q:Message, K:Nat)}) &
                         [C:CliName,S:ServName,Q:Message,K:Nat,mt] .                                                                                  
\end{verbatim}
}
\noindent Note that the check fails due to the narrowable subterm {\tt enc(Q:Message, K:Nat)} occurring in the left-hand side of the considered rule.\\

\noindent {\bf Iterative 
specialization.}
 \Presto\ can be automatically re-applied to any generated specialization in an iterative manner. After completing a given specialization for a rewrite theory, the user can perform a further specialization process to re-specialize the given outcome by using a distinct unfolding operator.  This is achieved by simply pressing the {\sf Specialize} button again, which is  directly displayed after the specialization and 
avoids the need to reload the initial  web input form to manually enter the intermediate program to be re-specialized. 
As shown in \cite{ABEMS20-Festschrift}, this feature is particularly useful in the case when a given input rewrite theory $\cR$ must be specialized using $\cU_{\ol{\mathit{FVP}}}$, since its equational theory does not satisfy the FVP, while the  resulting specialization $\cR'$ hosts an equational theory that does satisfy the FVP so that $\cR'$ can be further specialized by using the  $\cU_{\mathit{FVP}}$ operator.  For a detailed session with \Presto\ that illustrates  this feature, we refer to the \Presto\/  quick start guide.\\

\begin{figure}[t]
\centering
\includegraphics[width=\textwidth]{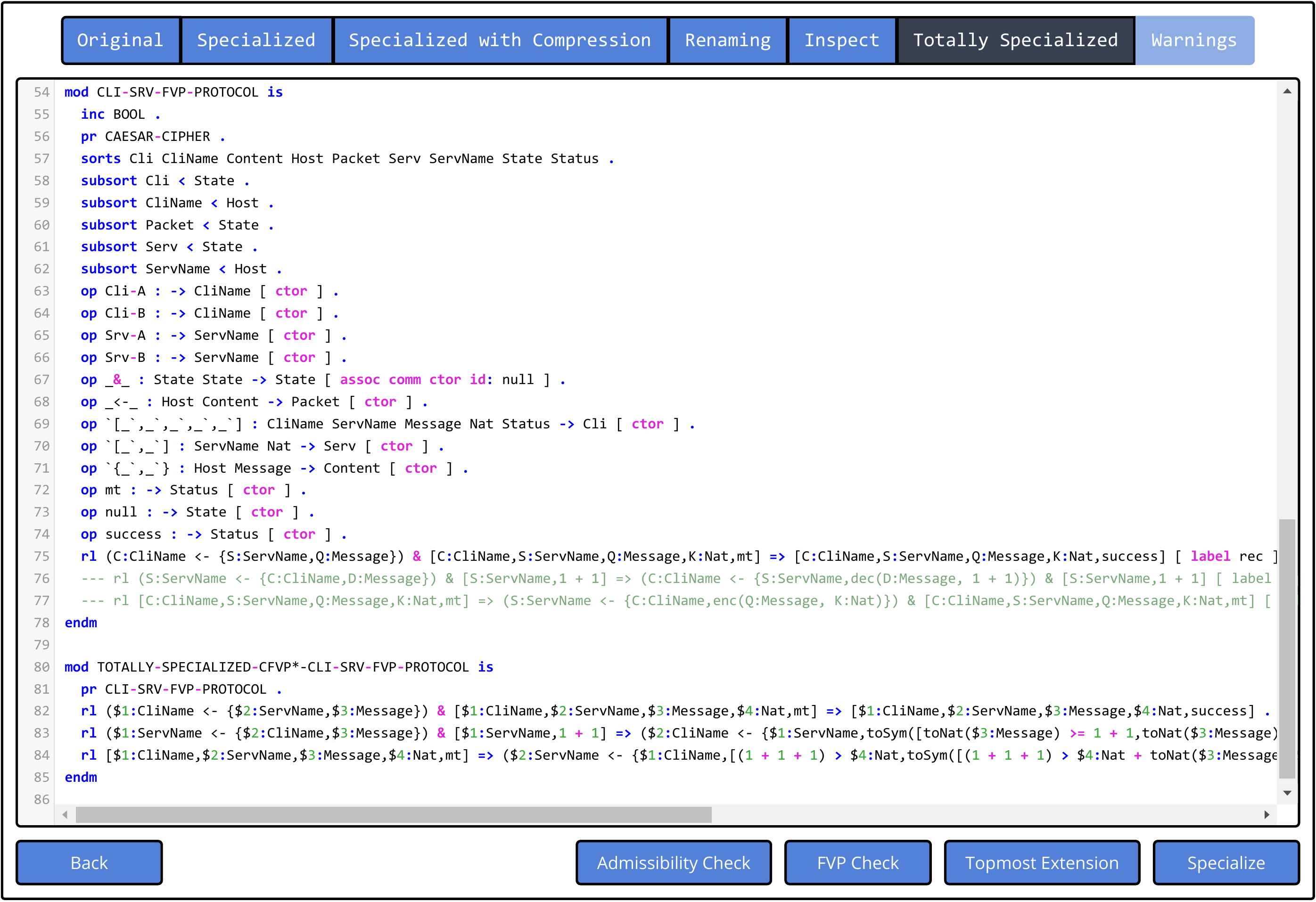}
\caption{ Total evaluation of  the Client-Server Communication Protocol with FVP.}\label{fig:totaleval}
\end{figure}

\noindent {\bf Total evaluation}. This feature provides a slightly modified version of the unfolding operator $\cU_{\mathit{FVP}}$ that implements the total evaluation transformation of \cite{Meseguer20-jlamp} for 
sufficiently complete equational theories that have the FVP. By means of a suitable {\em sort downgrading}, which is described in  \cite{Meseguer20-jlamp}, 
function calls always generate values and cannot be stuck in an intermediate, partial result.
In this scenario, unfolding can safely rule out variants that are not constructor terms. This means that all of the specialized calls get totally evaluated and the maximum compression is achieved, thereby dramatically reducing the search space for the construction of the specialized theories as shown in \cite{ABEMS20-Festschrift}.

For instance, the specialized rewrite theory {\it Client-Server Communication Protocol with FVP} can be totally evaluated into the extremely compact rewrite theory of Figure \ref{fig:totaleval}, which does not include any equation.  Actually, all calls to the encryption and decryption functions have been totally evaluated, and their values have been  hard-coded into their place in the specialized rewrite rules.

%% file: 5-experiments.tex
\section{Experimental Evaluation of \Presto}\label{sec:exp}

\begin{table}[]
\centering
\footnotesize
\resizebox{\textwidth}{!}{
\begin{tabular}{c r r r c r r c }
\hline
\multicolumn{1}{ c }{\it Program}
& \multicolumn{1}{l}{${Size}$}      
& \multicolumn{1}{c}{$\#Rews_{\cR}$} 
& \multicolumn{1}{c}{$\#Rews_{\cR'}$} 
& \multicolumn{1}{c}{\it Reduction} 
& \multicolumn{1}{c}{$T_{\cR}$ (ms)} 
& \multicolumn{1}{c}{$T_{\cR'}$ (ms)} 
& \multicolumn{1}{c}{\it Speedup} \Tstrut\Bstrut\\
\hline
\multicolumn{1}{c}{ \cellcolor{white}}                    &25K      &650,012        &100,002     &84.62\%     &47         &15         &3.13   \Tstrut\\
\multicolumn{1}{c}{ \cellcolor{white}\it Cli-Srv}         &100K     &2,600,115      &400,002     &84.62\%     &261        &87         &3.00   \\
\multicolumn{1}{c}{ \cellcolor{white}}                    &1M       &26,000,100     &4,000,002   &84.62\%     &5,630      &2,434      &2.31   \Bstrut\\
\hline
\multicolumn{1}{c}{ \cellcolor{white}}                    &25K      &8,900,012      &100,002     &98.88\%     &780        &14         &55.71  \Tstrut\\
\multicolumn{1}{c}{ \cellcolor{white}\it Cli-Srv-Mod}     &100K     &35,600,115     &400,002     &98.88\%     &3,723      &87         &42.79  \\
\multicolumn{1}{c}{ \cellcolor{white}}                    &1M       &356,000,100    &4,000,002   &98.88\%     &56,087     &2,399      &23.38  \Bstrut\\
\hline
\hline
\multicolumn{1}{c}{ \cellcolor{white}}                    &1M       &302,000,000    &2,000,000   &99.34\%     &21,163     &584        &36.24  \Tstrut\\
\multicolumn{1}{c}{ \cellcolor{white}\it Diffie-Hellman}  &5M       &1,510,000,000  &10,000,000  &99.34\%     &104,959    &3,087      &34.00  \\
\multicolumn{1}{c}{ \cellcolor{white}}                    &10M      &3,020,000,000  &20,000,000  &99.34\%     &207,189    &6,361      &32.57  \Bstrut\\
\hline
\multicolumn{1}{c}{ \cellcolor{white}}                    &1M       &12,000,000     &4,000,000   &66.67\%     &976        &339        &2.88   \Tstrut\\
\multicolumn{1}{c}{ \cellcolor{white}\it Cli-Srv-FVP}       &5M       &60,000,000     &20,000,000  &66.67\%     &4,866      &1,680      &2.90   \\
\multicolumn{1}{c}{ \cellcolor{white}}                    &10M      &120,000,000    &40,000,000  &66.67\%     &10,305     &3,310      &3.11   \\
\hline
\multicolumn{1}{c}{ \cellcolor{white}}                    &1M       &1,575,500,000  &1,500,000   &99.90\%     &67,646     &546        &123.89  \Tstrut\\
\multicolumn{1}{c}{ \cellcolor{white}\it Handshake-KMP}   &5M       &7,877,500,000  &7,500,000   &99.90\%     &336,147    &3,042      &110.50  \\
\multicolumn{1}{c}{ \cellcolor{white}}                    &10M      &15,755,000,000 &15,000,000  &99.90\%     &655,962    &6,077      &107.94  \Bstrut\\
\hline
\hline
\multicolumn{1}{c}{ \cellcolor{white}}                    		&25K       &1,500,000  &200,000  &86.67\%   &118      &36    &3.28  \Tstrut\\
\multicolumn{1}{c}{ \cellcolor{white}\it Cond-Cli-Srv-OO} 		&100K       &6,000,000  &800,000  &86.67\%     &405    &125    &3.24  \\
\multicolumn{1}{c}{ \cellcolor{white}}                     		&1M      &60,000,000  &8,000,000  &86.67\%     &4,017    &1,250    &3.21  \Bstrut\\
\hline
\multicolumn{1}{c}{ \cellcolor{white}}                    		 &1M      &695,000,000  &1,000,000  &99.86\%     &48,048    &822    &58.45  \Tstrut\\
\multicolumn{1}{c}{ \cellcolor{white}\it Cond-Diffie-Hellman-OO} &5M      &3,475,000,000  &5,000,000  &99.86\%     &249,551    &4,060    &61.47  \\
\multicolumn{1}{c}{ \cellcolor{white}}                     		&10M      &6,950,000,000  &10,000,000  &99.86\%     &526,527    &8,107    &64.95  \Bstrut\\
\hline
\multicolumn{1}{c}{ \cellcolor{white}}                    		&1M       &3,833,996,167   &1,000,000    &99.97\%     &189,779	&873   	  &217.39  \Tstrut\\
\multicolumn{1}{c}{ \cellcolor{white}\it Cond-Handshake-KMP-OO} &5M       &19,169,996,167  &5,000,000    &99.97\%     &923,511	&4,330    &213.28  \\
\multicolumn{1}{c}{ \cellcolor{white}}                    		&10M      &38,339,996,167  &10,000,000  &99.97\%     &1,886,552	&8,898    &212.02  \Bstrut\\
\hline

\end{tabular}
}
\caption{\small Experimental results for the specialization of rewrite theories with \Presto.}\label{tab:expo}
\end{table}
Table~\ref{tab:expo} contains  the experiments that we have performed with \Presto\  using an
		Intel Xeon E5-1660 3.3GHz CPU with 64 GB RAM running Maude v3.0              and considering the average of ten executions for each test. 
		These experiments, together with the source code of all of the examples  are also publicly available at \Presto's website. 
		
		We have considered the following benchmark programs:   {\em Cond-Cli-Srv-OO},  the  conditional, object-oriented, client-server communication protocol of Example \ref{ex:handshakeNOFVP-specialized}; {\em Cli-Srv},  a re-encoding of  {\em Cond-Cli-Srv-OO} that does not use either Maude built-in object-oriented features or conditional rules;
		 {\em Cli-Srv-Mod}, a variant of {\em Cli-Srv} where we introduce an extra function (i.e.,\ the {\tt mod} function  that computes the remainder of the integer division) in the underlying equational theory that is commonly used  in protocol specification and   makes the key 
		 generation heavier; {\em Cli-Srv-FVP}, a variant of {\em Cli-Srv} for one-symbol messages whose equational theory meets the FVP  and differs from the one in Example \ref{ex:NoFVP2FVP} in using Presburger's encoding for natural numbers;   {\em Diffie-Hellman}, the well-known network protocol for safely sharing keys between two nodes over an untrusted channel; {\em Cond-Diffie-Hellman-OO}, a conditional, object-oriented specification of the Diffie-Hellman protocol; {\em Handshake-KMP},  an unconditional rewrite theory that specifies a simple handshake protocol in which a client sends an arbitrary long and  noisy message {\tt M} to a server. The handshake succeeds if the server can recognize a secret handshake sequence {\tt P} inside the client message {\tt M} by matching {\tt P} against {\tt M} via the well-known KMP string matching algorithm; and
 {\em Cond-Handshake-KMP-OO}, a conditional,  object-oriented version of {\em Handshake-KMP}.	    
		  
As for the experiments with  {\em Cli-Srv}, {\em Cli-Srv-Mod} and {\em Cond-Cli-Srv-OO}, we considered input messages 
of three different sizes: from twenty-five thousand symbols to 
 one million symbols. 
 This cannot be done for {\em Cli-Srv-FVP} since it encodes one-symbol messages, and, hence, we cannot use this parameter for benchmarking purposes.
Similarly,  {\em Handshake-KMP and {\em Cond-Handshake-KMP-OO} consider} fixed-size messages of 350 symbols, while {\em Diffie-Hellman} and  {\em Cond-Diffie-Hellman-OO} are specialized w.r.t. fixed-size keys  (modelled as natural numbers obtained though exponentiation). 
For these experiments, we use as benchmark parameter the number of client requests sent over a time-bounded period of time. We considered one, five, and  ten millions  requests. 
The column {\it Size} in Table \ref{tab:expo} indicates either the message length parameter or the number of client requests according to the  benchmark program considered.

For each experiment, we executed the original specification $\cR$ and the specialized one $\cR'$ on the very same input states, and we recorded the following data: 
the execution times (in ms)  $T_\cR$ and  $T_\cR'$, the total number of rewrites $\#Rews_{\cR}$  and  $\#Rews_{\cR'}$, the percentage of {\it reduction} in terms of number of rewrites, and the specialization {\it speedup} computed as the ratio $T_{\cR} / T_{\cR'}$.

		Our figures show that the specialized rewrite theories achieve a significant improvement in  execution time when compared to the original  theory,
with an average speedup for these benchmarks of $59.24$. Particularly remarkable is  the performance improvement of the {\it Cond-Handshake-KMP-OO} which reaches two orders of magnitude for the case when the specialized maximal  calls within the rewrite theory   partially instantiate  the input pattern and  input message. 
Such an impressive performance is largely due to the compression phase of the specialization algorithm that, in this case, completely deconditionalizes the rewrite rules, thereby yielding as outcome a faster, unconditional rewrite theory. This does not happen for {\it Cond-Cli-Srv-OO} which exhibits a smaller speedup. For this benchmark, complete 
deconditionalization cannot be achieved as shown in Example \ref{ex:handshakeNOFVP-specialized}, hence some heavy conditional computations are kept in the specialized
program as well.

We note that none of these specializations could be performed by using our earlier partial evaluator {\sf Victoria},
 which cannot handle rewrite theories but only equational theories. Thus, the specialization of rewrite theories supported by \Presto\ could not be achieved in  {\sf Victoria} unless a complex hack is introduced not only at the level of the theory signature but also by providing a suitable program infrastructure that simulates rewrite rule nondeterminism through deterministic, equational evaluation.

We do not benchmark the specialization times since they are almost negligible (for most cases $<$100 ms).  Nonetheless, we compared the performance of the unfolding operators $\cU_{\mathit{FVP}}$ and $\cU_{\ol{FVP}}$, when both can be applied in the specialization process (i.e.,\ when the rewrite theory to be specialized includes a finite variant equational theory). The most significant difference appears in the  {\it Cli-Srv-FVP}  theory, where {\sc NPER}$^{\cU_{\mathit{FVP}}}$  is  56\% faster (on average) than {\sc NPER}$^{\cU_{\ol{FVP}}}$.
 This figure is expected and favors the use of $\cU_{\mathit{FVP}}$ whenever it is possible since it avoids any overhead caused by   homeomorphic embedding checks performed by $\cU_{\ol{FVP}}$. 	Finally, we benchmarked the symbolic execution of  long-winded  {\em reachability goals} in rewrite theories that include a  finite variant equational theory such as the  pathological theory 
of Example \ref{ex:mkEven}. The achieved improvement is also striking, with an average   speedup of $16.46$.

Let us finally discuss how {\sf Presto} compares in practice with existing on-line specialization systems with related transformation power.
A narrowing-based partial evaluator for the lazy functional logic language Curry is described in \cite{HP14,Peemoller17} whose implementation can be seen as an instance of the generic narrowing-based partial evaluation framework of  \cite{AFV98}.
This system improves a former prototype in \cite{AHV02} by taking into account 
(mutually recursive) let expressions and 
non-deterministic operations, while the PE system of \cite{AHV02} was restricted to confluent programs.
Obviously, the protocol benchmarks in this paper cannot be directly specialized by using Curry's partial evaluator since neither evaluation modulo algebraic axioms nor  concurrency are  supported by Curry's partial evaluator; this would require artificially rewriting the program code so that any comparison would be  meaningless. 
In the opposite direction, \Presto\ cannot manage the specialization of higher-order functions that is achieved by \cite{HP14} since RWL is not a higher-order logic. 
Furthermore, our partial evaluator cannot deal with many other important Curry features such as monadic I/O or encapsulated search.
 In order to  provide some evidence of how \Presto\ performs on classical specialization examples, 
  we reproduced the standard {\em kmp-match} test\footnote{This test is often used to compare the
strength of specializers: the specialization of a semi-na\"ive pattern matcher for a fixed pattern into an 
efficient algorithm \cite{SGJ94}.}
 (i.e., without using any algebraic axioms), and we   benchmarked  the speed and specialization that is achieved by \Presto\/ on  
this program. This example is particularly interesting because it is a kind of transformation that neither
	(conventional) PE nor deforestation can perform automatically while conjunctive partial deduction\footnote{The transformation power  of conjunctive partial deduction is comparable to narrowing-driven partial evaluation as they both achieve unfold/fold-like program transformations such as tupling and deforestation within a fully automated  partial evaluation framework \cite{AFV98b}.}  of logic programs and positive
	supercompilation of functional programs can pass the test  \cite{AFV98b,JLM96}. The speedup achieved by \Presto\ on this test for an input list with 500,000 elements is $58$,
	 which is comparable to the speedup  of the last release of {\sf Victoria} (actually, the KMP specialized programs that are generated by \Presto\ and {\sf Victoria} are  essentially the same  and they both yield an optimal specialized  string matcher). 
Curry's partial evaluator also produces an optimal KMP specialization  and the achieved speedup  for  input lists of  $500,000$ elements is $12.22$, which is obtained by using an unlimited unfolding strategy that, however, may risk nontermination \cite{Peemoller17}. 
The corresponding specialization time  of Curry's partial evaluator reported in \cite{Peemoller17}  is $1.61$ seconds.

We   compared the  specialization times achieved by \Presto\ in comparison to {\sf Victoria} as follows.  
In order to reduce the  noise on the small  time for a single specialization, we benchmarked the  total specialization time 
 for a loop consisting of $1000$ specializations of the KMP program,  and then we divided the accumulated specialization time  by $1000$. 
The resulting specialization times of {\sf Victoria} and \Presto\   are   $6.6$ ms and  $5.6$ ms, respectively,
which shows that  \Presto\ outperforms {\sf Victoria} with a
performance improvement of  $15\%$.
This is noteworthy since the implementation of \Presto\ is considerably more ambitious, and it provides much 
greater coverage compared to the simpler approach of {\sf Victoria}.

%% file: 6-conclusion.tex
\section{Conclusion}\label{sec:conc}

The use of logic programming and narrowing for 
 system analysis has been advocated since  the narrowing-based NRL protocol analyzer \cite{Meadows96}, which was developed in Prolog. In this article, we have shown that   concurrent software analysis  is a promising  new application area for narrowing-based program specialization. We believe that Maude's partial  evaluator  \Presto\  is  not only relevant for 
the (multi-paradigm) logic programming community working on source code optimization, but  it can also be  a valuable tool  for anyone who is interested in 
the symbolic analysis and verification of concurrent systems. 
The main reason why  \Presto\ is so effective in this area
is that it   not only  achieves huge speedup    for  important classes 
of rewrite theories, but it can also cut down an infinite (folding variant)
narrowing space to a finite one for the underlying equational theory $\cE$. By doing this, any $\cE$-unification problem can be finitely solved and  symbolic, narrowing-based analysis with rules $R$ modulo $\cE$ can be effectively performed. Moreover, in many cases, the specialization process transforms a rewrite theory whose operators obey algebraic axioms, such as associativity, commutativity, and unity, 
into a much simpler rewrite theory with the same semantics and no structural axioms so that it can be run in an  independent rewriting infrastructure that does not support rewriting or narrowing modulo axioms. 
Finally, some costly    analyses that  may  require significant (or even unaffordable) resources  in both time and space, can  now be safely and effectively performed.
  
 The current specialization system goes beyond the scope of the original {\sf NPER} 
 framework of  \cite{ABES22-jlamp}, 
  which only works on unconditional rewrite theories.
 \Presto\ is  now  ``feature complete'' and stable. As such, it is not only suitable for experimental use but also for production use. 
Since \Presto\ is a fully automatic online specializer, its usage is simple and accessible  for  Maude users, 
and its interactive features make the  specialization process fully 
 checkable. For instance, the user can simply click on a specialized call to see its unfolding tree and all of the derived resultants.

Besides the applications outlined in this article, further applications could benefit from the optimization of variant generation that is achieved by  \Presto. For instance, an important number of applications (and tools) are currently based on narrowing-based variant generation: for example, the protocol analyzers Maude-NPA \cite{EMM09} 
 and Tamarin ~\cite{MSCB13}, Maude debuggers and program analyzers ~\cite{ABFS16-jlamp,ABFR12-lpar,ABR13-scp},
 termination provers, 
 variant-based satisfiability checkers, 
 coherence and confluence provers, 
 and different applications of symbolic reachability analysis 
 \cite{DEEM+20}.
Also,  as stated in \cite{LL00}, there are big advantages  for using the partial deduction approach to model checking  due to the built-in support of logic programming for non-determinism and unification. This approach might also be  of great value in the optimization of Maude's logical 
model checkers for infinite-state systems, where narrowing, equational unification, and equational abstractions play a primary role \cite{EM07,BM15}. We plan to investigate this in future research.

We also plan to implement new unfolding operators in \Presto\ that can achieve the total evaluation of  \cite{Meseguer20-jlamp} for the case of equational theories that are sufficiently complete modulo axioms   but do not have the FVP. This requires significant research effort since total evaluation relies on the  notion of constructor  variant
and there does not exist in the literature  a finite procedure to compute a finite, minimal, and complete set of  most general  constructor  variants for  theories that do not have the FVP, even if this set exists. Another interesting strand for further research  is dealing with the recently proposed Maude strategy language \cite{DEEM+20} in program specialization.

%% file: appendix.tex
\appendix
\section{Proofs of Technical Results}\label{app:appendix}

\begin{ftheorem}[strong completeness of topmost extension]{\ref{theo:sc}}
Let  $\cR=(\Sigma,E\uplus B,R)$ be a topmost modulo $Ax$ rewrite theory, with $Ax\in B$, and let $\cE=(\Sigma,E\uplus B)$
be  a convergent, finite variant equational theory where $B$-unification is decidable.  Let $\hat{\cR}=(\hat{\Sigma},E\uplus B,\hat{R})$
be the topmost extension of $\cR$ and let $G=(\exists X)\; t \longrightarrow^{*} t'$ be a reachability goal in $\cR$. 
If $\sigma$ is a solution for $G$ in $\cR$, then there exists a term $t''$ such that $\{t\}\leadsto_{\hat{R},\cE}^* \{t''\}$ with computed substitution $\theta$ and
$\sigma=_\cE\theta\eta$, where $\eta$ is an $\cE$-unifier of $t''$ and $t'$.
\end{ftheorem}

\begin{proof}
Let  $\cR=(\Sigma,E\uplus B,R)$ be a topmost modulo $Ax$ rewrite theory and $Ax\in B$, and let $\cE=(\Sigma,E\uplus B)$
be  a convergent, finite variant equational theory where $B$-unification is decidable.  

Let $\hat{\cR}=(\hat{\Sigma},E\uplus B,\hat{R})$
be the topmost extension of $\cR$.
Let $G=((\exists X)\; t \longrightarrow^{*} t')$ be a reachability goal for $\cR$. 
If $\sigma$ is a solution for $G$ in $\cR$, then $t\sigma\rightarrow_{R,\cE}^*t'\sigma$. By Proposition \ref{pr:TopmoduloAXTransf}, we have
$\{t\}\sigma\rightarrow_{\hat{R},\cE}^*\{t'\}\sigma$, which means that $\sigma$ is a solution in $\hat{\cR}$ 
for $\hat{G}=((\exists X)\; \{t\} \longrightarrow^{*} \{t'\})$. 

Since $\cE=(\Sigma,E\uplus B)$
is  a convergent, finite variant equational theory where $B$-unification is decidable, $\cE$-unification is decidable. Furthermore,
since   $\hat{\cR}$ is topmost and $\cE$-unification is decidable, $\leadsto_{\hat{R},\cE}$ is complete and thus there exist
$\{t\}\leadsto_{\hat{R},\cE}^* \{t''\}$ with a computed substitution $\theta$ and an $\cE$-unifier $\eta$ such that $\theta\eta=_\cE \sigma$. 
\end{proof}

\begin{fproposition}[correctness and completeness of the topmost extension]{\ref{pr:TopmoduloAXTransf}}
Let $\cR=(\Sigma,E\uplus B,R)$, with $\cE=(\Sigma, E\uplus B)$, be a topmost modulo $Ax$ theory and  let $\hat{\cR}$ be
the topmost extension of $\cR$.
For any terms $t_i$ and $t_f$ of sort $\mathit{Config}$, $t_i\rightarrow_{R,\cE}^{*}t_f$ iff  
$\{t_i\}\rightarrow_{\hat{R},\cE}^{*}\{t_f\}$.
\end{fproposition}

\begin{proof}

\noindent The case when $Ax=$ACU has been stated in Lemma 5.3 of \cite{MT07}. 
Here, we prove the case when $Ax=$AC, which involves two extension rewrite rules, namely, $(\{X\otimes \lambda\} \Rightarrow \{X\otimes \rho\}\mbox{ if } C)$ and $(\{\lambda\} \Rightarrow \{\rho\}\mbox{ if } C)$. The remaining cases are 
straightforward
adaptations of the following proof scheme.

\noindent 
We have to prove the following two implications for the case when $Ax=$AC: 
\begin{description}
\item[$(\rightarrow)$] for any term $t_i$ and $t_f$ of sort $\mathit{Config}$, $t_i\rightarrow_{R,\cE}^{*}t_f$ implies $\{t_i\}\rightarrow_{\hat{R},\cE}^{*}\{t_f\}$;
\item[$(\leftarrow)$] for any term $t_i$ and $t_f$ of sort $\mathit{Config}$, $\{t_i\}\rightarrow_{\hat{R},\cE}^{*}\{t_f\}$ implies $t_i\rightarrow_{R,\cE}^{*}t_f$.
\end{description}
\begin{description}
\item[$(\rightarrow)$] Assume that $t_i\rightarrow_{R,\cE}^{*}t_f$, where $t_i$ and $t_f$ are arbitrary terms of  sort $\mathit{Config}$. Then, $t_i\rightarrow_{R,\cE}^{*}t_f$ has the form
$$ t_i=t_0 \rightarrow_{R,\cE} \ldots \rightarrow_{R,\cE} t_{n-1}\rightarrow_{R,\cE} t_n=t_f, \mbox{for some natural number } n\geq 0.$$
We proceed by induction on the length  $n$ of the rewriting sequence $t_i\rightarrow_{R,\cE}^{*}t_f$.
\\\\
\noindent $n=0.$ Immediate since there are no rewrite steps.
\\\\
\noindent $n>0.$ By induction hypothesis, we have 
\begin{equation}\label{proof:ii} 
t_i=t_0\rightarrow_{R,\cE}^{*}t_{n-1} \mbox{ implies } \{t_i\}=\{t_0\}\rightarrow_{\hat{R},\cE}^{*}\{t_{n-1}\}.
\end{equation}
Thus, in order to prove $(\rightarrow)$, we just need to show 
\begin{equation}\label{proof:rs}
\{t_{n-1}\}\rightarrow_{\hat{R},\cE}\{t_{n}\},
\end{equation}
whenever  $t_{n-1} \rightarrow_{R,\cE}t_{n}$.
The computation step $t_{n-1} \rightarrow_{R,\cE}t_{n}$ in the rewrite theory $\cR$ can be expanded into the following rewrite sequence
$$t_{n-1} \stackrel{r,\sigma,w}{\rightarrow}_{\!\!R,B} \tilde{t}_{n-1} \rightarrow_{\vec{E},B}
^*\tilde{t}_{n-1}\!\downarrow_{\vec{E},B} = t_n$$
where $r = (\lambda\: \Rightarrow\: \rho\ \mbox{if}\ C)\in R$. 
Here, we distinguish two cases according to the value of the position $w\in\pos(t_{n-1})$: $w=\Lambda$ and $w\neq\Lambda$. 
\\\\
\noindent $(w=\Lambda)$ In this case,  $t_{n-1}=_B\lambda\sigma$ and $\tilde{t}_{n-1}=_B\rho\sigma$ by the definition of ${\rightarrow}_{R,B}$. Furthermore, since $\cR$ is topmost modulo $Ax$,  $\lambda\sigma$ and $\rho\sigma$ have sort $\mathit{Config}$.
From these facts, it immediately follows that
$$\{t_{n-1}\} =_B \{\lambda\sigma\} \stackrel{\hat{r},\sigma,\Lambda}{\rightarrow}_{\!\!\hat{R},B}  \{\rho\sigma\}=_B \{\tilde{t}_{n-1}\} \rightarrow_{\vec{E},B}
^*\{\tilde{t}_{n-1}\}\!\downarrow_{\vec{E},B} = \{t_n\}$$
with $\hat{r}=\{\lambda\} \Rightarrow \{\rho\} \mbox{ if } C\in \hat{R}$.
Hence, $\{t_{n-1}\} \rightarrow_{\hat{R},\cE} \{t_n\}$ when $w=\Lambda$.  
\\\\
\noindent $(w\neq\Lambda)$ Since $\cR$ is topmost modulo $Ax$ and $w\neq\Lambda$, there exist
 $u_i\in \cT_{\Sigma}(\cX)$ of sort $\mathit{Config}$, $i=1,\ldots,k$, with $k>1$ such that 
$$t_{n-1} = u_1 \otimes \ldots \otimes u_k$$ 
and  $t_{n-1}\in \cT_{{\Sigma}}(\cX)$ of sort $\mathit{Config}$.
Now, since $t_{n-1} \stackrel{r,\sigma,w}{\rightarrow}_{\!\!R,B} \tilde{t}_{n-1}\rightarrow_{\vec{E},B}
^* t_n
$, with $Ax=AC$ and $r=(\lambda\Rightarrow\rho\mbox{ if } C)$,
\begin{align*}
t_{n-1} = u_1 \otimes \ldots \otimes u_k &=_{AC} u_{\pi(1)}\otimes \ldots\otimes u_{\pi(m)}\otimes 
\lambda\sigma\\
&\stackrel{r,\sigma,w}{\rightarrow}_{\!\!R,B} u_{\pi(1)}\otimes \ldots\otimes u_{\pi(m)}\otimes 
\rho\sigma  \rightarrow_{\vec{E},B}
^* t_n
\end{align*}
where $1\leq m \leq k-1$, and $\pi\colon\{1,...,m\}\to\{1,\ldots,k\}$ is an injective function that selects a permutation of $m$ $u_i$'s within $u_1 \otimes \ldots \otimes u_k$. 
Hence, we can build the following rewrite sequence
\begin{align*}
\{t_{n-1}\} = \{u_1 \otimes \ldots \otimes u_k\} &=_{AC} \{u_{\pi(1)}\otimes \ldots\otimes u_{\pi(m)}\otimes 
\lambda\sigma\}\\
&\stackrel{\hat{r},\hat{\sigma},\Lambda}{\rightarrow}_{\!\!\hat{R},B} \{u_{\pi(1)}\otimes \ldots\otimes u_{\pi(m)}\otimes \rho\sigma\}  \rightarrow_{\vec{E},B}^* \{t_n\}
\end{align*}
with $\hat{\sigma}=\sigma\cup\{X\mapsto u_{\pi(1)}\otimes \ldots\otimes u_{\pi(m)}\}$, and $\hat{r} =  (\{X\otimes \lambda\} \Rightarrow \{X\otimes \rho\}\mbox{ if } C) \in \hat{R}$.
This proves that $\{t_{n-1}\}\rightarrow_{\hat{R},\cE}\{t_{n}\}$ also in the case when $w\neq\Lambda$.

Finally, by using the induction hypothesis \ref{proof:ii} and the rewrite step \ref{proof:rs}, we
easily derive the implication $(\rightarrow)$. 
\item[$(\leftarrow)$] Assume that $\{t_i\}\rightarrow_{\hat{R},\cE}^{*}\{t_f\}$, where $t_i$ and $t_f$ are arbitrary terms of  sort $\mathit{Config}$. Then, $\{t_i\}\rightarrow_{\hat{R},\cE}^{*}\{t_f\}$ is of the form
$$ \{t_i\}=\{t_0\}\rightarrow_{\hat{R},\cE}\ldots \rightarrow_{\hat{R},\cE} \{t_{n-1}\}\rightarrow_{\hat{R},\cE} \{t_n\}=\{t_f\}$$ for some natural number  $n\geq 0.$
We proceed by induction on  the length $n$ of the computation $\{t_i\}\rightarrow_{\hat{R},\cE}^{*}\{t_f\}$.
\\\\
\noindent $n=0.$ Immediate, since there are no rewrite steps.
\\\\
\noindent $n>0.$ This case is analogous to the proof of the inductive step of Case $(\rightarrow)$.
By induction hypothesis, we have 
\begin{equation}\label{proof:ii2} 
\{t_i\}=\{t_0\}\rightarrow_{\hat{R},\cE}^{*}\{t_{n-1}\} \mbox{ implies } t_i=t_0\rightarrow_{R,\cE}^{*} t_{n-1}.
\end{equation}
Therefore, it suffices to show that $t_{n-1}\rightarrow_{R,\cE}^{*} t_{n}$ and combine this result with the induction hypothesis to finally prove Case $(\leftarrow)$. 

By hypothesis,  $\{t_{n-1}\} \rightarrow_{\hat{R},\cE}\{t_{n}\}$, which can  be expanded into the following rewrite sequence
\begin{equation}\label{proof:mStep}\{t_{n-1}\} \stackrel{\hat{r},\hat{\sigma},\Lambda}{\rightarrow}_{\!\!\hat{R},B} \{\tilde{t}_{n-1}\} \rightarrow_{\vec{E},B}
^*\{\tilde{t}_{n-1}\}\!\downarrow_{\vec{E},B} = \{t_n\}
\end{equation}
where $\hat{r}\in\hat{R}$, and $\{t_{n-1}\},\{\tilde{t}_{n-1}\},\{t_{n}\}\in\cT_{\hat{\Sigma}}(\cX)$  of sort $\mathit{State}$. Observe that the first rewrite step of the rewrite sequence (\ref{proof:mStep}) must occur at position $\Lambda$, since the rewrite theory
$\hat{\cR}$ is topmost.

Here, we distinguish two cases according to the form of the rewrite rule $\hat{r}\in\hat{R}$
applied in $\{t_{n-1}\} \stackrel{\hat{r},\hat{\sigma},\Lambda}{\rightarrow}_{\!\!\hat{R},B} \{\tilde{t}_{n-1}\}$.

By Definition \ref{def:transfTopmost}, $\hat{r}$ is either $\{\lambda\} \Rightarrow \{\rho\}\mbox{ if } C$ or $\{X\otimes \lambda\} \Rightarrow \{X\otimes \rho\}\mbox{ if } C$, as $Ax=AC$ and $\{t_{n-1}\},\{\tilde{t}_{n-1}\}\in\cT_{\hat{\Sigma}}(\cX)$  of sort $\mathit{State}$.
\\\\
\noindent \textbf{Case} $(\hat{r}=(\{\lambda\} \Rightarrow \{\rho\}\mbox{ if } C))$. In this case, $\{t_{n-1}\} \stackrel{\hat{r},\hat{\sigma},\Lambda}{\rightarrow}_{\!\!\hat{R},B} \{\tilde{t}_{n-1}\}$ assumes the following form:
$$
\{t_{n-1}\} =_{AC} \{\lambda\sigma\}\stackrel{\hat{r},\sigma,\Lambda}{\rightarrow}_{\!\!\hat{R},B} 
\{\rho\sigma\}=_{AC} \{{\tilde{t}_{n-1}}\}.$$
Now, by Definition \ref{def:transfTopmost}, $\lambda\sigma$ and $\rho\sigma$ are terms of sort $\mathit{Config}$; thus, we can also apply $r=(\lambda \Rightarrow \rho\mbox{ if } C)\in R$ to 
$\lambda\sigma$, thereby obtaining the following computation
\begin{align*}
t_{n-1} =_{AC} \lambda\sigma\stackrel{r,\sigma,\Lambda}{\rightarrow}_{\!\!R,B} \rho\sigma 
\rightarrow_{\vec{E},B}^*(\rho\sigma\downarrow_{\vec{E},B}) = t_n
\end{align*}
which corresponds to $t_{n-1}\rightarrow_{R,\cE} t_n$ when $\hat{r}=(\{\lambda\} \Rightarrow \{\rho\}\mbox{ if } C)$.
\\\\
\noindent \textbf{Case} $(\hat{r}=(\{X\otimes \lambda\} \Rightarrow \{X\otimes \rho\}\mbox{ if } C))$. In this case,
$\{t_{n-1}\} \stackrel{\hat{r},\hat{\sigma},\Lambda}{\rightarrow}_{\!\!\hat{R},B} \{\tilde{t}_{n-1}\}$ must have the following form:
\begin{align*}
\{t_{n-1}\} =_{AC} \{u \otimes \lambda\hat{\sigma}\} \stackrel{\hat{r},\hat{\sigma},\Lambda}{\rightarrow}_{\!\!\hat{R},B} \{u \otimes \rho\hat{\sigma}\} =_{AC} \{\tilde{t}_{n-1}\} 
\end{align*}
where $u, \lambda\hat{\sigma}, \rho\hat{\sigma}\in\cT(\Sigma,\cV)_{\mathit{Config}}$, and 
$\hat{\sigma}=\{X\mapsto c\}\cup\sigma$, for some substitution $\sigma$.

Now, by Definition \ref{def:transfTopmost}, variable $X$ does not occur in either  $\lambda$ or $\rho$; this implies that $\lambda\hat{\sigma}=\lambda\sigma$ and $\rho\hat{\sigma}=\rho\sigma$. Therefore, we can construct the following computation:
\begin{align*}
t_{n-1} =_{AC} u \otimes \lambda\hat{\sigma} =  u \otimes \lambda\sigma\stackrel{r,\sigma,w} {\rightarrow}_{\!\!R,B} u \otimes \rho\sigma = u\otimes \rho\hat{\sigma} =_{AC} \tilde{t}_{n-1} \rightarrow_{\vec{E},B}^* t_n
\end{align*}
where $r=(\lambda\Rightarrow\rho\mbox{ if } C)\in R$ and $w\in\pos(u \otimes \lambda\sigma)$ is the position of the term $\lambda\sigma$ inside $u \otimes \lambda\sigma$.
Hence, $t_{n-1}\rightarrow_{R,\cE} t_n$ even in the case when $\hat{r}=(\{X\otimes \lambda\} \Rightarrow \{X\otimes \rho\}\mbox{ if } C)$.
\end{description}
\end{proof}